\RequirePackage{fix-cm}
\documentclass[twocolumn,epjc3]{svjour3}  
\smartqed  % flush right qed marks, e.g. at end of proof
\RequirePackage{graphicx}
\RequirePackage[colorlinks,citecolor=blue,urlcolor=blue,linkcolor=blue]{hyperref} 

\usepackage{subfigure} 
\usepackage{braket}     
\usepackage{amssymb}
\usepackage{upgreek}     
\usepackage{siunitx}
\usepackage{color}
\usepackage{ifthen}
\usepackage{multirow}
\newboolean{makefigures}                %   declaration
\setboolean{makefigures}{true}          %   definition     on == true
\usepackage{amsmath}
\usepackage{pdflscape}
\usepackage{varioref}
\usepackage{fancyhdr}
\journalname{Eur. Phys. J. C}
\usepackage{lineno}
\begin{document}

\title{Characterization of 30 $^{76}$Ge enriched Broad Energy Ge detectors for \textsc{GERDA} Phase II}

\titlerunning{thirty enriched BEGe detectors}   % if too long for running head

\author{
The \mbox{\protect{\sc{Gerda}}} collaboration\thanksref{corrauthor}
        \and  \\[4mm]
M.~Agostini\thanksref{TUM} \and
A.M.~Bakalyarov\thanksref{KU} \and
E.~Andreotti\thanksref{GEEL,nowAndre} \and
M.~Balata\thanksref{ALNGS} \and
I.~Barabanov\thanksref{INR} \and
L.~Baudis\thanksref{UZH} \and
N.~Barros\thanksref{DD,nowBarros} \and
C.~Bauer\thanksref{HD} \and
E.~Bellotti\thanksref{MIBF,MIBINFN} \and
S.~Belogurov\thanksref{ITEP,INR,alsoMEPHI}
% ...
}
%%%%%%%%%%%%%%%%%%%%%%%%%%%%%%%%%%%%%%%%%%%%%%%%%%%%%%%%%%%%%%%%%%%%%%%%%%%%%

\authorrunning{the \textsc{Gerda} collaboration}
\thankstext{corrauthor}{INFN Laboratori Nazionali del Gran Sasso, Italy.\\
\emph{Correspondence},
                                email: gerda-eb@mpi-hd.mpg.de}
\thankstext{nowAndre}{\emph{present address:} University College Leuven-Limburg,
  Expertisecel Art of  Teaching, Diepenbeek, Belgium}
\thankstext{nowBarros}{\emph{present address:} Dept. of Physics and Astronomy,
  U. of Pennsylvania, Philadelphia, Pennsylvania, USA} 
\thankstext{alsoMEPHI}{\emph{also at:} NRNU MEPhI, Moscow, Russia}
\thankstext{nowBenatoMarco}{\emph{present address:} University of California,
  Berkeley, USA}
\thankstext{nowBoro}{\emph{also at:} The Henryk Niewodniczanski Institute of
  Nuclear Physics PAS, Krakow, Poland}
\thankstext{alsoLev}{\emph{also at:} Moscow Inst. of Physics and Technology,
  Russia}
\thankstext{nowIKZ}{\emph{present address:} Leibniz-Institut f{\"u}r
  Kristallz{\"u}chtung, Berlin, Germany}
\thankstext{alsoKlimenko}{\emph{also at:} Int. Univ. for Nature, Society and
    Man ``Dubna'', Russia} 
\thankstext{nowCarlton}{\emph{present address:} Carleton University, Ottawa,
  Canada}
\thankstext{nowMacolino}{\emph{present address:} LAL, CNRS/IN2P3,
  Universit{\'e} Paris-Saclay, Orsay, France} 
\thankstext{nowVici}{\emph{present address:} CEA, Saclay,  IRFU,
  Gif-sur-Yvette, France}
%%%%%%%%%%%%%%%%%%%%%%%%%%%%%%%%%%%%%%%%%%%%%%%%%%%%%%%%%%%%%%%%%%%%%%%%%%%%%%%
\institute{%
INFN Laboratori Nazionali del Gran Sasso, LNGS, Assergi, Italy\label{ALNGS} \and
INFN Laboratori Nazionali del Gran Sasso and Universit{\`a} degli Studi
dell'Aquila, L'Aquila, Italy\label{AQU} \and
INFN Laboratori Nazionali del Sud, Catania, Italy\label{CAT} \and
Institute of Physics, Jagiellonian University, Cracow, Poland\label{CR} \and
Institut f{\"u}r Kern- und Teilchenphysik, Technische Universit{\"a}t Dresden,
      Dresden, Germany\label{DD} \and
Joint Institute for Nuclear Research, Dubna, Russia\label{JINR} \and
European Commission, JRC-Geel, Geel,  Belgium\label{GEEL} \and
Max-Planck-Institut f{\"u}r Kernphysik, Heidelberg, Germany\label{HD} \and
Dipartimento di Fisica, Universit{\`a} Milano Bicocca,
     Milano, Italy\label{MIBF} \and
INFN Milano Bicocca, Milano, Italy\label{MIBINFN} \and
Dipartimento di Fisica, Universit{\`a} degli Studi di Milano e INFN Milano,
    Milano, Italy\label{MILUINFN} \and
Institute for Nuclear Research of the Russian Academy of Sciences,
    Moscow, Russia\label{INR} \and
Institute for Theoretical and Experimental Physics, NRC ``Kurchatov Institute'',
    Moscow, Russia\label{ITEP} \and
National Research Centre ``Kurchatov Institute'', Moscow, Russia\label{KU} \and
Max-Planck-Institut f{\"ur} Physik, M{\"u}nchen, Germany\label{MPIP} \and
Physik Department and Excellence Cluster Universe,
    Technische  Universit{\"a}t M{\"u}nchen, Germany\label{TUM} \and
Dipartimento di Fisica e Astronomia dell'Universit{\`a} di Padova,
    Padova, Italy\label{PDUNI} \and
INFN  Padova, Padova, Italy\label{PDINFN} \and
Physikalisches Institut, Eberhard Karls Universit{\"a}t T{\"u}bingen,
    T{\"u}bingen, Germany\label{TU} \and
Physik Institut der Universit{\"a}t Z{\"u}rich, Z{\"u}rich,
    Switzerland\label{UZH}
}

\date{}

\maketitle

\begin{abstract}
The GERmanium Detector Array (\textsc{Gerda}) is a low background experiment located at the Laboratori Nazionali del Gran Sasso in Italy, which searches for neutrinoless double beta decay of $^{76}$Ge into $^{76}$Se+2e$^-$. \textsc{Gerda} has been conceived in two phases. Phase~II, which started in December 2015, features several no\-velties including 30 new Ge detectors. These were manu\-factured according to the Broad Energy Germanium (BEGe) detector design that has a better background discrimination capability and energy resolution compared to formerly widely-used types. Prior to their installation, the new BEGe detectors were mounted in vacuum cryostats and characterized in detail in the \textsc{Hades} underground laboratory in Belgium. This paper describes the properties and the overall performance of these detectors during operation in vacuum. The cha\-racterization campaign provided not only direct input for \textsc{Gerda} Phase~II data collection and analyses, but also allowed to study detector phenomena, detector correlations as well as to test the strength of pulse shape simulation codes.
\end{abstract}

%%%%%%%%%%%%%%%%%%%%%%%%%%%%%%%%%%%%%%%%%%%%%%%%%%%%%%%%%%%%%%%%%%%%%%%%%%%%%%%%
\section{Introduction}
 \label{sec:introduction} 
 
The search for neutrinoless double beta (0$\nu\beta\beta$) decay of $^{76}$Ge with germanium (Ge) detectors has a 50-year-long tradition. While the former experiments that were concluded in 1967 \cite{bib:1967_fiorini}, 2002 \cite{bib:2003_igex} and 2003 \cite{bib:2004_hdm}, exclusively used the widespread semi-coaxial detector design, the more recent \textsc{Gerda} \cite{bib:gerda_intro1}\cite{bib:gerda_intro2} and \textsc{Majorana} \cite{bib:majorana_intro2} experiments have intensively searched for new Ge detector designs aiming at improving the background suppression compared to the semi-coaxial type. This was partly possible due to a strong cooperation with leading Ge detector manufacturers worldwide. We selected a slightly mo\-dified version of the Broad Energy Ge (BEGe) design \cite{bib:luke_firstbegeprod} offered by the company Canberra~\cite{bib:canberra}, now part of Mirion Technologies~\cite{bib:mirion}. Compared to the semi-coaxial type, the average BEGe mass is typically by a factor 2-3 smaller, but its design was found to lead to an improved energy resolution and superior background rejection capability via pulse shape analysis of the detector signals \cite{bib:deplBEGe_firstpaper}. After a test phase based on BEGe detectors of natural isotopic composition and made from material with reduced $^{76}$Ge isotope fraction
\cite{bib:deplBEGe}, 30 new BEGe diodes made from Ge with enriched $^{76}$Ge isotope fraction were produced in two batches. Prior to their installation at \textsc{Gerda}'s experimental site at the Laboratori Nazionali del Gran Sasso (LNGS) in Assergi, Italy, the detectors underwent extensive acceptance and characterization tests in the \textsc{Hades} underground laboratory in Mol, Belgium \cite{bib:hades}. This is located only 20\,km from the detector manufacturer site in Olen and it provided underground storage whenever the detectors were not processed. This was required to avoid cosmic-ray activation of the 
Ge material. For the detector survey, a proper infrastructure called \textsc{Heroica} has been installed \cite{bib:heroica_infra} capable of testing several BEGe detectors at the same time, partly in an automatized scanning modus.\\
\indent A previous publication \cite{bib:first-hades-paper} discussed the production of the first batch of enriched BEGe detectors with focus on the isotopic enrichment process, detector production, activation history and operation in vacuum as well as in the \textsc{Gerda} liquid argon cryostat. This was achieved by means of five BEGe detectors belonging to the first batch that had already been operated test-wise during Phase~I of the experiment.\\
\indent The present paper provides a full description of the characterization test results obtained with the 30 new BEGe detectors during their operation in vacuum cryo\-stats in \textsc{Hades}. Results already presented in \cite{bib:first-hades-paper} were revised and partly improved. Chapter~\ref{sec:detector-production-and-geometries} describes the main properties of the crystals used for the manufacturing of the diodes. It also provides an introduction to pulse shape simulation codes that are useful not only to optimize the Ge crystal slice cut, but also for a better understanding of different phenomena observed in this work. Chapter~\ref{sec:depletion-and-operational-voltage} is dedicated to the electrical depletion behavior of the detectors, including some peculiarities and observed parameter correlations. It also introduces a useful methodology to refine the nominal ope\-rational voltage values demonstrating the advantages for \textsc{Gerda} data collection. Chapter~\ref{sec:energy-resolution} describes the \linebreak energy resolution of the detectors and searches for dependencies on other detector quantities. Chapter~\ref{sec:full-charge-collection-depth-and-active-volume} contains the results of a high precision study of the full charge collection depths and active volumes of the new BEGe detectors. These are essential ingredients for \linebreak \textsc{Gerda}'s exposure calculation. Chapter~\ref{sec:pulse-shape-behavior} deals with the pulse shape performance of the BEGe detectors. It examines the capability to reject $\gamma$-radiation and the possibility of fine-grained surface scans to see local effects that are partly affected by the crystal lattice of Ge. Summary and conclusions are given in Chapter~\ref{sec:summary}.

%%%%%%%%%%%%%%%%%%%%%%%%%%%%%%%%%%%%%%%%%%%%%%%%%%%%%%%%%%%%%%%%%%%%%%%%%%%%%%%%
\section{Detector crystals: selection and properties}
 \label{sec:detector-production-and-geometries} 

\subsection{Production of \textsc{Gerda} Phase~II BEGe detectors}
\label{sec:Manufacturing-GERDA-BEGe-detectors}

\begin{figure*}[t]
\begin{center}
\includegraphics[scale=0.50]{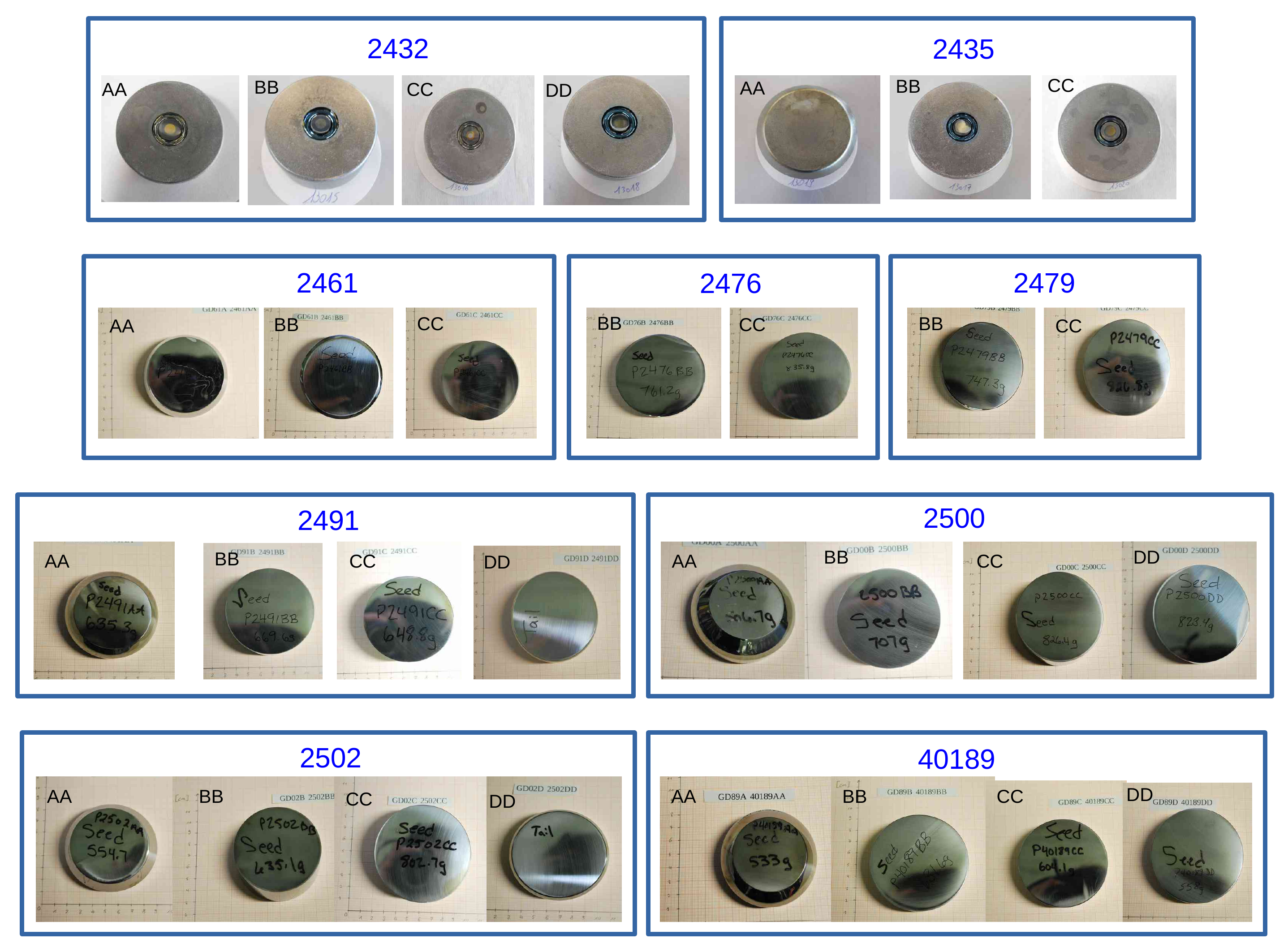}
\caption{Full inventory of crystal slices / diodes belonging to the \textsc{Gerda}  Phase~II BEGe detector production. Crystals / diodes obtained from the same ingot are framed in blue. The GD32 and GD35 detector series belonging to the 1$^{st}$ batch are depicted in their final diode form (row 1), while the other 7 series from the 2$^{nd}$ batch are shown as crystal slices prior to diode conversion (rows 2 to 4).}
\label{fig:30BEGe-collage_2}
\end{center}
\end{figure*}

\paragraph{\bf Manufacturing:} The company Canberra Industries \linebreak Inc.~\cite{bib:canberra_usa} in Oak Ridge (TN), USA (short: Canberra Oak Ridge), was selected for the Ge crystal growing process. Before that the Ge was enriched to 88\% in $^{76}$Ge at the Joint Stock Company ``Production Association Electrochemical Plant'' (ECP) in Zelenogorsk, Russia, and purified at PPM Pure Metals in Langelsheim, Germany, reaching 35.5\,kg of 6N grade material purified Ge (cf. \cite{bib:first-hades-paper}). Using different pul\-lers, nine crystal ingots with a typical length of $\sim$(18-25)\,cm were grown. Out of these, 30 crystal slices were successfully cut, totaling a mass of 20.77\,kg. The crystal slice cutting was optimized following two criteria: firstly, obtaining the largest possible diodes out of one ingot, and secondly producing the lowest possible amount of residual material. For a given ingot the first point was obtained by selecting the largest possible diode height while avoiding an excessive net impurity concentration gradient from top to bottom. The second criterion was achieved by considering conical tail and seed ends of the ingots as well. As a result, 21 crystal slices are cylindrical, whereas 9 are conical or even double-conical.\\
\indent The company Canberra Semiconductors N.V.~\cite{bib:canberra_belgium} in Olen, Belgium (short: Canberra Olen), was assigned to convert the crystal slices into working diodes following the BEGe design. The crystals were processed in two batches consisting of 7 and 23 slices each. In ge\-neral, the obtained diodes conserved the overall crystal slice dimensions. Only a small mass loss was induced by the creation of the insulating groove that separates the read-out p+ electrode from the n+ contact. Only in two problematic cases the mass loss was larger (cf. Section \ref{sec:Dimensions-and-masses-of-BEGe-detectors}). In the end, the 30 diodes amounted to 20.02\,kg.
\paragraph{\bf Nomenclature:} The full inventory of the 30 \textsc{Gerda} Phase~II BEGe detectors is depicted in Figure~\ref{fig:30BEGe-collage_2}. As indicated by the blue frames, 2 to 4 slices were obtained from one single ingot. For each slice, Canberra Oak Ridge provided a unique identifier consisting of two parts: the 4-digit serial number of a growth process with a certain puller and the relative seed- to tail-end position of a slice in terms of AA, BB, CC or DD. A few examples are 2432AA, 2476CC and 40189AA. We formed new distinct names that include both pieces of information, i.e. GD32A, GD76C and GD89A for the mentioned cases. This nomenclature will be adopted in all following chapters.
\paragraph{\bf Net impurity concentrations:} The manufacturer \linebreak cut thin slices at around the seed- and tail-end of each single crystal ingot and measured the impurity concentrations $N_{a-d}:=$ $|N_a$-$N_d|$ via the Hall effect. Herein, $N_a$ and $N_d$ are the acceptor and donor concentrations. Further, at several axial positions of an ingot resistivity measurements were performed to determine the gradient of $N_{a-d}$ and to establish the appro\-ximate cut positions. The overall measurement precision of the $N_{a-d}$ values was quoted with $\pm$10\% \cite{bib:patrick-vermeulen}. Within a non-dis\-closure agreement, we received the $N_{a-d}$ values for all crystal slices and used them for the studies presented below. In general, the $N_{a-d}$ values lie in the range [0.5,3]$\cdot10^{10}$/cm$^{3}$, which is ideal for high purity Ge \linebreak(HPGe) detector fabrication \cite{bib:knoll}. Only in the case of GD02D, the $N_{a-d}$ value was not fully satisfactory. \linebreak Therefore, the electric field strength inside this detector is expected to be deteriorated and need special attention.\\
\indent Even though the $N_{a-d}$ values vary from ingot to ingot, their absolute values typically increase from seed to tail of a single ingot. Thus, crystal slices of the same position in two ingots might differ in $N_{a-d}$, while slices of different positions in the same two ingots can have very similar $N_{a-d}$ values and gradients. As a consequence, one should not expect correlations of the electric field strength and depletion voltage with the relative slice position, but -- if at all -- only with the $N_{a-d}$ values. Among other items, the following chapters will address this issue. 

\subsection{Dimensions and masses of the BEGe detectors}
\label{sec:Dimensions-and-masses-of-BEGe-detectors}

\paragraph{\bf Dimensions:} The dimensions of the 30 \textsc{Gerda}\linebreak Phase~II BEGe diodes were measured by Canberra Olen. In all cases, the diodes were treated as completely symmetric cylinders and accordingly only one height and one diameter per detector were quoted. Even though it could not be directly measured after diode production, the manufacturer stated, that the groove between the p$+$ and n$+$ electrodes is equal for all detectors, with an inner and outer diameter of 15 and 21\,mm and a depth of $\approx$2.0\,mm (cf. Figure \ref{fig:diode-shapes}).\\ 
\indent We performed a precise re-measurement, which included multiple measurements of diameters and heights at typically 4-5 different azimuth angles with respect to the $z$-axis (height) of a diode. Table~\ref{table:dimensions-masses} summarizes the mean values of the BEGe diode outer dimensions including their uncertainties as measured and used by us. The underlying terminology is explained in Figure~\ref{fig:diode-shapes}. The average diameter D1 and average height H1 of all 30 diodes are 72.8 and 29.6\,mm with a standard deviation (SD) of $\pm$3.9 and $\pm$3.2\,mm, respectively.\\
\indent We considered detectors with conical shape separately. However, our classification distinguishes only between perfect symmetric cylindrical and conical diodes. This simplification facilitates the implementation of the individual diode geometry in Monte Carlo (MC) simulation models (cf. Chapter \ref{sec:full-charge-collection-depth-and-active-volume}). By doing so, however, MC simulations omit a few existing facts:
\begin{figure}[t]
\begin{center}
\includegraphics[scale=0.40]{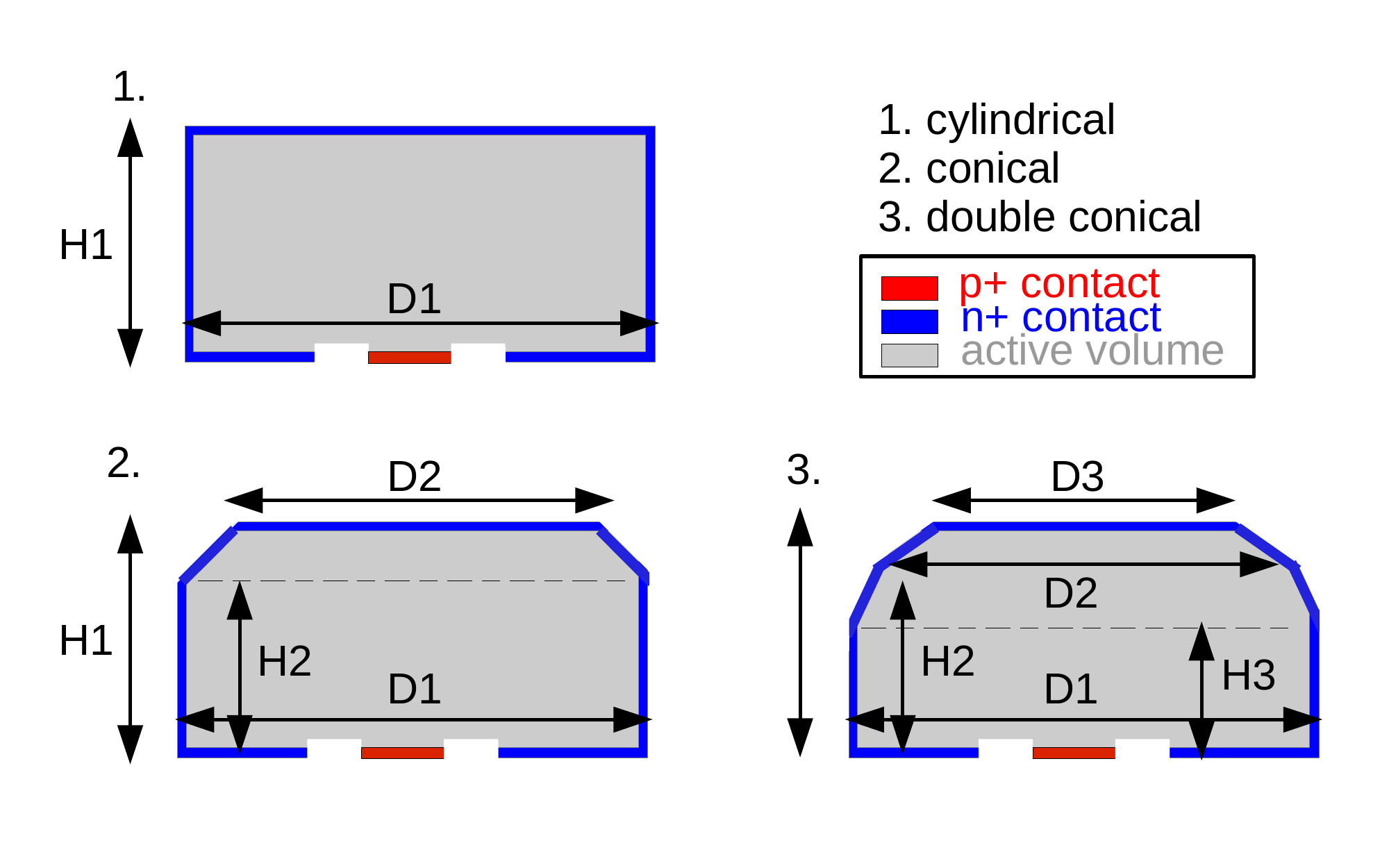}
\caption{\textsc{Gerda} Phase~II BEGe diodes are either 1.) cylindrical, 2.) single-conical, or 3.) double-conical. Overall heights are denoted with H1, while corner heights with H2 and H3. The overall diameter is distinguished from corner diameters by D1 vs. D2, D3.}
\label{fig:diode-shapes}
\end{center}
\end{figure}
\begin{itemize}
\item Some detectors have a slightly oval base and/or a small variation in diameter or in height. The extreme cases are detectors GD79B (diameter variation up to 0.4\%) and GD89A (height variation up to 4\%). 
\item Detectors GD61B, GD91D and GD32D are classified as cylindrical shaped, even though the original crystal slices had a slightly conical shape.
\item Detectors GD61A, GD91A and GD00A, which are classified as conical shaped, are based on double-conical crystal slices. 
\item Detector GD89D, which is classified as cylindrical shaped, has a deformed shape (chopped-off edge and different heights).
\end{itemize}
All these detectors are asterisked in Table~\ref{table:dimensions-masses} and had to be treated with caution in analyses depending on the geometry of the detector volume.
\paragraph{\bf Masses:} We determined the masses of the diodes with a precision of $\pm$1\,g. The results are reported in Table~\ref{table:dimensions-masses}. The average diode mass is 667\,g and the SD of the detector mass distribution is $\pm$115\,g. The detector mass of all 30 \textsc{Gerda} Phase~II BEGe detectors $M=\sum_{i=1}^{30} M_i$ is (20.024$\pm$0.030)\,kg. Herein, the $\pm 1 $\,g uncertainty from weighing was assumed to be correlated for all detectors. Neglecting the problematic detector GD02D (cf. Section \ref{sec:results-active-volume}), the total detector mass $M=\sum_{i=1}^{29} M_i$ is reduced to (19.362$\pm$0.029)\,kg.\\
\indent The measurements of the single diode masses $M_m$ were also useful to compensate the geometry simplifications proposed for MC simulations. For this purpose, the analytical mass $M_a$=V$\cdot\rho$ was calculated using the mean dimensions and the independently determined density of the Ge crystals enriched in $^{76}$Ge, which is $\rho=$ 5.55\,g/cm$^{3}$ \cite{bib:first-hades-paper}. Then the difference $\Delta M := (M_a-M_m)/M_a$ was calculated. From $\Delta M$ one deduces the volume difference $\Delta V$ and from here a correction on the diameter and height needed to fulfill the condition $\Delta M\rightarrow \Delta M'\approx 0$. That way, it was possible to minimize the systematic uncertainty in MC simulations arising from the diode dimension simplification. As shown in Chapter \ref{sec:full-charge-collection-depth-and-active-volume}, this will be of major importance for the determination of the full charge collection depths and active volumes of the detectors.

\subsection{Pulse shape simulations}
\label{sec:pulse-shape-simulations} 

The crystal slice cutting applied by the manufacturer was done in close cooperation with us. We used the net impurity concentrations $N_{a-d}$ provided by Canberra Oak Ridge and simulated the expected charge drift and signal generation on the read-out electrodes for slices of different thicknesses. The optimized cuts were done after feedback from our calculations. These were based on the Multi Geometry Simulation (\texttt{MGS}) software \cite{bib:MGS-software}. \texttt{MGS} has been also used for the prototype BEGe detector measurement campaign \cite{bib:diss_agostini_2013}.\\
\indent Within the characterization campaign of the 30\linebreak \textsc{Gerda} Phase II BEGe detectors, we looked for alternative field calculation and pulse shape simulation codes able to combine requirements with several advantages: easy and user-friendly adaptation of different geometries, a correct description of the field distribution inside a detector, fast processing, usage of up-to-date libraries, and the possibility to combine with \textsc{Gerda} related analysis software tools, i.e. the \textsc{Root}-based \textsc{Gelatio} \cite{bib:gelatio} for spectral analysis, and the \linebreak \textsc{Geant4}-based MC simulation package \textsc{Mage} \cite{bib:mage}. 
\paragraph{\bf ADL3:} The \textsc{Agata} Detector Library {\texttt{ADL3}} \cite{adl3-description_2015} is an open-source code written in the programming language C. The original code had limited field calculation possibilities, partly based on a commercial software \cite{adl3-aux-software_2007}, but then optimized by new algorithms and physics models \cite{adl3-own-algorithms} providing a complete pulse shape simulation framework, once the fields are calculated. The code is easily extensible and flexible enough to allow adaptation to any detector geometry and detector segmentation.
\paragraph{\bf mjd\_fieldgen/mjd\_siggen:} The code {\texttt{mjd\_fieldgen / mjd\_siggen}} (short: \texttt{siggen}) \cite{siggen-code} is an open-source code written in C. It provides an electric and weighting potential calculation and powerful pulse shape simulation for energy depositions at specific locations inside the detector. With respect to \texttt{ADL3}, however, it was not so flexible and required editing of the existing programs for the implementation of more complex geometries at that time.\\
\indent We started with the {\texttt{ADL3}} code and implemented the potential calculation algorithm used in {\texttt {siggen}} into {\texttt{ADL3}} to complete the software into a full detector simulation library \cite{bib:2015_diss_msalathe}. The following modifications were applied: 
\begin{itemize}
\item Description of variable permittivity in a medium (important for groove simulation),
\item Implementation of an electrically non-depleted region (n+ surface; later also transition layer),
\item Optional 2D field calculation in cylindrical coordinates,
\item Extension for optional implementation of electronic response either with or without noise.
\end{itemize}
All these features were implemented in C, so that the library can be used in the \textsc{Gelatio}/\textsc{Mage} framework. The code turned out to be very useful for diagnostics and the description of observed effects in the BEGe characterization data. Several examples are included and described in the following chapters. Besides that, the modified code has become a useful tool within and outside \textsc{Gerda}. For instance, it was used for the cha\-racterization and optimization of the standard BEGe detector design \cite{{bib:2015_diss_msalathe},{bib:2014_liverpool-agata-group}}, for pulse shape simulations of semi-coaxial and BEGe detectors in \textsc{Gerda} Phase~I and II, and more recently for pulse shape studies of novel inverted semi-coaxial detectors installed in a second upgrade of \textsc{Gerda} Phase~II \cite{bib:inverted1,bib:inverted2}.

%%%%%%%%%%%%%%%%%%%%%%%%%%%%%%%%%%%%%%%%%%%%%%%%%%%%%%%%%%%%%%%%%%%%%%%%%%%%%%%%
%%%%%%%%%%%%%%%%%%%%%%%%%%%%%%%%%%%%%%%%%%%%%%%%%%%%%%%%%%%%%%%%%%%%%%%%%%%%%%%%
\section{Depletion and operational voltage}
\label{sec:depletion-and-operational-voltage}
%%%%%%%%%%%%%%%%%%%%%%%%%%%%%%%%%%%%%%%%%%%%%%%%%%%%%%%%%%%%%%%%%%%%%%%%%%%%%%%%
%%%%%%%%%%%%%%%%%%%%%%%%%%%%%%%%%%%%%%%%%%%%%%%%%%%%%%%%%%%%%%%%%%%%%%%%%%%%%%%%

\subsection{Methodology: manufacturer and \textsc{Gerda}}
\label{sec:methodogy-depletion-and-operational-voltage}
\begin{figure}
\begin{center}
    \includegraphics[width=0.50\textwidth, height=0.70\textwidth]{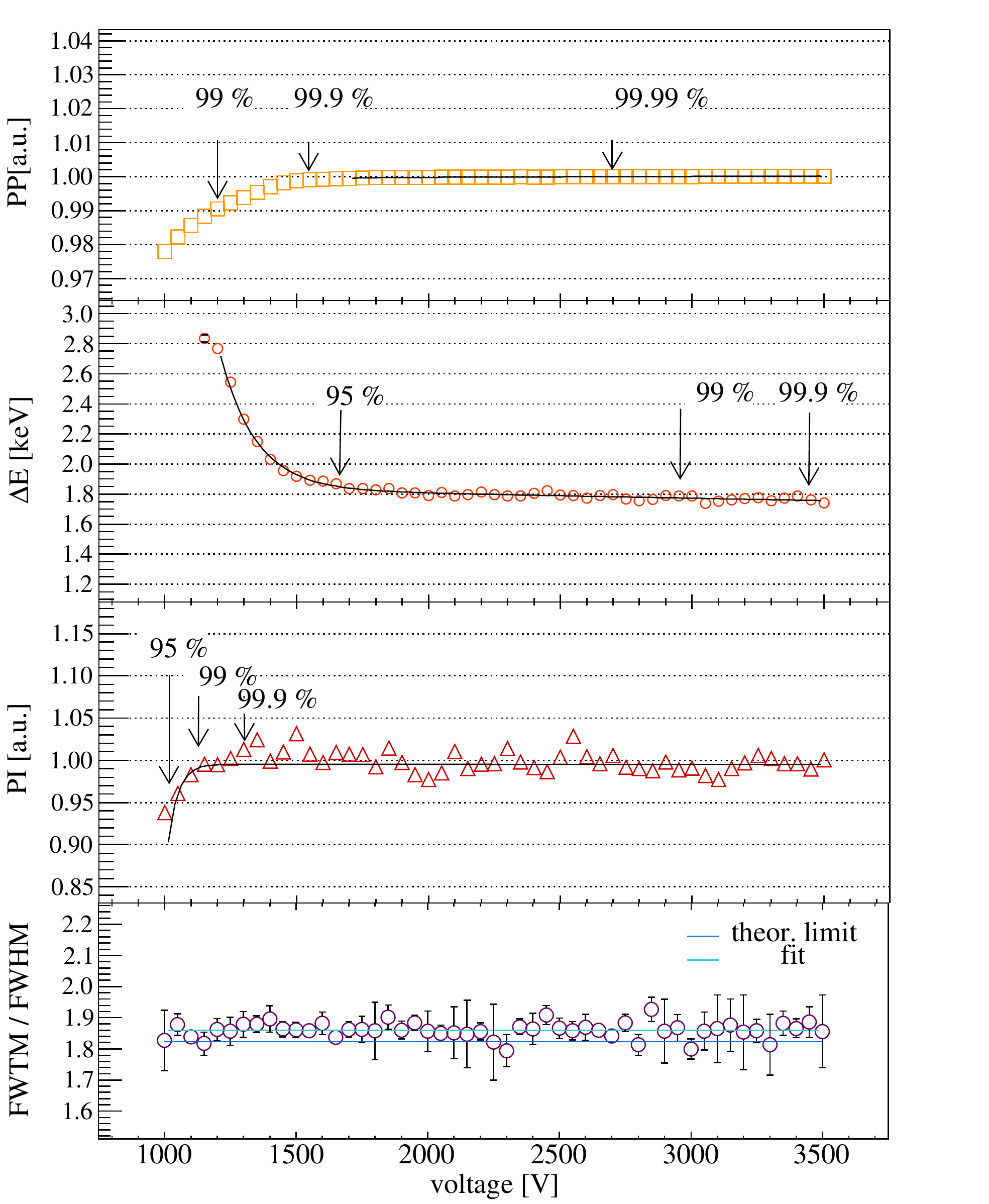} 
  \caption{Characteristic HV scan curves of detector GD00B based on the evaluation of the 1333\,keV $\gamma$-line from a $^{60}$Co calibration source. The three curves of the peak position, energy resolution and peak integral are used for the definition of the full depletion and operational voltage in \textsc{Gerda}. Additionally, the peak asymmetry curve in the bottom canvas demonstrates that the Gaussian peak form is conserved over a large voltage interval. More explanations are included in the text.}
   \label{fig:GD00B_merged-curves}  
  \end{center}
\end{figure}

\paragraph {\bf Full depletion voltage:} The manufacturer Canberra Olen determines the electrical depletion voltage $V_d^C$ of a detector in a two-step approach. First, it operates the diode in a liquid nitrogen bath and measures the leakage current as well as the capacitance as a function of the applied voltage. When the capacitance reaches a `constant' value, the detector is depleted by definition. This allows for a first estimation of the full depletion voltage. In a second step, Canberra Olen installs the diode in a vacuum test cryostat and irradiates it with a $\gamma$-ray source. The spectral positions of characteristic $\gamma$-ray peaks are monitored, while the voltage is increased. When the peak position of an individual $\gamma$-line stops to shift, the detector is expected to be depleted \cite{2016_wim_private}.\\
\indent In contrast, we use a multi-parameter approach \linebreak which monitors the detector properties that are relevant for the physics goals of the experiment. The diodes installed in vacuum cryostats are irradiated with $\gamma$-ray sources, too. During a high voltage (HV) scan, which typically starts at 500\,V and increases in 100\,V steps up to the Canberra recommended voltage $V_r^C$, the following quantities of prominent $\gamma$-ray peaks are monitored: the peak position ($PP$), the energy resolution ($\Delta E$) and the peak integral ($PI$). In some cases, the peak asymmetry and pulse shape parameters are also registered. An example is shown in Figure~\ref{fig:GD00B_merged-curves}, which depicts the corresponding curves for detector GD00B. Depending on the number of peak fit parameters, the data points of the curves fluctuate more or less. Hence, the peak integral curve (depending on the correct peak shape mo\-deling and background subtraction) typically fluctuates stronger than that for the energy resolution and peak position. In the specific case of the peak position only one parameter of a $\gamma$-line, the peak maximum, has to be extracted. The three curves are fitted with a polynomial function. Herein the plateaus encountered beyond the full depletion voltage knees are always fitted with a linear term. Based on the fit parameters, several reference depletion voltage points are extracted, at which the peak position ($PP$) reaches {99\%, 99.9\% and 99.99\%} of its highest fit value obtained at $V_r^C$, the energy re\-solution $(\Delta E)$ {95\%, 99\% and 99.9\%} of its smallest fit value at $V_r^C$, and the peak integral ($PI$) {95\%, 99\% and 99.9\%} of its largest fit value at $V_r^C$. The corresponding voltage points are denoted with $V_d(99\%\,PP$) etc.

\paragraph {\bf Operational voltage:} The operational voltages $V_r^C$ recommended by Canberra Olen typically lie 500-1000\,V\\above their estimated full depletion voltages. These relatively high $V_r^C$ are still below a critical break-down voltage, but are driven by the fact that the energy resolution can still improve at the percent level within the full depletion plateau and that most customers are mainly interested in achieving the best possible energy resolution \cite{2016_wim_private}. For the \textsc{Gerda} experiment, however, slightly lower operational voltages might be more advantageous, for instance to keep leakage currents low or to attract less ions present in the liquid argon towards the detector surface. We defined new operational voltages $V_r^G$ which fulfill the following three criteria:
\begin{itemize}
\item The volume, in which a charge collection efficiency $\epsilon$ of 1 (cf. Figure~\ref{fig:fccd-scheme}) can be achieved, has to be electrically fully depleted to guarantee a correct determination of the active volume and the exposure during the experimental phase. Thus, the peak integral has to be close to its maximum value and a limit of {$>$99\%} is required.
\item The energy resolution has to be close to the optimum fit value to guarantee an optimum sensitivity for the $0\nu\beta\beta$ decay of $^{76}$Ge, which scales in the presence of background as $\sqrt{(1/\Delta E)}$. By default, we require {$>$95\%} compared to the best fit value. In the realistic scenario of a 3\,keV full-width at half-maximum for a peak at Q$_{\beta\beta}$($^{76}$Ge)=2039\,keV, this would correspond to a tolerable increase by 0.15\,keV.
\item Finally, the peak position should be stable, but does not necessarily have to be at the maximum fit value. By default, we ask for a limit better than $>$99.9\%.
\end{itemize}
Typically, the following inequality holds:
\vspace{-0.7 cm}
\begin{center}
\begin{equation}
V_d(99\% PI) < V_d(99.9\% PP)
\end{equation}
\end{center}
with $V_d(95\%\,\Delta E)$ being similar to $V_d(99.9\%\,PP)$. The new operational voltage $V_r^G$ is defined as the voltage, which adds 500\,V to the full depletion voltage  $V_d$(99.9\% $PP$). This fulfills all three introduced criteria and provides enough margin to stay well above the `depletion knee'. The latter represents the transition region between the slopes and the plateaus of the curves.\\
\indent An incomplete electronic depletion or incomplete charge collection efficiency in the Ge detector can result not only in a broader peak width, but also in the formation of peak tails to an extent that the ideal Gaussian form of a $\gamma$-ray peak might not be observed. A way of quantifying the effect is to measure the full-width of the peak at different heights and to calculate the related ratios $\rho$ of such widths, e.g. for the full-width at half-maximum (FWHM) and the full-width at tenth-maximum (FWTM). The ratio $\rho_{10}$ = FWTM/FWHM, which is 1.823 for a pure Gaussian peak, was calculated for voltage scans applied on individual detectors. In all examined detector cases, the experimental ratios for all voltages applied above $V_r^G$ were found to be very close to the theoretical best value. Moreover, our stu\-dies confirmed that any detector operated at a voltage above $V_r^G$ has already reached the optimum pulse shape performance (cf. Chapter~\ref{sec:pulse-shape-behavior}).

\subsection{Results}
\label{sec:results-depletion-and-operational-voltage}
\paragraph {\bf Full depletion and operational voltages:} The determined values by Canberra Olen as well as by us are summarized in Table~\ref{table:depl-voltages-all-detectors}.\\
\indent The operational voltages recommended by us are typically higher than the ones needed to reach 99\% of the optimum energy resolution, i.e. $V_d(99\%\,\Delta E$). In all but two cases, the $V_r^G$ values are below 3.7\,kV. In the case of GD91D, the applied voltage should be as high as possible. GD02D is the only detector that does not deplete completely (cf. Section \ref{sec:Manufacturing-GERDA-BEGe-detectors} and \ref{sec:results-active-volume}). The average of our recommended operational voltage is 3.1\,kV. This is 0.6\,kV lower than the average of the operational voltages recommended by Canberra Olen.\\
\indent The new definition of operational voltages turned out to be truly useful for the \textsc{Gerda} experiment. We agreed to operate the new BEGe detectors at $V_r^C$. In the case of instabilities or a prohibitive increase of leakage current, however, a detector can be operated at lower voltages as long as it will not be less than $V_r^G$. In \textsc{Gerda} Phase~II, more than four BEGe detectors have been operated at least temporarily at voltages between $V_r^G$ and $V_r^C$. Additionally, one detector was operated below the $V_r^G$ benchmark: GD02D. More details about this problematic detector are reported in Section \ref{sec:results-active-volume}.

\begin{figure}
\begin{center}
   \includegraphics[width=0.45\textwidth, height=0.55\textwidth]{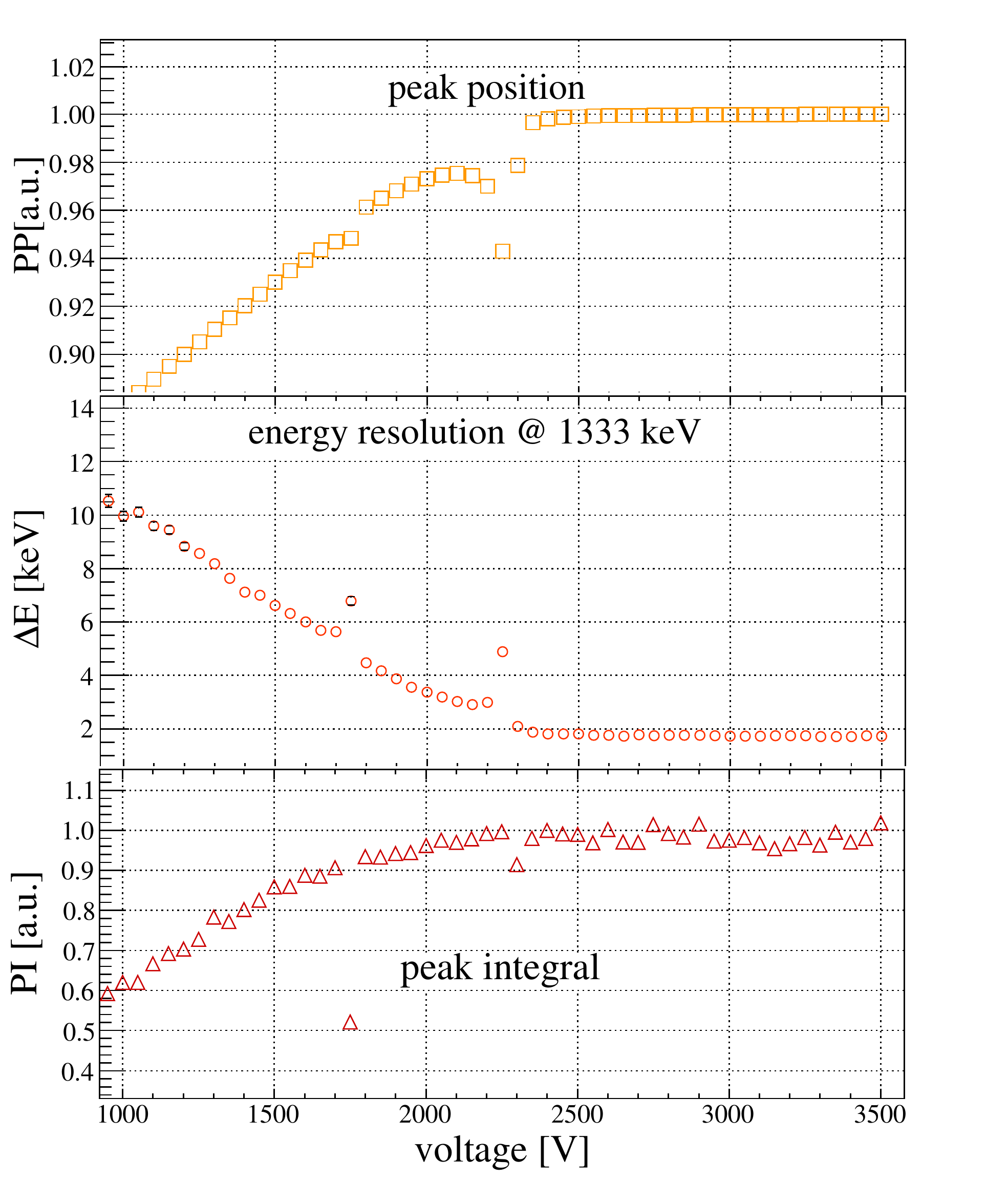} 
  \caption{Characteristic HV scan curves of detector GD00D based on the evaluation of the  1333\,keV $\gamma$-line from a $^{60}$Co calibration source. For a given curve, the distance between two values are 100\,V. In all three curves of the peak position, energy resolution and peak integral the presence of two `bubbles' becomes visible.}
  \label{fig:GD00D_merged-curves1}
  \end{center}
\end{figure}

\paragraph {\bf Detectors with discontinuities in the HV scans:}
A closer look at the normally smooth $PP$ curves of the 30 \textsc{Gerda} Phase~II BEGe detectors sometimes reveals dips that appear around the depletion voltage knees. In a few cases, the discontinuities are also observed in the corresponding $\Delta E$ and in rare cases in the $PI$ curves. The discontinuity behavior has been attributed to the so-called `bubble depletion' \cite{bib:bubble-effect} or `pinch-off' effect \cite{bib:majorana_intro2}: For some combinations of detector geometries and net impurity concentrations  the total electric field strength consisting of the applied voltage and the one from the intrinsic charge concentration can become zero in a vo\-lume around the center of the detector. This occurs for voltages just below depletion. As a consequence, the charge collection behavior changes in the sense that the holes are largely trapped or slowed down locally near the center. This leads to a reduction in the observed pulse amplitude (peak position) and potentially to a worse energy resolution and a reduction of the peak efficiency (peak count rate). Around 40\% of the \textsc{Gerda} Phase~II BEGe detectors were found to have one or two discontinuities in the HV scans. They are listed in Table~\ref{table:bubble-effect}. Three subgroups are identified:
\begin{itemize}
\item Detectors with one small `bubble': The weak discontinuity is seen only in the $PP$ curve, but not in the energy resolution or peak integral curves. Four detectors belong to this class.
\item Detectors with one large `bubble': The discontinuity is clearly seen in the $PP$ as well as in the $\Delta E$ curve, but not in the $PI$ curve. Seven detectors belong to this class.
\item Detectors with two independent discontinuities: Two discontinuities at different voltages are found. They are seen in the $PP$ and $\Delta E$ curves, to some extent also in the $PI$ curves. The discontinuity at higher voltage which is closer to the depletion knee is typically enhanced, i.e. deeper and broader. The two instabilities are separated by approximately 500\,V. Two detectors belong to this class. The curves of detector GD00D are shown exemplarily in Figure~\ref{fig:GD00D_merged-curves1}.
\end{itemize}

\begin{table}[t]
\begin{center}
	\caption{\rm{List of \textsc{Gerda} Phase~II BEGe detectors that are affected by the single or double `bubble depletion' effect. Parentheses around the number of discontinuities denote a less intense effect.}}
	\vspace{0.3 cm}
	\begin{tabular}{|c|cc|}	
	   \hline	
\multicolumn{1}{|c|}{detector} & \multicolumn{2}{c|}{discontinuities in $PP$ curve} \\
ID			& number [\#] 	  		& voltage [kV] 	\\	   
	   \hline	
GD32B		&1						&2.7				\\
GD35A		&1						&2.4			    \\
GD35B		&2						&2.1; 2.7			\\
GD61A		&1						&2.9				\\
GD61C		&1						&2.3				\\
GD76C		&1						&1.9				\\
GD79B		&(1)					&2.2				\\
GD89B		&(1)					&2.2				\\
GD89D		&(1)					&2.3				\\
GD91A		&1						&2.4				\\
GD91B		&(1)					&2.5				\\
GD00C		&1						&2.5				\\
GD00D		&2						&1.8; 2.3			\\
		\hline
    \end{tabular}
	\label{table:bubble-effect}	
\end{center}
\end{table}

\begin{figure}
\begin{center}
   \includegraphics[width=0.48\textwidth, height=0.65\textwidth]{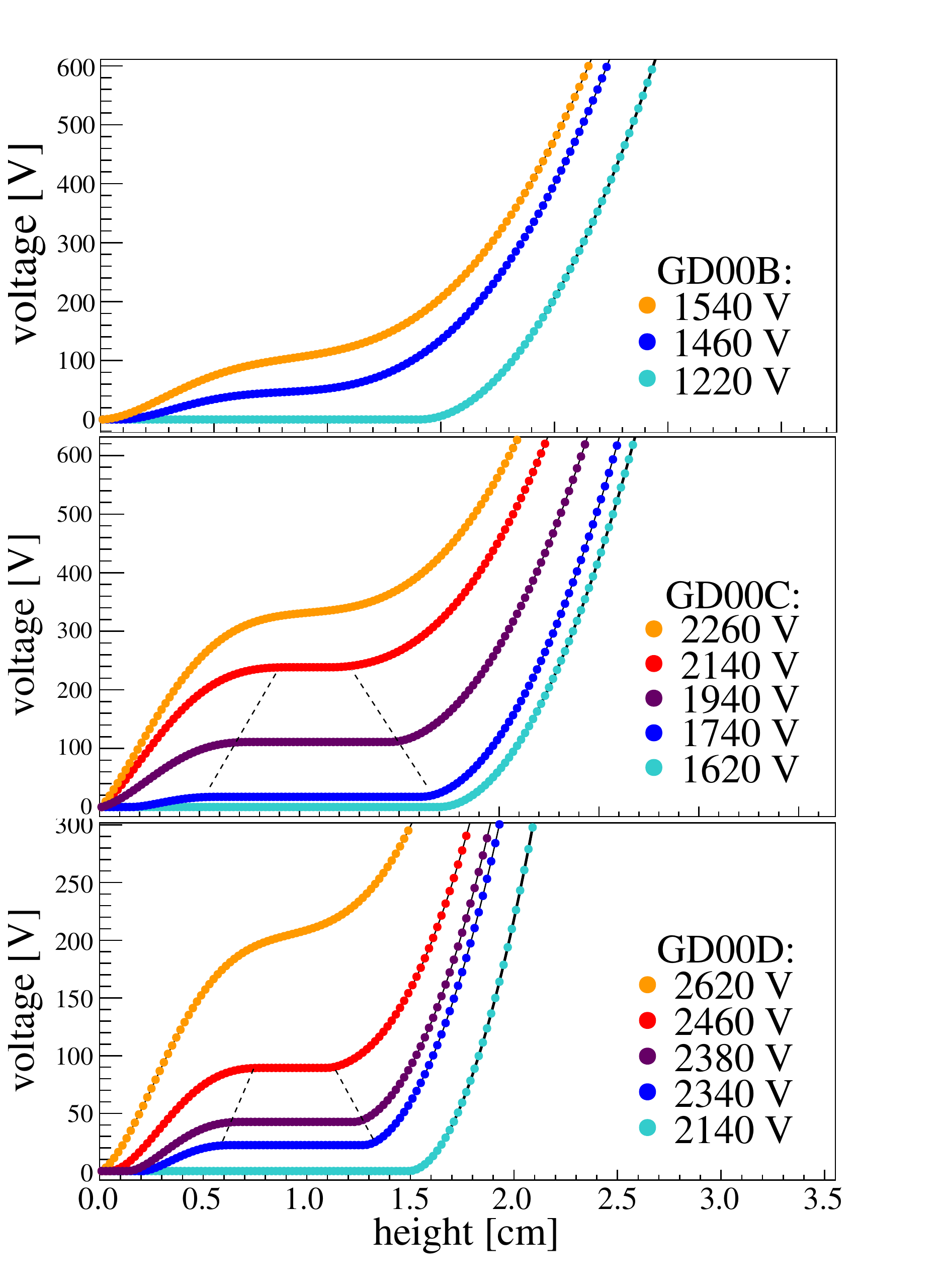} 
  \caption{\texttt{ADL3}-simulation of the electrical depletion process of the detectors GD00B, GD00C and GD00D: the voltage along the central axis starting at the p+ electrode read-out is depicted as function of the voltage applied externally to the n+ contact. The curves in orange represent the voltage when the full depletion is reached. Curves with an intermediate constant interval are marked with a dotted line and correspond to voltages, at which a `bubble' persists. For experimental data see Figures \ref{fig:GD00B_merged-curves} and \ref{fig:GD00D_merged-curves1}.}
  \label{fig:bubble_GD00-series_plot}
  \end{center}
\end{figure}
To our knowledge, two discontinuities at different voltages in one individual BEGe detector have been observed for the first time within this survey. In order to reproduce this scenario, we performed \texttt{siggen} and \texttt{ADL3} simulations for detectors GD00B, GD00C and GD00D, which were produced from the same crystal ingot. Besides the exact crystal dimensions, the simulations included the net impurity concentration values provided by the manu\-facturer assuming a linear gradient. Figure~\ref{fig:bubble_GD00-series_plot} depicts exemplarily the \texttt{ADL3} calculated voltage at different heights above the read-out electrodes, i.e. along the central axes of the three diodes, when different HVs are applied to the n+ electrodes. The corresponding \texttt{siggen} curves are not shown, but behave very similarly. In the case of detector GD00B, the si\-mulated curves for voltages above 1000\,V have no intermediate constant interval, even though hardly visible in Figure~\ref{fig:bubble_GD00-series_plot}, and thus -- like in reality -- no `bubble'. The full depletion voltage simulated with \texttt{siggen} and \texttt{ADL3} are in the range [1.5,1.7]\,kV, which agrees well with $V_d(99.9\%\,PP)$. For detector GD00C the two codes predict the existence of a `bubble' around [1.7,2.2]\,kV and a full depletion voltage at [2.0,2.3]\,kV. However, the measured `bubble depletion' and full depletion vol\-tage $V_d(99.9\%\,PP)$ are larger by $\sim$500\,V. Finally, in the case of GD00D the codes predict a full depletion voltage around [2.5,2.8]\,kV. This matches well with \linebreak $V_d(99.9\%\,PP)$. But both codes foresee only one single `bubble' occurring in the central bulk region when the bias voltage of [2.3,2.6]\,kV is reached. The second independent discontinuity appearing at a lower voltage in experimental data is not reproduced by the codes. A prediction of such a `bubble' might require a more detailed knowledge and implementation of the radial and axial variation of the net impurity concentration. Beside this deficit, both \texttt{siggen} and \texttt{ADL3} codes have meanwhile demonstrated to be a reliable tool for the prediction of the full depletion voltage and the appearance of single `bubbles' also in other HPGe detector designs (e.g. for inverted semi-coaxial Ge detectors \cite{bib:inverted1}). 

\paragraph {\bf Dependence of the full depletion voltage on detector parameters:}
This paragraph addresses the \linebreak question how the full depletion voltages $V_d$ of the 29 well operating \textsc{Gerda} Phase~II BEGe detectors depend on other detector parameters. In order to find such dependencies, one should solve the Poisson equation for the non-symmetric electric field in a BEGe detector. However, an analytical expression cannot be deduced easily and requires typically numerical calculations. On the contrary, an analytical expression for the depletion thickness $h$ of a planar detector geometry can be found \cite{bib:knoll}:
\begin{equation}
h = \sqrt{\frac{2\cdot V_d \cdot \epsilon}{e\cdot N_{a-d}}}
\label{eq:planar_depletion-voltage-dependency}
\end{equation}
Here, $e$ stands for the elementary electric charge and $\epsilon$ for the dielectric permittivity in the medium. The latter is defined as $\epsilon = \epsilon_0\cdot\epsilon_r$ with $\epsilon_0$ being the permittivity in vacuum and $\epsilon_r$ the relative dielectric susceptibility. In the case of Ge, $\epsilon_r$=16.0 at 295\,K \cite{ge-suszeptibility}. Since  $\epsilon_r$ has only a small temperature dependence, the value is still valid for HPGe detectors operated at the boiling point of liquid argon/nitrogen at 87\,K resp. 77\,K  \cite{ge-suszeptibility-temp-dep}.\\ 
\indent Starting from Equation \ref{eq:planar_depletion-voltage-dependency}, the following {\it{ansatz}} for the BEGe design is introduced:
\begin{equation}
k=\frac{V_d}{h^2 \cdot N_{a-d}}=\frac{e}{\epsilon}\cdot a
\label{eq:bege_depletion-voltage-dependency}
\end{equation}
with $a$ being a free parameter that still has to be determined. In the case of a planar detector geometry, $1/a$ is 2.
\begin{figure}[t]
\begin{center}
  \includegraphics[width=0.50\textwidth]{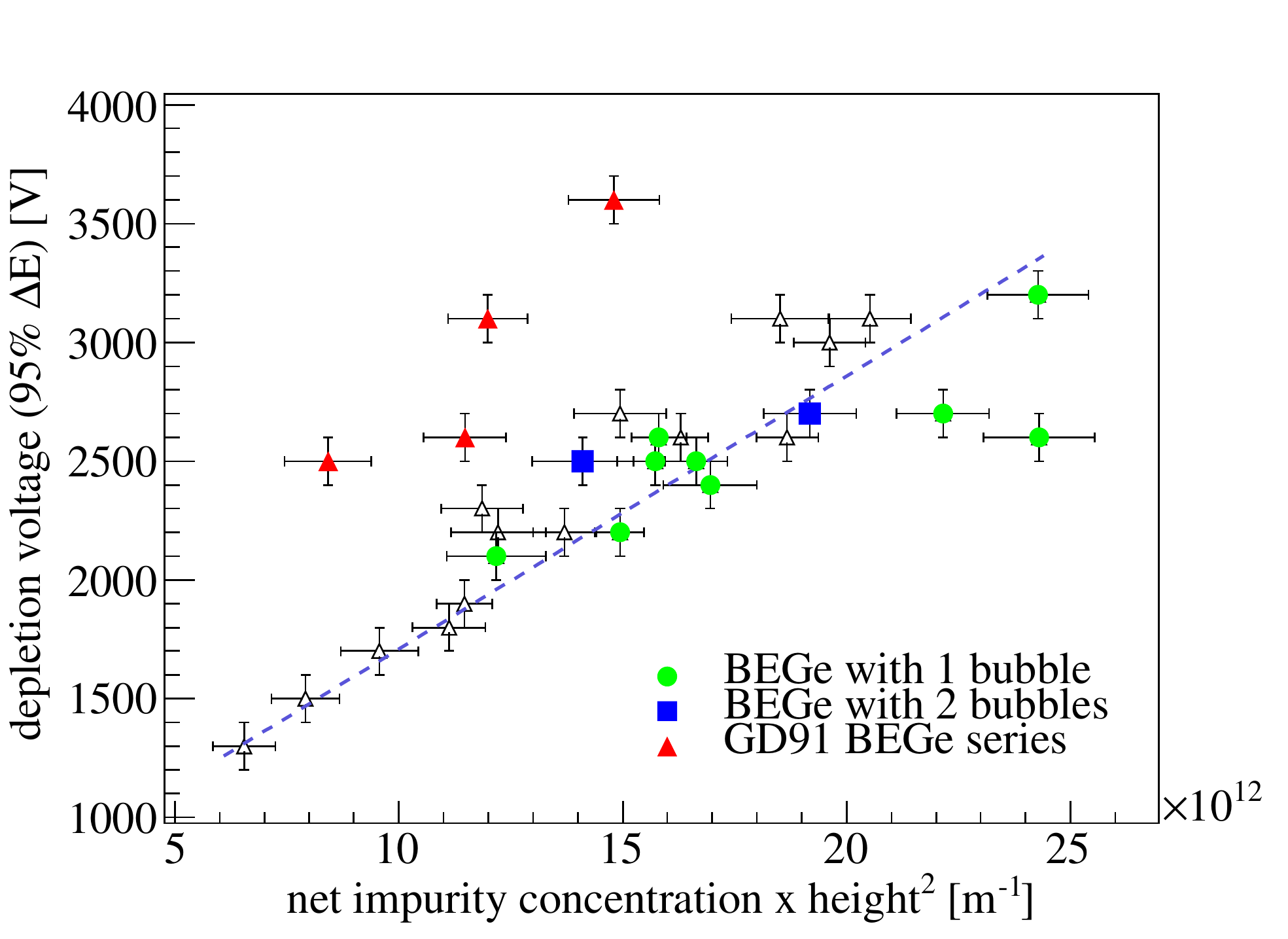}  
  \caption{Correlation between the full depletion voltage $V_d(95\%$ $\Delta E)$ and the product of the net impurity concentration and the squared height for all 30 \textsc{Gerda} Phase~II BEGe detectors except for GD02D. Open symbols pertain to detectors with no appearance of the `bubble depletion' effect or which do not belong to the GD91 series.}
  \label{fig:Vd-er_vs_c-d-d_for-paper}
  \end{center}
\end{figure}
Figure~\ref{fig:Vd-er_vs_c-d-d_for-paper} depicts ${V_d}$ vs. ${h^2 \cdot N_{a-d}}$. Herein,\\$V_d(95\%\Delta E)$ has been selected as ${V_d}$. The parameter $N_{a-d}$ corresponds to the net impurity concentration deduced from the crystal slice seed and tail measurements by Canberra Oak Ridge. A proportional dependence of $V_d(95\%\Delta E)$ becomes evident, independent of the strength of the $N_{a-d}$ gradient. A linear trend exists also for those situations in which the full depletion voltages defined from the $PP$ and $PI$ curves are used. Four detectors are particularly off-line and marked in red. The detectors are GD91A, GD91B, GD91C and GD91D and belong to the same crystal ingot. A potential error of factor $\sim$2 in the net impurity concentration determination by the underlying Hall effect measurement would be able to explain the offset.\\
\indent A linear fit of the remaining 25 points in the ${V_d}$ vs. ${h^2 \cdot N_{a-d}}$ representation has been performed. The fit parameters $k$ and $1/a$ are reported in Table \ref{table:depl-voltages-dependency-fit}. Contrary to a planar geometry, the $1/a$ value for the examined BEGe detectors is close to 10 and thus $\sim$5 times larger.\\
\begin{table}[h]
\begin{center}
	\caption{\rm{Dependence of the full depletion voltage on detector parameters following Equation~\ref{eq:bege_depletion-voltage-dependency}. The best parameters of the linear fit are reported.}}
	\vspace{0.3 cm}
	\begin{tabular}{|c|c|c|c|}	
	\hline	
detector			&${V_d}$   					&$k$ 										&$1/a$		\\
design				&[V]						&[10$^{9}$\,V$\cdot$m] 	 					&				\\
	\hline
planar				& 							& 1.13 (= $e$/(2$\epsilon$))	& 2				\\	
	\hline	
BEGe				&$V_d(95\% \Delta{E}$)		& (10.5$\pm$1.0)			& (10.8$\pm$1.0)	\\	
					&$V_d(99\%$ PP$)$			& (12.6$\pm$1.2)			& (9.0$\pm$0.9)	\\	
					&$V_d(95\%$ PI$)$			& (10.6$\pm$1.0) 			& (10.7$\pm$1.0)	\\	
	\hline
    \end{tabular}
	\label{table:depl-voltages-dependency-fit}	
\end{center}
\end{table}
\indent Moreover, the possibility to find detectors affected by the `bubble depletion' effect in a certain region of the ${V_d}$ vs. ${h^2 \cdot N_{a-d}}$ representation was investigated. Figure~\ref{fig:Vd-er_vs_c-d-d_for-paper} points towards detectors with one or two `bubbles'. There is no statistical significant tendency that detectors affected by this `bubble depletion' prefer a certain region of the parameter space.

%%%%%%%%%%%%%%%%%%%%%%%%%%%%%%%%%%%%%%%%%%%%%%%%%%%%%%%%%%%%%%%%%%%%%%%%%%%%%%%%
\section{Energy resolution}
\label{sec:energy-resolution}
 
\subsection{General remarks}
The energy resolution $\Delta E$ of a HPGe detector is defined as the width of one characteristic $\gamma$-line at a given energy $E$ and consists of three sub-components:
\begin{equation}
\Delta E^2 = \Delta E_{sf}^2 + \Delta E_{cc}^2 + \Delta E_{el}^2 .
\label{eq:ene-res-definition}
\end{equation}
$\Delta E_{sf}$ corresponds to the statistical fluctuation in the charge release and depends on the material-dependent Fano factor $F$, on the energy ${\cal E}=2.96$\,eV needed for the production of one electron-hole pair in Ge at 77\,K and the absorbed $\gamma$-ray energy $E$. $\Delta E_{cc}$ corresponds to the charge carrier collection efficiency, which depends on the concentration of defects/vacancies in the bulk of the Ge crystal. It is relevant for detectors of large size and/or with low electric field strength. Finally, $\Delta E_{el}$ corresponds to the electronics and environmental noise term.\\ 
\indent In order to obtain a reproducible and robust determination of $\Delta E$ at a given $\gamma$-ray energy, one has to specify the measurement and analysis procedure:
\begin{itemize}
\item {Operational detector conditions}: the voltage at which $\Delta E$ is measured has to be quoted. As observed in Chapter \ref{sec:depletion-and-operational-voltage}, the energy resolution can still improve within the depletion plateau at the level of a few percent. One has to minimize and quantify the noise contribution, e.g. via adequate pulser measurements. In our case, we further specify if the detectors are operated in cooled vacuum cryostats or bare inside a cryogenic liquid.
\item {Energy reconstruction:} the filter type (Gaussian, tra\-pezoidal, cusp etc.) and shaping time applied for the reconstruction of the energy variable have to be defined. It should be stated whether a ballistic deficit correction, which becomes important at \linebreak energies above $\mathcal{O}$(1\,MeV) \cite{geana-paper}, is applied.
\item {Fit and $\Delta E$ definition:} the fit procedure of the $\gamma$-ray peak has to be specified. Within the depletion plateau, the peaks have often an almost perfect \linebreak Gaussian shape. Thus, they can be fitted with a three-component function consisting of a Gaussian, a linear and a step-like term. The latter two describe the shape of the energy spectrum underlying the peak (background). The energy resolution $\Delta E$ is then defined in terms of the variance $\sigma$ or the full-width at half-maximum (FWHM) of the background-sub\-tracted Gaussian fit component of the $\gamma$-ray peak. If a detector has a bad crystal quality, radiation damage or cannot be fully depleted, the $\gamma$-line shape might deviate from the pure Gaussian form. The appearance of a low energy tail might be a consequence. In such cases, the fit function has to be adopted accordingly.
\end{itemize}

\subsection{Methodology: manufacturer and \textsc{Gerda}}
\label{sec:methodogy-energy-resolution} 
The manufacturer Canberra Olen determines the energy resolution $\Delta E$ of a detector in the following way: the diode is mounted in a vacuum cryostat and operated at the recommended voltage $V_r^C$. Then the detector is irradiated with non-collimated $^{57}$Co and $^{60}$Co $\gamma$-ray sources. The $\Delta E$ of the two $\gamma$-lines at 122\,keV and 1333\,keV are typically expressed in terms of FWHM, whereas a potential peak distortion from the pure Gaussian shape is quantified via the measurement of the FWTM and the ratio $\rho_{10}$. The measurement is performed with a standard Canberra multi-channel analyzer. The energy reconstruction is carried out applying a constant shaping time of 4\,$\upmu$s \cite{2016_wim_private}.\\
\indent Within the \textsc{Gerda} detector characterization campaign, the $\Delta E$ of the BEGe detectors operated in vacuum are evaluated mainly for the $^{60}$Co $\gamma$-line at \linebreak 1333\,keV, but also for other peaks originating from other sources. In general, the detectors are irradiated with non-colli\-ma\-ted sources at a distance of typically $\sim$20\,cm from the diode`s top surfaces. In the case of 1333\,keV, the $^{60}$Co calibration has been subdivided into:
\begin{itemize}
\item {Standard approach}: 10-60 minutes measurements\linebreak with a $^{60}$Co source of several 100\,kBq activity are performed at the voltage $V_r^C$. $\Delta E$ is extracted from this single measurement.
\item {Alternative approach}: In order to exclude eventual temporary instabilities due to e.g. microphony from other ongoing work on-site, data collected during the HV scans described in Chapter \ref{sec:depletion-and-operational-voltage} are used. The energy resolution value at $V_r^C$ is extrapolated from the polynomial fit of the $\Delta E$ curve.
\end{itemize}
Data are collected with Multi Channel Analyzer (MCA) modules by ORTEC (926) and Canberra (Multiport II NIM), and with Struck SIS3301 VME Flash Analog-to-Digital-Converters (FADC). The latter ones allow a sampling-rate of 100\,MHz with a 14-bit resolution per sample. Up to 128\,k samples with a maximum trace length of 1.28\,ms can be registered. The energy of an event is reconstructed with a shaping time of 8\,$\upmu$s. No ballistic deficit correction is applied. The $\gamma$-peaks are fitted with the following fit function $f(E)$:
\begin{equation}
f(E) =  \frac{A}{\sqrt{2\pi}\sigma}\cdot e^{-\frac{(E-E_0)^2}{2\sigma^2}} + \frac{B}{e^\frac{2(E-E_0)}{\sigma}+1} + C\cdot E,
\label{eq:main-fit-fcn}
\end{equation}
with $A$, $B$, $C$ being normalizations and $\sigma$ the variance of the Gaussian distribution. The second term corresponds to a Fermi-like step function. The effect of including other step and low-side energy tail functions, as proposed in literature (see e.g. \cite{bib:helmer-lee}), was investigated for different extensions and tested on BEGe data. The impact of the fit function diversity on $\Delta E$ for a fixed $V_r^C$ was estimated to be $\pm$0.01\,keV. Only in the case of detector GD02D, the peak shape has a larger low-energy-tail even at $V_r^C$=4\,kV and needs an adequate fit model extension.

\subsection{Results}
\label{sec:results-energy-resolution}
\begin{figure}
\begin{center}
\includegraphics[width=0.35\textwidth, angle=-90 ]{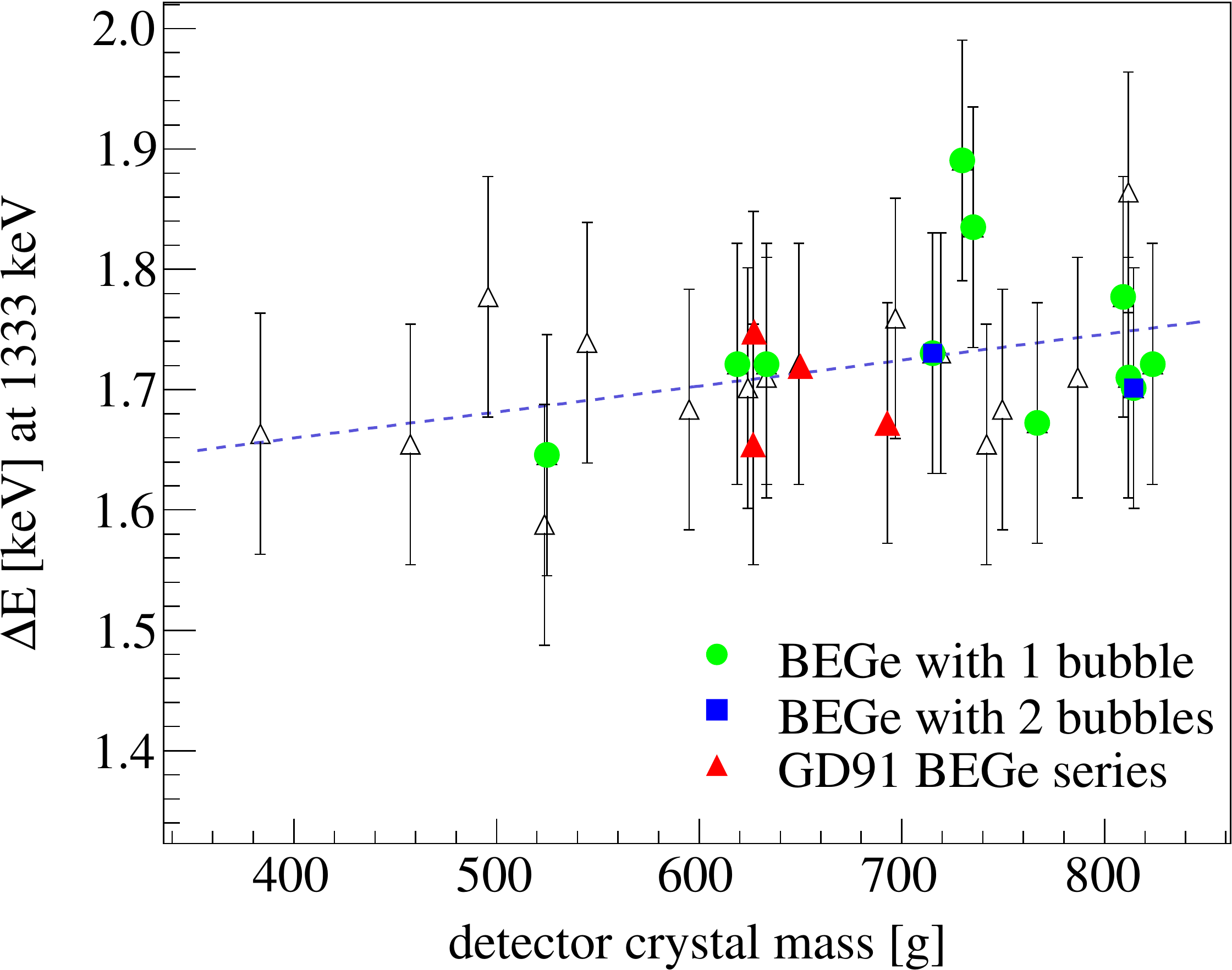}

\caption{Detector energy resolution dependence on the detector mass of all 30 \textsc{Gerda} Phase~II BEGe detectors but GD02D. Open symbols are used like in Figure~\ref{fig:Vd-er_vs_c-d-d_for-paper}.}
 \label{fig:er_vs_mass_0p05keVerror}
\end{center}
\end{figure}

\begin{figure}
\begin{center}
\includegraphics[width=0.5\textwidth]{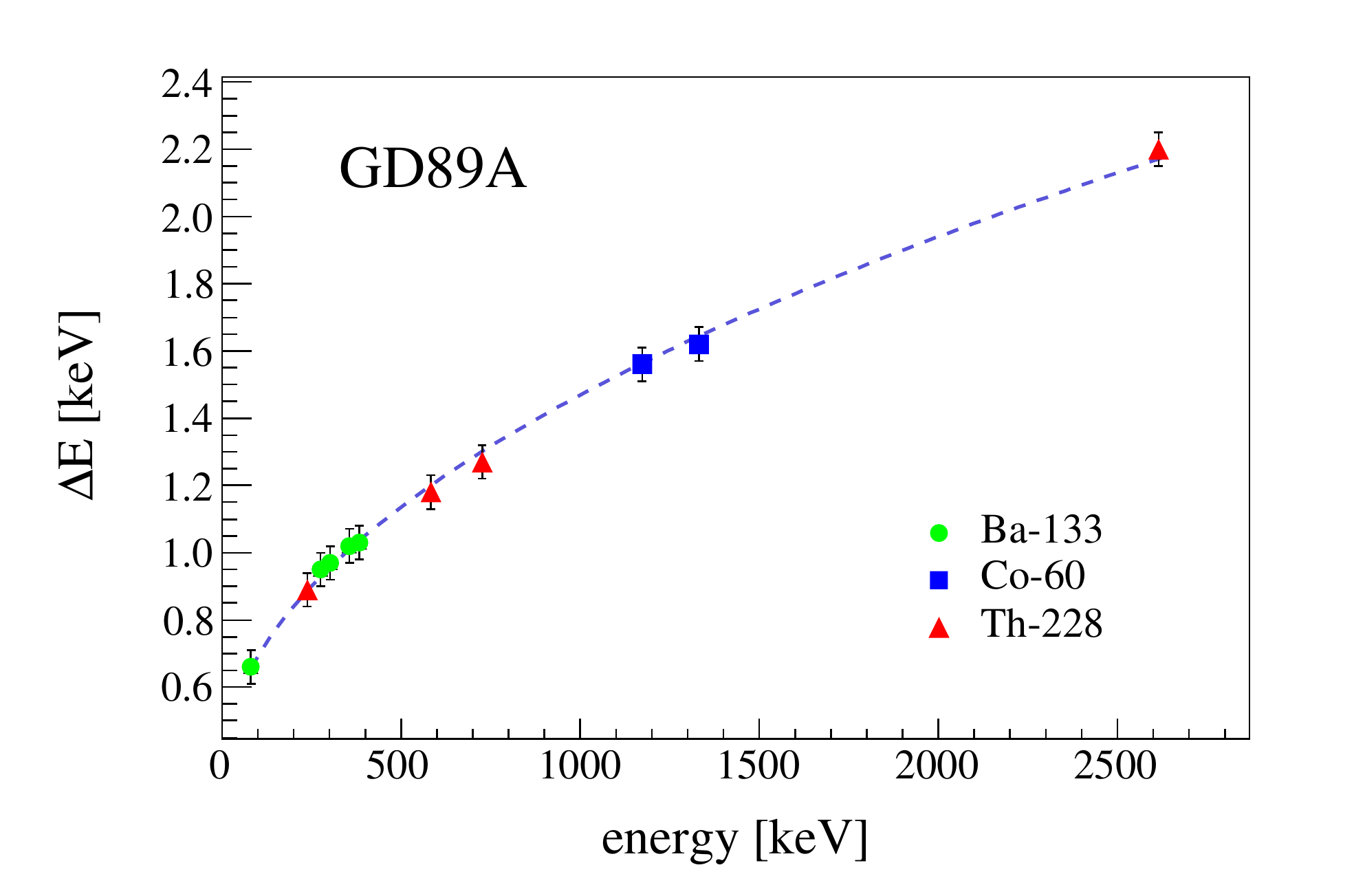}
\caption{Energy resolution (in terms of FWHM) as function of energy for detector GD89A. For the fit function only the two energy resolution terms $\Delta E_{sf}$ and $\Delta E_{el}$ were considered, while $\Delta E_{cc}$ was neglected.}
 \label{fig:er_vs_ene}
\end{center}
\end{figure}

\paragraph {\bf Energy resolution at 1333\,keV:} The energy resolutions $\Delta E$ of all 30 \textsc{Gerda} Phase~II BEGe detectors were examined according to the procedure described in Section \ref{sec:methodogy-energy-resolution}. The determined values by Canberra Olen as well as by us are summarized in Table~\ref{tab:ene-res-all-results}. The second column contains our values obtained with the  method based on the HV scans. The third column shows the \textsc{Gerda} values obtained with the classic method based on one single measurement at $V_r^C$. The values based on the two methods sometimes disagree by $\sim$0.05\,keV due to fit and experimental instabilities that are not considered in the total uncertainty budget. The fourth column reports the results obtained by Canberra Olen. Only in rare cases, they differ by more than $\sim$0.1\,keV from the \textsc{Gerda} values. The average of all mean values quoted by \textsc{Gerda} and Canberra Olen are in very good agreement.\\
\indent In general, the \textsc{Gerda} BEGe detectors have excellent energy resolutions. According to the HV scan based \textsc{Gerda} analysis, the average value is 1.72\,keV with a SD of 0.07\,keV. Further, the best detector is GD89A with (1.59$\pm$0.01)\,keV and the worst GD61A with (1.89$\pm$0.01)\,keV. Detector GD02D has an accep\-table resolution of (1.84$\pm$0.11)\,keV, but a strong low-side energy tail due to incomplete charge collection (cf. Figure \ref{fig:GD02D-vs-GD91B_Co60-full-spec_2-BEGe-paper}).
\paragraph {\bf Dependence of energy resolution on detector parameters:} This section raises the question whether the energy resolution of the 29 well working \textsc{Gerda} Phase~II BEGe detectors is correlated to other detector parameters or not.\\
\indent The $\Delta E$ value was investigated separately for conical and cylindrical shaped detectors.  No evidence was found that they would differ from each other. This further supports the decision taken during crystal production to optimize the slice cut towards a maximum mass yield. $\Delta E$ turned out to be also not strongly correlated to the electronics noise term $\Delta E_{el}$ in Equation~\ref{eq:ene-res-definition}, which partly depends on the detector capacitance and the latter in turn on the net impurity concentration $N_{a-d}$ and dielectric permittivity $\epsilon$ in the medium.\\
\indent Finally, $\Delta E$ was plotted against the detector mass: charge collection deficits and bulk leakage currents \linebreak might scale with the volume and thus with the detector mass. Figure~\ref{fig:er_vs_mass_0p05keVerror} shows that a small correlation in the investigated mass range from 384\,g to 824\,g exists. The distribution was fitted with a linear function leading to the following relation: 
\begin{equation}
\Delta E(m) = 1.57(6)\,\text{keV} + m \cdot 2.2(8)\cdot 10^{-4}\,\text{keV/g}
\label{eq:eneres-fnc-mass}
\end{equation}
with $m$ being the detector mass in units of gram.
Furthermore, detectors affected by the `bubble depletion' effect do not appear in a clearly confined region of the parameter space. 
\paragraph {\bf Dependence of energy resolution on energy:} For each \textsc{Gerda} Phase~II BEGe detector $^{241}$Am, $^{133}$Ba, $^{60}$Co and $^{228}$Th calibration data were collected. 
This allowed to analyze for each detector the resolution for a dozen of $\gamma$-lines over energy and to deduce the \linebreak energy resolution dependence from it. 
Figure~\ref{fig:er_vs_ene} illustrates the measured $\Delta E$ points for detector GD89A, which was identified to have the best energy resolution of all 30 BEGe detectors. The curve follows a $\sqrt{E}$ dependence, which arises from the charge carrier statistics term $\Delta E_{sf}=2.355\cdot\sqrt{F\cdot {\cal E} \cdot E}$. This gives the opportunity to estimate the poorly known Fano factor $F$. By neglecting an expected tiny loss in energy resolution due to incomplete charge collection, one gets the following equation:
\begin{equation}
F = \frac{\Delta E^2 - \Delta E_{el}^2}{2.355^2\cdot {\cal E} \cdot E}
\label{eq:fano-factor}
\end{equation}
The fit in Figure~\ref{fig:er_vs_ene} gives a noise contribution of $\Delta E_{el}$=\\(331$\pm$36)\,eV, which coincides well with pulser resolution measurements performed on single BEGe detectors. The fitted value of the Fano factor is $F$=(0.079$\pm$\\0.006). This is comparable with recently published va\-lues of $F$, which lie in the range [0.05,0.11] \cite{bib:fano-publications}. For a more precise determination of $F$, a ballistic deficit correction at higher energies, a precise measurement of the noise term via a very stable test pulse generator, and a potential energy dependence of $F$ (visible especially at lower energies) should be considered. 
\paragraph {\bf Comparison with P-PC detectors of Majorana:} The \textsc{Majorana} collaboration disposes of a large number of HPGe detectors that are similar to BEGe detectors as used in \textsc{Gerda}. Thus, it is interesting to compare the energy resolutions obtained in vacuum conditions for the two samples. There are 35 \textsc{Majorana} detectors also enriched in $^{76}$Ge to 88\% and produced by the company AMETEK/ORTEC in Oak Ridge, TN, USA \cite{bib:ametek-ortec}, utilizing the p-type point contact (P-PC) detector design. With an average of 840\,g per detector, they amount to a total mass of 29.7\,kg \cite{2015_MJD}.\\
\indent The characterization tests of the \textsc{Majorana} detectors were performed in the Sanford Underground Research Facility (SURF) and are similar to the ones in \textsc{Gerda}. The average energy resolution at 1333\,keV of the \textsc{Majorana} P-PC detectors is 1.9\,keV which is in good agreement with the vendor's specification. The average value is by 10\% larger than the one obtained for the \textsc{Gerda} Phase~II BEGe detectors but compatible within their SDs. For the case of a 840\,g detector, which corresponds to the average \textsc{Majorana} P-PC detector mass, the linear dependence of the energy resolution on the detector mass established for the \textsc{Gerda} Phase~II BEGe detectors (see previous paragraph) would lead to a FWHM of 1.76\,keV for the latter detector design. At lower energies the \textsc{Majorana} P-PC detectors perform also similarly: the FWHM of the $^{57}$Co 122\,keV $\gamma$-line of the \textsc{Gerda} and \textsc{Majorana} detectors were measured by the manufacturer Canberra Olen as well as AMETEK/ORTEC. In the case of the 30 \textsc{Gerda} Phase~II BEGe detectors, the average value is 641\,eV with a SD of 26\,eV, while in the case of the \textsc{Majorana} P-PC detectors the average value was quoted to be 624\,eV \cite{2015_mertens}.

%%%%%%%%%%%%%%%%%%%%%%%%%%%%%%%%%%%%%%%%%%%%%%%%%%%%%%%%%%%%%%%%%%%%%%%%%%%%%%%%
\section{Full charge collection depth and active volume}
\label{sec:full-charge-collection-depth-and-active-volume}

\subsection{General remarks}
\label{sec:remarks-active-volume}

\begin{figure}[h]
\begin{center}
\includegraphics[width=0.35\textwidth]{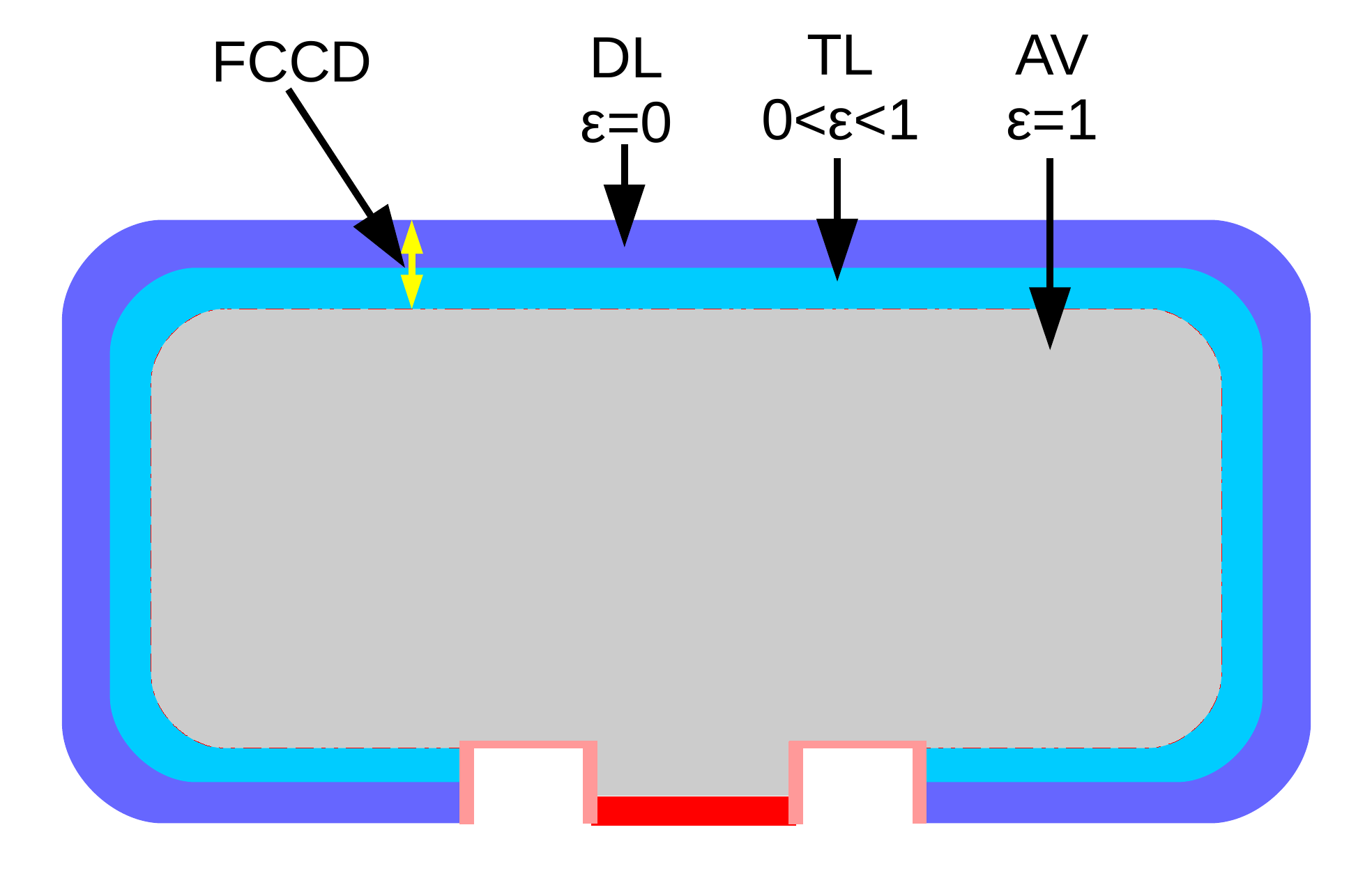}
\caption{Conceptual representation of the full charge collection depth (FCCD) and active volume (AV) of the \textsc{Gerda} Phase~II BEGe detectors. The charge collection efficiency in the dead layer (DL), transition layer (TL) and AV is denoted with $\epsilon$. Moreover, the read-out p+ electrode is depicted in red, while the inactive wrapped around n+ contact is drawn in blue. The insulating ring between the two electrodes is shown in pink. For further details see the text.}
 \label{fig:fccd-scheme}
\end{center}
\end{figure}

This chapter is devoted to the active volume (AV) and the full charge collection depth (FCCD) results of the p-type \textsc{Gerda} Phase~II BEGe detectors. A conceptual representation of the two non-standardly named quantities is depicted in Figure~\ref{fig:fccd-scheme}. While in the first case it corresponds to the detector volume with complete charge collection efficiency (CCE), in the second case it is a one-dimensional parameter describing the thickness of a dead layer (DL) with zero CCE plus a transition layer (TL) with partial CCE. Only $\gamma$-rays depositing their entire energy in the AV can end up in a respective full-energy peak, which is mandatory for the identification of the hypothetical 0$\nu\beta\beta$ decay. This explains why a correct determination of the AV is important for a precise exposure calculation in \textsc{Gerda}. Under the assumption that the FCCD is equally thick across the entire surface and there are no less efficient subregions, the AV should be equal to the crystal vo\-lume minus the volume of the surrounding layer with a thickness corresponding to the FCCD. This allows one to use either surface-sensitive low energy $\gamma$-ray probes to measure the FCCD directly, or bulk-sensitive high energy sources to directly probe the AV. Within this work, both types of sources have been used to deduce the FCCD and AV of the BEGe detectors. The metho\-dology and results are presented in the following Sections \ref{sec:methodogy-active-volume} and \ref{sec:results-active-volume} respectively.\\
\indent In a complementary study, that will not be further described here, the same calibration data were used to model the TL alone and to simulate background events that partly deposit their energy in the TL.
Due to the lack of an electric field in the TL, charges have to diffuse from the TL to the AV. Since the diffusion velocity is typically smaller than the drift velocity, events ge\-nerated in the TL have a longer rise time. Such characteristic `slow pulses' can be efficiently rejected via pulse shape analyses techniques (cf. Chapter \ref{sec:pulse-shape-behavior}). For details see \cite{bib:diss_lehnert2013}.

\subsection{Methodology}
\label{sec:methodogy-active-volume}

The basic principle behind the FCCD and AV determination is a spectral comparison of a calibration source measurement with a MC simulation, which simulates the same experimental setup and varies the FCCD \linebreak around the expected one. In order to achieve the highest possible precision, several prerequisities have to be fulfilled:
\begin{itemize}
\item Optimized experimental setup,
\item Different source types with complementary observables,
\item Exact description of the experimental setup in the MC simulation,
\item Careful investigation of all potential systematic effects arising from the experiment and from the MC.
\end{itemize}
\indent In order to accomplish the first two criteria, two surface-sensitive type of sources, $^{241}$Am and $^{133}$Ba, and one bulk-sensitive source based on $^{60}$Co were selected. The $^{241}$Am and $^{133}$Ba sources typically had activities of se\-veral tens of kBq, while the $^{60}$Co sources had activities of $\sim$(6-14)\,kBq. For data collection, the calibration devices were then installed at a distance of $\sim$20\,cm from the cryostat end caps inside an optimized lead-copper shield as described in \cite{bib:heroica_infra}.\\
\indent For the third criterion, the geometries of the setup, of the detectors and of the sources were implemented very accurately in the MC. The chemical composition and density of each component were investigated and re-evaluated. Especially metal components tur\-ned out to have sometimes wrong specifications. For instance, the used cryostats turned out to be made not of pure Al, but of an Al alloy with Mg, Si, Cu and Cr additions, which notably affects the absorption length of low-energy $\gamma$-rays. For the simulation part, the MC framework \textsc{Mage} \cite{bib:mage} was used. The simulations included a fine-grained scan of the FCCD in 150 equidistant steps between [0,1.5]\,mm.\\
\indent To fulfill the fourth requisite, the impact of 34 potential systematic effects was investigated. These can originate from the MC physics processes, the radioactive sources, the properties of the cryostat and the included diode, from data collection and data analysis. A partial list containing the most relevant effects has already been reported in Table~8 of \cite{bib:first-hades-paper}. The final total uncertainty budget was divided into detector correlated and non-correlated parts. An example for the first ca\-tegory is the usage of the same calibration source for each detector, which --in case of an offset-- would translate into an asymmetric shift in one direction for all FCCD/AV mean values. Both terms are considered in \textsc{Gerda} Phase~II data analyses.\\
\indent For the analysis of the energy spectra, the $\gamma$-ray peaks in the measured and MC simulated spectrum were evaluated via a fitting and a counting method. In the case of the surface-sensitive measurements, two groups of $\gamma$-lines were evaluated for each source: \linebreak 59.5\,keV and 99.0\,keV (summed with 103.0\,keV) for \linebreak $^{241}$Am, 79.6\,keV (summed with 81.0\,keV) and 356.0\,keV for $^{133}$Ba. Then the peak count ratios $\rho_{exp}$ were calculated separately for each source and compared with the variable MC ratio $\rho_{mc}$. The real FCCD of the detector was established when the two ratios converge, i.e. $\rho_{exp}=\rho_{mc}$. For the bulk sensitive $^{60}$Co measurements, the absolute count rate $I_{exp}$ of one single $\gamma$-line (either 1173\,keV or 1333\,keV) was evaluated, whereas the source activity and dead time have to be known with high precision. Correspondingly, a MC of the same source-detector configuration was performed for variable FCCD values. The intersection of the experimental value $I_{exp}$ with the simulated curve $I_{mc}$(FCCD) is expected to agree with the real FCCD of the detector. An illustration of the two approaches can be found in Figure~10 and 11 of \cite{bib:first-hades-paper}.

\subsection{Results}
\label{sec:results-active-volume}

\begin{figure}[h]
\begin{center}
\includegraphics[width=0.53\textwidth]{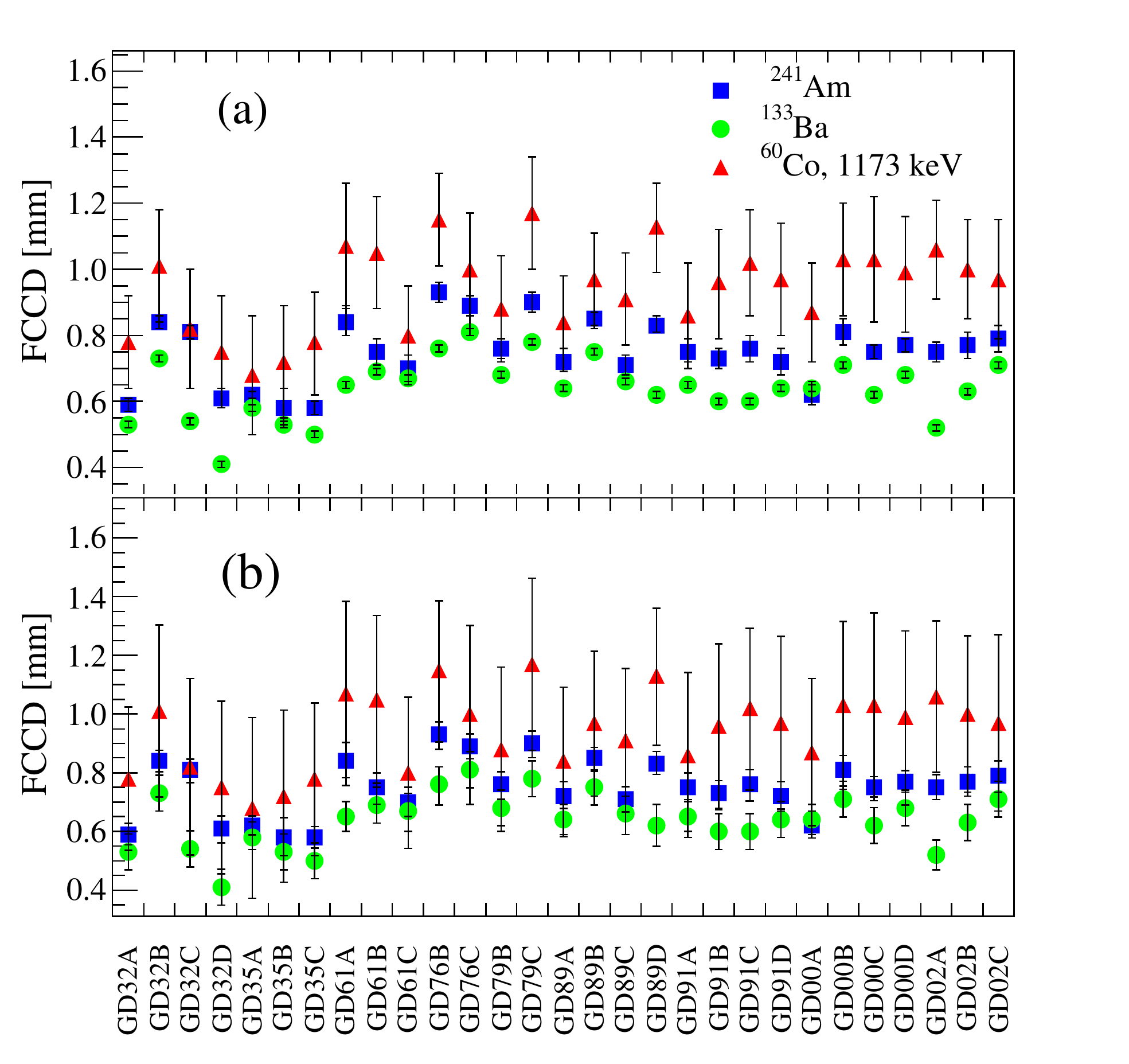}

\caption{FCCD values of \textsc{Gerda} all 30 \textsc{Gerda} Phase~II BEGe detectors except of GD02D. The FCCD values in plot (a) contain only the uncorrelated uncertainties, while in plot (b) they contain the combined correlated and uncorrelated uncertainties.}
 \label{fig:30_fccd_2-BEGe-paper}
\end{center}
\end{figure}

\paragraph{\bf FCCD and AV from different source measurements:} The FCCD results of the 29 well working \linebreak  \textsc{Gerda} Phase~II BEGe detectors are reported in Table~\ref{tab:fccd-single-source-measurements}. The results are based on different calibration source irradiations. The first two columns summarize the FCCD values obtained from the surface sensitive $^{241}$Am and $^{133}$Ba $\gamma$-lines, while the third and fourth column represent the outcome from the two bulk sensitive $^{60}$Co $\gamma$-rays. The detectors for which the systematic effects in the determination of the FCCD were kept small are denoted with a (+) sign. The corresponding FCCD values are more reliable and can be used as reference detectors in \textsc{Gerda} Phase~II physics data analyses. Vice versa, there are detectors with less reliable FCCD values, which are marked with (-). Detectors with e.g. an asymmetric shape and large mass difference $\Delta M$ (cf. Section~\ref{sec:Dimensions-and-masses-of-BEGe-detectors}) belong to this class, since the applied mass correction might compensate the mass discrepancy, but still does not agree with the real shape. All the mentioned FCCD values are also represented in Figure~\ref{fig:30_fccd_2-BEGe-paper}.\\
\indent By comparing the four sets of FCCD values obtained with three different calibration sources by \textsc{Gerda} and with one source by Canberra Olen, it is possible to conclude: 
\begin{itemize}
\item The $^{241}$Am-based values determined by \textsc{Gerda} are in good agreement with the manufacturer's specifications. The difference is typically not larger than $\pm$0.1\,mm. The $^{241}$Am-based average FCCD values of the first (GD32 to GD35 series) and second batch (GD61 to GD02 series) are 0.66\,mm and 0.77\,mm, respectively. These match the targeted values of 0.5 and [0.7,1.0]\,mm well. 
\item The FCCD values from different source measurements on a single detector agree within the combined correlated and uncorrelated uncertainties, however the mean values typically fulfill the inequality:
\begin{equation}
\text{FCCD}(^{133}\text{Ba})<\text{FCCD}(^{241}\text{Am})<\text{FCCD}(^{60}\text{Co})
\label{eq:av-discrepancy-inequality}
\end{equation}
This reappears in the average FCCD values of all 29 working detectors, which are summarized in Table~\ref{table:fccd-discrepancy}. Only the two $^{60}$Co-based results agree well within uncertainties. For a BEGe detector with the average mass of 667\,g, the two average FCCD values from the $^{133}$Ba and $^{60}$Co calibrations would translate into an AV fraction of 89.0 and 91.5\%, resulting in a difference of 2.5\%. None of the 34 investigated systematic effects was able to explain the discrepancy. One of few remaining, but not in deep investigated possibilities might be related to the simulated energy-dependent electron-hole cloud size induced by energy depositions in $\gamma$-ray source irradiations. If the \textsc{Geant4} description is correct, than the observed Inequation \ref{eq:av-discrepancy-inequality} is real and would mean that the FCCD/AV is an energy-dependent quantity. However, if the description in the MC simulation (e.g. the thermal diffusion processes on crystals and condensed matter physics) was incomplete at that time, then the Inequation \ref{eq:av-discrepancy-inequality} might be due to this artifact and probably more pronounced at higher energies.
\begin{table}
\begin{center}
\caption{\rm{\textsc{Gerda} Phase~II BEGe detectors: averaged FCCD values based on different source types. The uncertainties include the correlated and uncorrelated averaged terms. The relative difference corresponds to the difference of $^{133}$Ba and $^{60}$Co based FCCD averaged values from the $^{241}$Am based result.}}
	\vspace{0.3 cm}
	\begin{tabular}{|l|cc|}
	\hline
	source type 				&average FCCD 							&rel. difference 	\\
	\hline
	$^{241}$Am				&0.75$^{+0.04}_{-0.05}$\,mm				&0\%				\\
	$^{133}$Ba				&0.64$^{+0.06}_{-0.06}$\,mm				&-15\%			\\	
	$^{60}$Co, 1173\,keV 	&0.94$^{+0.28}_{-0.28}$\,mm				&+25\%			\\
	$^{60}$Co, 1333\,keV	&0.95$^{+0.28}_{-0.28}$\,mm				&+27\%			\\
	\hline
    \end{tabular}
	\label{table:fccd-discrepancy}	
\end{center}
\end{table}
\end{itemize}
By comparing the values of one specific FCCD set obtained with one source measurement, one observes a certain variation on a detector-by-detector basis. In the case of the first detector batch, the SD from the average FCCD is $\pm$0.11\,mm, while it is $\pm$0.07\,mm in the case of the second batch. The following effects were excluded:
\begin{itemize}
\item Long-term environmental conditions: 
The first batch was produced from February-March 2012, the second batch from August 2012 to February 2013. No seasonal correlation was found. 
\item Detector related properties: The detector-dependent net impurity concentrations $N_{a-d}$ seem not to influence the penetration depth of lithium (Li) in Ge. Also no correlation with the slice positions was found.
\end{itemize}
Daily environmental changes and not fully stable production conditions might be the most probable explanation for the observed detector-by-detector variations.

\begin{figure}
\begin{center}
\includegraphics[scale=0.42]{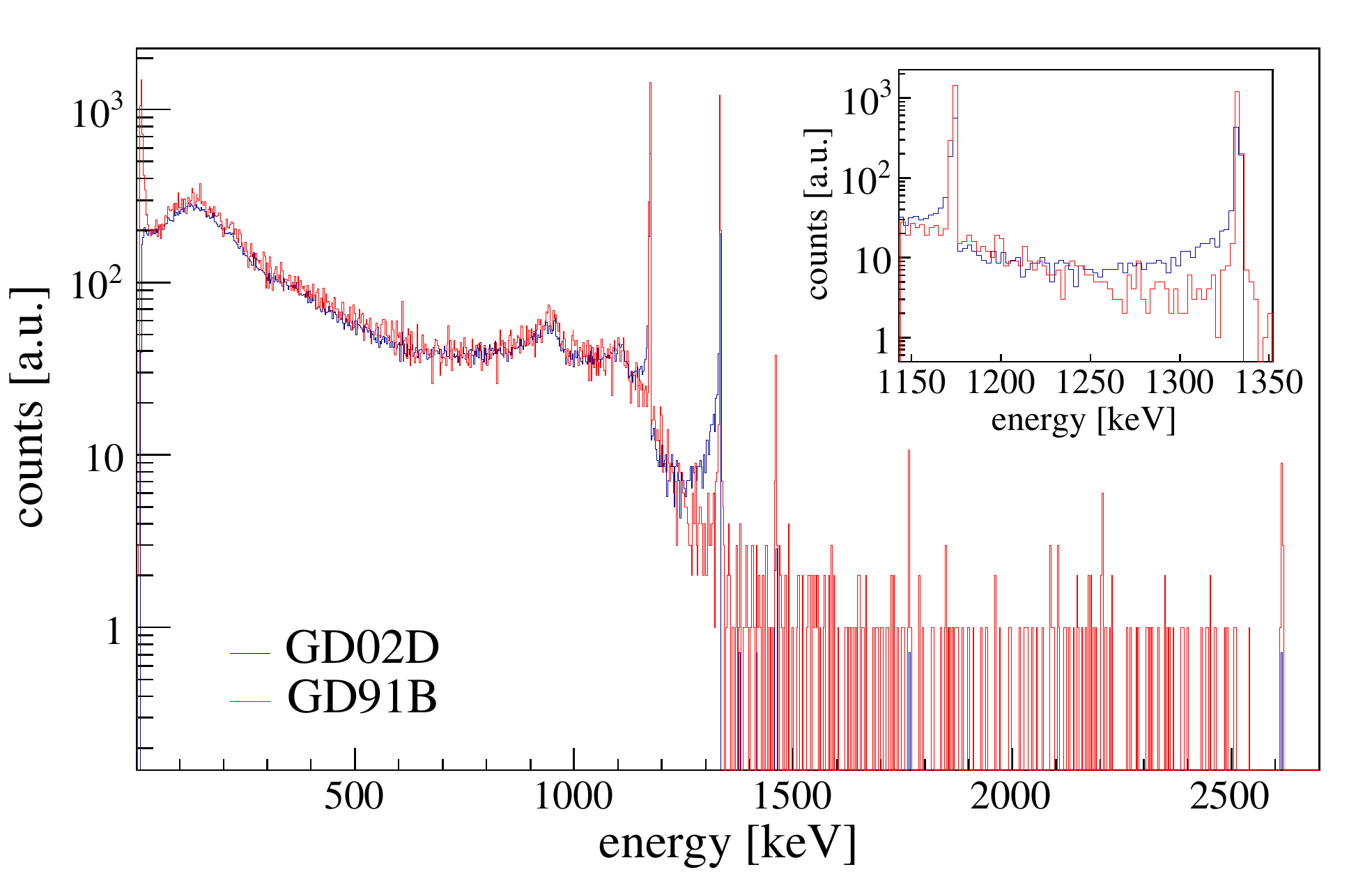}
\caption{Energy spectra of the $^{60}$Co source measurements performed with the detectors GD02D and GD91B.}
\label{fig:GD02D-vs-GD91B_Co60-full-spec_2-BEGe-paper}
\end{center}
\end{figure}

\paragraph{\bf GD02D, a special detector:} Out of 30 delivered BEGe detectors, GD02D is the only one with a non-satisfactory net impurity concentration.\\
\indent In order to characterize and quantify the electrically depleted volume, we first irradiated the detector with the bulk-sensitive $^{60}$Co source. Then the FCCD procedure was applied as done before for the well performing detectors. That way, the measured FCCD of GD02D is 7.2\,mm, which translates into an AV fraction of $\sim$30\% only.\\
\indent The question was whether the residual volume is purely dead or partly also semi-active. Events depositing energy in a semi-active volume would not contribute to the full-energy peak (FEP) count rates but would be shifted to lower energies. This hypothesis was inve\-stigated by comparing GD02D with the well performing detector GD91B. The two detectors GD02D and GD91B have a very similar mass, i.e. 662\,g vs. 650\,g, and their diameter and height are comparable: 74.6\,mm vs. 70.6\,mm, and 27.9\,mm vs. 30.3\,mm. The two detectors were calibrated with $^{60}$Co sources under almost identical conditions. The measured spectra are shown in Figure~\ref{fig:GD02D-vs-GD91B_Co60-full-spec_2-BEGe-paper}. Herein, the GD02D spectrum was normalized by the measuring time of detector GD91B. No correction due to the GD02D mass surplus of 2\% and a slightly different solid angle was applied. The count rates in different ranges of the energy spectra are summarized in Table~\ref{table:diana-vs-calypso_count-rates}. The results demonstrate that the count rates in the FEPs (ID: \#1 and \#3) in GD02D are by factor of 2.2-2.3 lower than in GD91B. A notable part of the missing FEP events are detected in the energy intervals directly below the FEP (ID: \#2 and \#4). Over the entire spectrum (ID: \#5 and \#6), however, GD91B sees 10\% more events than GD02D. Hence, the sum of the active and semi-active volume fraction of GD02D results to be $\sim$80\%, out of those $\sim$50\% are semi-active. Thus, within \textsc{Gerda} Phase~II the detector GD02D is used in detector-detector anti-coincidence mode only. 

\begin{table}[h]
\begin{center}
	\caption{\rm{Comparison of GD02D with GD91B: Count rates in different peaks and energy regions. The measurement time of GD02D was normalized to the measurement time of GD91B, and the count rates scaled accordingly.}}
	\begin{tabular}{|ll|ccc|}
	\hline
ID	&	energy region 			&GD02D	 				&GD91B						&factor 	\\
	\hline
\# 1&	(1333$\pm$3)\,keV		&693					&1429						&2.2		\\	
\# 2&	[1178,1328]\,keV		&683					&406						&0.6		\\	
\# 3&	(1173$\pm$3)\,keV		&758					&1730						&2.3		\\	
\# 4&	[1013,1168]\,keV		&2204					&2030						&0.9		\\	
\# 5&	[100,1336]\,keV			&34144					&38101						&1.1		\\	
\# 6&	[100,1000]\,keV			&29599					&32275						&1.1		\\	
	\hline	
    \end{tabular}
    \label{table:diana-vs-calypso_count-rates}
\end{center}
\end{table}

\paragraph{\bf FCCD and AV values used in GERDA Phase~II:} Even though the FCCD values obtained with different calibration sources agree within the total uncertainty budgets, it is important to clarify which set of FCCD mean values are more suitable for \textsc{Gerda} Phase~II data analyses. By balancing pros and cons, we agreed to use the combined FCCD values from the independent $^{241}$Am and $^{133}$Ba source measurements (see the second column in Table~\ref{tab:241Am133BaFCCDcombined-FCCDgrowing}) for the following reasons:
\begin{itemize}
\item The $^{241}$Am- and $^{133}$Ba-based FCCD values are determined with higher accuracy than the $^{60}$Co-based values.
\item The $^{241}$Am- and $^{133}$Ba-based FCCD mean values agree better among to each other than with the $^{60}$Co-based ones (cf. Table~\ref{table:fccd-discrepancy}).
\item Assuming that the hypothesis of a non-fully correct charge cloud size generation in \textsc{Geant4} is true, this would mainly affect the $^{60}$Co-based values.
\item Background events depositing energy in the dead and transition layer of the Li-diffused detector surfaces can be reproduced better in MC simulations, when using the $^{241}$Am- and $^{133}$Ba-based calibration data (cf. Chapter 8 in \cite{bib:diss_lehnert2013}).
\end{itemize} 
\indent Between the $^{241}${Am}, $^{133}${Ba} and $^{60}${Co} source measurements of the BEGe detectors in vacuum cryostats, and the final detector deployment in \textsc{Gerda} in 2015, the detectors were stored at room temperature for a period of nearly three years. Under such conditions, the Li-diffused FCCD of HPGe detectors can increase. According to several authors \cite{bib:fccd-growth} using p-type HPGe detectors from different vendors, the FCCD growth at room temperature has an average speed of $\sim$0.1\,mm/yr, with a variance of $\pm$0.04\,mm/yr.
\indent Guided by these FCCD growth speed values, a correction on the combined $^{241}${Am} and $^{133}${Ba} FCCD and AV values was applied including a systematic uncertainty of $\pm$50\%. The new FCCD values, AV fractions and active masses are reported in the third to fifth column of Table~\ref{tab:241Am133BaFCCDcombined-FCCDgrowing}. The average FCCD of all BEGe detectors but GD02D increased from 0.70\,mm to 0.98\,mm. Its total uncertainty increased from $^{+0.04}_{-0.07}$\,mm to $^{+0.15}_{-0.16}$\,mm. The average AV fraction $f_{av}$ and active mass $M_{av}$ of all 29 fully operational \textsc{Gerda} Phase~II BEGe detectors become:
\begin{align}
f_{av} = & 0.885^{+0.016}_{-0.015} \text{(uncorr)}^{+0.006}_{-0.003}\text{(corr)} \\
M_{av} = & 17.132^{+0.315}_{-0.294} \text{(uncorr)}^{+0.123}_{-0.063}\text{(corr)}\, \text{kg}	
\end{align}
Compared to the original total active mass of 17.791\,kg, the new value after storage at room temperature for an averaged time of 3\,yr represents an AV reduction of $\sim$4\%.

%%%%%%%%%%%%%%%%%%%%%%%%%%%%%%%%%%%%%%%%%%%%%%%%%%%%%%%%%%%%%%%%%%%%%%%%%%%%%%%%
\section{Pulse shape behavior}
\label{sec:pulse-shape-behavior}

\subsection{General remarks}
\label{Intro_PSA}

Figure~\ref{fig:fccd-scheme} depicts the arrangement of the n+ and p+ electrodes for the BEGe detector design. As described in detail in Section 4.1 of \cite{bib:first-hades-paper} this configuration implies an electric field profile that is beneficial for pulse shape (PS) studies. By analyzing the time profile of the rising edge of the charge signal $Q(t)$ (or the time structure of the induced current signal $I(t)=dQ(t)/dt$) it is possible to identify the topology of energy depositions. Their release in either the bulk region or within the surface layers can be distinguished. The time structure of the signals are also sensitive to crystal lattice effects.\\
\indent In terms of topology, a 0$\nu\beta\beta$ event in the AV would manifest itself as absorption of two emitted $\beta$-particles within a small volume of $\mathcal{O}$(10 mm$^3$). This would correspond to a localized energy deposition, commonly named single-site event (SSE). In contrast, $\gamma$-ray background events of similar energy often undergo multiple Compton scattering leading to well separated energy depositions in multiple sites (MSE). In the point-contact design, the maximum amplitude $A$ of $I(t)$ for a SSE is directly proportional to the deposited energy, independently from the location inside the AV of the detector except for a small region around the read-out electrode \cite{bib:diss_agostini_2013}; in other words, the ratio $A/E$ remains constant for a given total energy $E$. A MSE induced signal, however, is a superposition of several SSE energy depositions which are shifted in time because of different drift times. This reduces the $A/E$ ratio compared to a pure SSE of the same total energy. Another type of background is given by $\beta$-particles that typically have small penetration depths and are partly absorbed in the DL and TL. Their $A/E$ ratio is typically also reduced compared to SSE and the rise time of the $Q(t)$ signal dilated (`slow pulse'). Finally, $\alpha$-particles can penetrate only very thin layers, i.e. the one of the p+ contact and the insulating groove ring between the n+ and p+ electrodes. Such events typically have an enhanced $A/E$ compared to SSE in the AV \cite{bib:diss_agostini_2013}. To summarize, the basic difference in $A/E$ can be used to define a one-parameter cut able to discriminate background from signal-like events, as demonstrated first in \cite{bib:deplBEGe_firstpaper}.\\
\indent Within the \textsc{Gerda} Phase~II detector characterization tests in vacuum, $^{228}$Th and $^{241}$Am $\gamma$-ray sources were used to test the corresponding PS response. The impact of $\alpha$- and $\beta$-emitting sources on the PS was not quantified here, since the used Al cryostats would block these particles from reaching the detector. The procedures applied for $^{228}$Th and $^{241}$Am calibrations are outlined in the next paragraphs.

\subsection{Methodology}
\label{Methodology_PSA}

\paragraph{\bf $^{228}$Th source measurements:} In order to measure the pulse shape discrimination (PSD) capability against $\gamma$-radiation, the BEGe detectors were irradiated for (8-18)\,hours with non-collimated $^{228}$Th sources of \linebreak (1-15)\,kBq activity placed in 20\,cm distance from the cryostat end caps. $^{228}$Th emitters are opportune, since they generate both classes of events. First, 0$\nu\beta\beta$-like SSEs lie in the double escape peak (DEP) of 2615\,keV $\gamma$-rays. Second, background MSEs are present in the full-energy peaks (FEP) at 1620\,keV and 2615\,keV, in the single escape peak (SEP) at 2105\,keV, and in Compton continua, e.g. in the energy interval [2004,2074]\,keV that corresponds to the region of interest (ROI) around Q$_{\beta\beta}$($^{76}$Ge).\\ 
\indent Following the `$A/E$ method', the calculated $A/E$ values of all events are plotted against the energy $E$. Then, DEP $A/E$ events are projected onto the $A/E$ axis (cf. Figures 12/13 in \cite{bib:first-hades-paper}). This DEP $A/E$ distribution $\Phi(A/E)$ is first back\-ground-sub\-tracted by looking at the $A/E$ event distributions in the neighboring energy intervals around the DEP. The resulting $\Phi(A/E)$ is normalized so that its maximum -- denoted here with $\zeta^{max}$ -- lies at $A/E=1$. Then $\Phi(A/E)$ is fitted with a tail and one or more Gaussian functions. The different cases will be discussed in Section \ref{228Th source results}. The width of $\Phi(A/E)$ is defined as:
\begin{equation}
b_{A/E} =\begin{cases}
 FWHM \equiv  2.35\cdot \sigma \,\,\,\,\,\,\,\text{for 1 Gaussian}\\
(\zeta_1^{max} - \zeta_N^{max}) + \frac{(FWHM_1 + FWHM_N)}{2}\\
 \hspace{2.2 cm} \text{for}\,\, N\,\, \text{Gaussians}, N>1.
\end{cases}
\label{eq:AEres-definition}
\end{equation}
where 1 and $N$ denote the two most distant Gaussian functions inside $\Phi(A/E)$. Finally, the fit is used to localize a cut at $A/E<1$ which keeps 90\% of the DEP events. This measure effectively excludes many background events with lower $A/E<1$, while keeping a large fraction of the SSEs.

\paragraph{\bf $^{241}$Am source measurements:} Fine-grained surface scans with low-energy $\gamma$-ray sources are not only useful to determine the exact diode position inside a cryostat end cap and to test the homogeneity of the FCCD (cf. Section \ref{sec:methodogy-active-volume}), but also to study the pulse shape response to local surface energy depositions. Within the \textsc{Gerda} Phase~II BEGe detector characterization tests, such scans with collimated 5\,MBq $^{241}$Am sources mounted on automatized robotic scanning arms were scheduled for all detectors. During the characterization of the first batch, however, the scanning tables had not yet been commissioned completely. Thus, high quality scans are reserved for the second batch of detectors, which will be presented below.\\
\indent Circular scans of the top surface were composed of up to 230 positions distributed over maximal 11 rings with radii of $<$35\,mm. Side surface scans comprised up to 380 positions distributed over a maximum of 11 rings at different heights. In each position, $\sim$10$^3$ events were collected.\\
\indent The pulse shape surface scan data were used for the calculation of the mean rise time $\tau_r$ of the $Q(t)$ signals for different time intervals of the rising edge. By comparing the local difference in $\tau_r$, it was possible to study hole mobilities in Ge. Moreover, the individual $A/E$ distributions of the locally absorbed 60\,keV $\gamma$-rays were analyzed focusing on the relative position of the maxima $\zeta^{max}$ and on the width $b_{A/E}$. These two parameters allow to study local dependencies on the $A/E$ parameter.

\begin{figure}[t]
\begin{center}
  	\includegraphics[scale=.45]{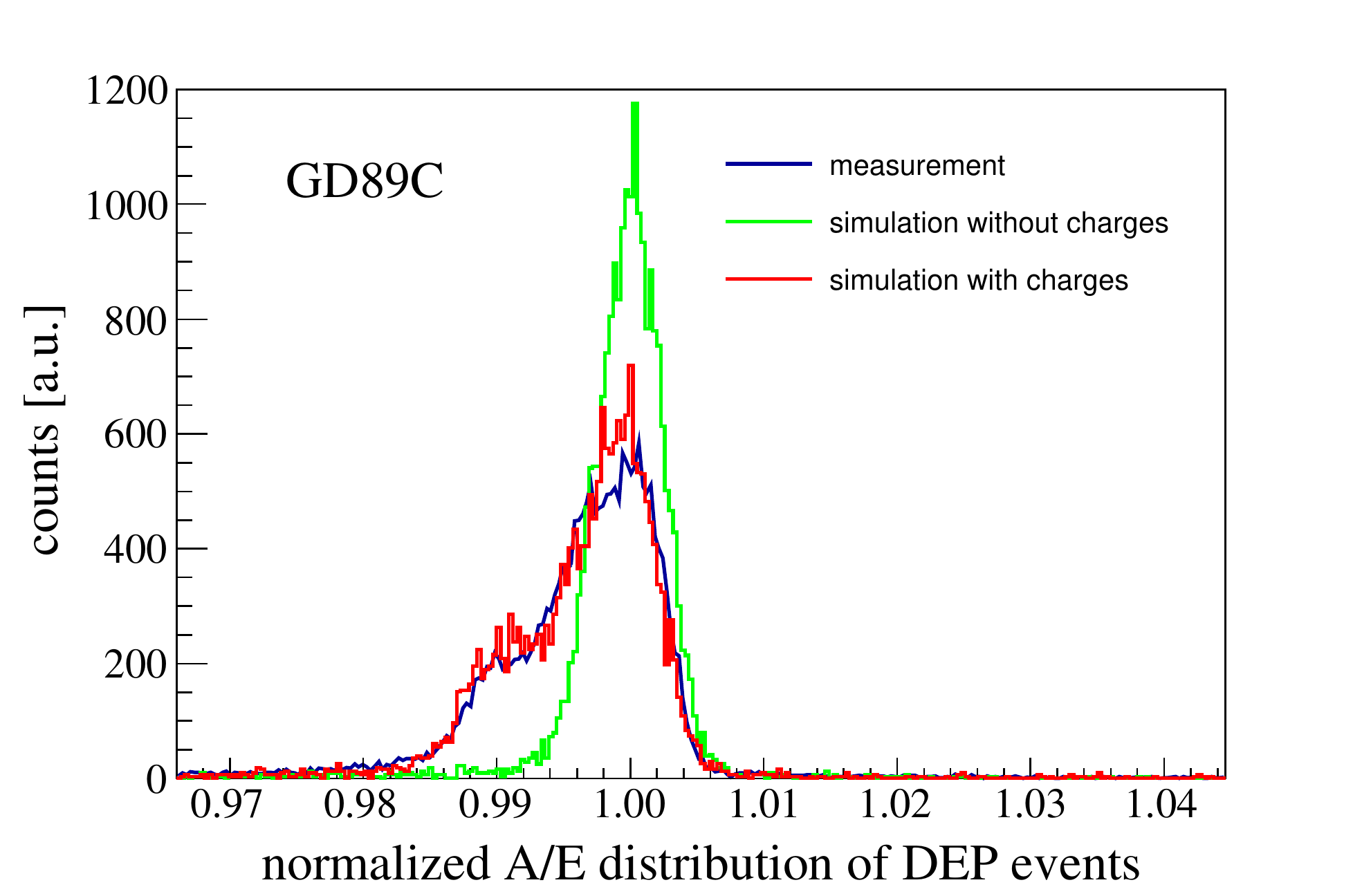}  	
  	\vspace{-0.0 cm}
 	\caption{Detector GD89C: Measurement vs. {\texttt{siggen}}-based simulation of the $A/E$ distribution of DEP from a non-collimated $^{228}$Th source irradiation.}
 	\label{fig:mjsiggen_AEdist_GD89C} 
\end{center}
\end{figure} 

\subsection{$^{228}${Th} source results}
\label{228Th source results}

\paragraph{\bf $A/E$ resolution of the $^{228}$Th double escape peak:}
This section is dedicated to the determined widths of the $A/E$ distributions in $^{228}$Th calibration data. These will be denoted in the following with $b_{A/E}$($^{228}$Th). The calculated $b_{A/E}$($^{228}$Th) values are summarized in Table~\ref{tab:th228-pulse-shape-results}. They lie in the range between 0.32\% and 3.58\%, with an average of 1.42\% and a SD of $\pm$0.90\%. The largest distance of two Gaussian peaks within an $A/E$ distribution is observed for detector GD02A, with a relative distance of 2.3\%. Looking at the individual detectors, the following situation appears:
\begin{itemize}
\item 11 detectors (37\%) have a narrow DEP $A/E$ resolutions, i.e. $b_{A/E}$($^{228}$Th)$<$1\,\%.
\item 13 detectors (43\%) have a half-way broad DEP $A/E$ resolution, with 1$\leq b_{A/E}$($^{228}$Th)$<$2\,\%.
\item 6 detectors (20\%) have a broad DEP $A/E$ resolution with multiple-structured distributions, where $b_{A/E}$($^{228}$Th)$\geq$2\,\%.
\end{itemize} 
\indent 
Quantities such as the energy resolutions are unaffected by a larger $b_{A/E}$($^{228}$Th) values, but the PSD capability can be deteriorated. Prototype BEGe detectors had typically widths in the range of $b_{A/E}$($^{228}$Th)$<$1\,\%. Therefore, the origin(s) of the here observed larger \linebreak $b_{A/E}$($^{228}$Th) values were thoroughly investigated to understand this `$A/E$ anomaly'. A deterioration of the electronics read-out, but also the potential impact of detector intrinsic properties such as the net impurity concentrations, were excluded. Finally a series of hints from detector reprocessing and thermal cycles of single detectors pointed towards a common origin: negatively charged compounds/particu\-lates accumulated inside the groove during diode production. In {\texttt {ADL3}}- and {\texttt{siggen}}-based simulations we tried to reproduce such DEP $A/E$ distributions. One example is shown in Figure~\ref{fig:mjsiggen_AEdist_GD89C}. By simulating ${25}\times$10$^{10}$/cm$^2$ negative charges (4\,$\upmu$C) deposited on the groove surface of GD89C,\linebreak {\texttt{siggen}} succeeds to reproduce the measurement very well. However, it is not yet understood how such a large amount of charge was able to stick to the small groove surface during manufacturing.

\begin{figure}[h]
\begin{center}
\includegraphics[scale=0.44]{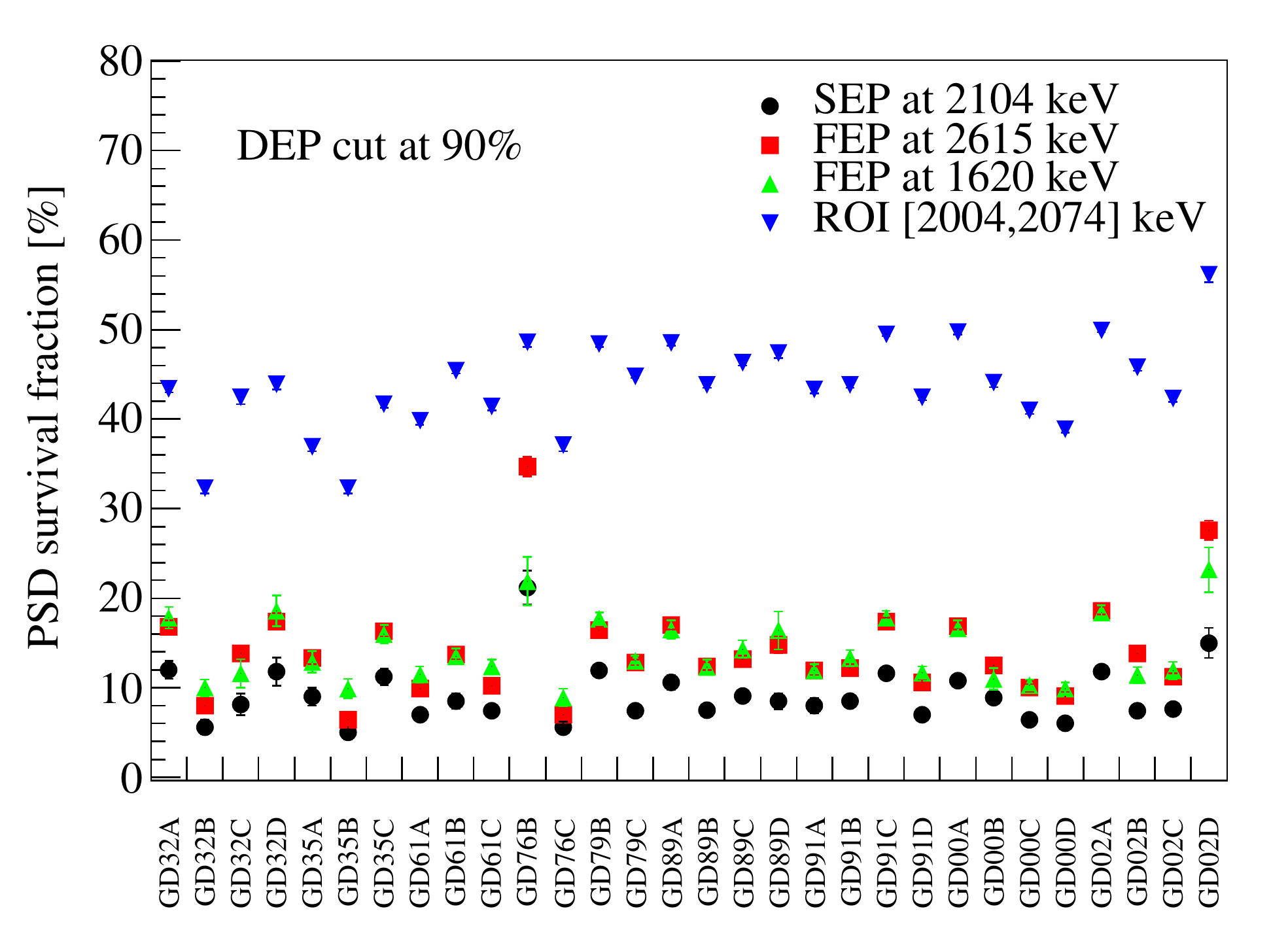}
\caption{Illustration of the PSD efficiencies of different $^{228}${Th}-induced $\gamma$-ray background  in all 30 \textsc{Gerda} Phase~II BEGe detectors operated under vacuum conditions.}
\label{fig:30-BEGe-PSD_eff_dec}
\end{center}
\end{figure}

\begin{figure}[h]
\begin{center}
\includegraphics[scale=0.42]{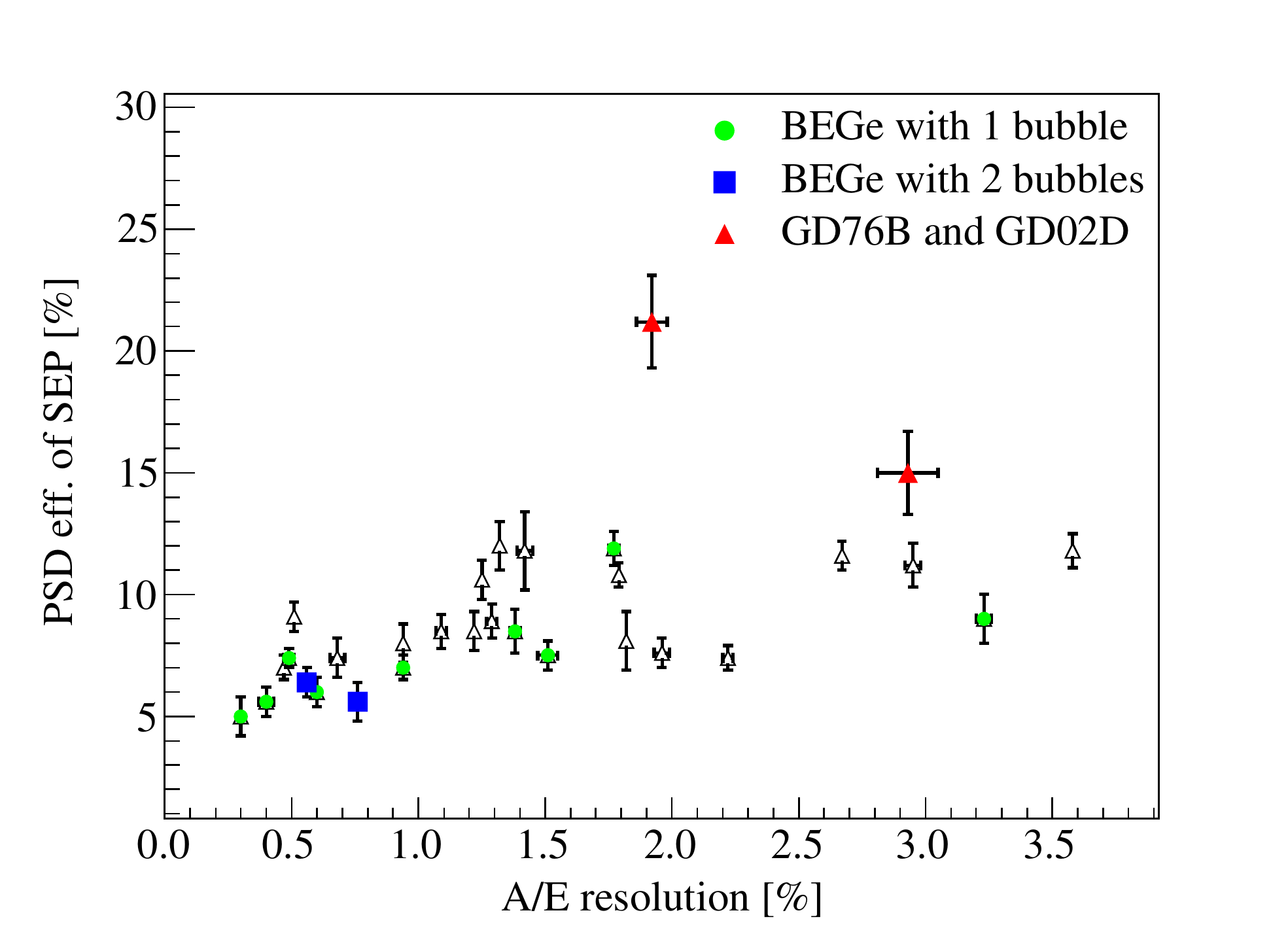}
\caption{PSD efficiency of $^{228}$Th SEP vs. the DEP $A/E$ resolution for all 30 \textsc{Gerda} Phase~II BEGe detectors operated under vacuum conditions. Open symbols are used like in Figure~\ref{fig:Vd-er_vs_c-d-d_for-paper}.}
\label{fig:30-BEGe-PSD_eff-vs-AEres_corr}
\end{center}
\end{figure}

\paragraph{\bf PSD efficiencies of Phase~II BEGe detectors:}
This paragraph describes the capability of the \textsc{Gerda} Phase~II BEGe detectors to discriminate SSE from MSE that are generated from $\gamma$-ray background sources. As described in Section \ref{Methodology_PSA}, the survival probabilities of several FEPs and Compton continua were deduced with 90\% of the DEP events being kept. The obtained PSD efficiencies are plotted in Figure~\ref{fig:30-BEGe-PSD_eff_dec}. The PSD efficiency values including their statistical and systematical uncertainties are summarized in Table~\ref{tab:th228-pulse-shape-results}. The values of the survival probabilities of the SEP, 2615\,keV and \linebreak 1620\,keV FEPs, and in the ROI around Q$_{\beta\beta}$($^{76}$Ge), lie in the range of [5,21]\%, [6,35]\%, [9,23]\%, and [32,56]\%, respectively. The detector with the best performance is GD35B. The detectors with the worst performance are GD76B and GD02D. However, the latter two have already been found earlier to be problematic. Thus, by excluding these two problematic detectors, the relatively large ranges of survival probabilities shrink to [5,12]\%, [6,19]\%, [9,19]\% and [32,48]\%, respectively.
\paragraph{\bf Correlations of the PSD efficiencies with other parameters:}
This paragraph compares the obtained PSD efficiencies with other PS quantities and detector parameters.\\
\indent First, they were compared with the $A/E$ resolution $b_{A/E}$ ($^{228}$Th). In the scatter plot in Figure~\ref{fig:30-BEGe-PSD_eff-vs-AEres_corr} a quantitative view of the PSD efficiencies of the $^{228}$Th SEP as a function of $b_{A/E}$($^{228}$Th), as obtained for all 30 BEGe detectors, is given. As one can see, a dependency exists especially for $b_{A/E}<$1\%, while at larger values this trend is less pronounced. The situation is very similar for the FEPs at 1620\,keV and 2615\,keV as well as for the energy interval [2004,2074]\,keV. A similar trend is found for $A/E$ values of single detectors before and after they were reprocessed or underwent thermal cycles. This can be seen in Table~\ref{tab:th228-pulse-shape-results} for detector GD32D, which was reprocessed and had an improved pulse shape performance afterwards. It is evident that the lower $b_{A/E}$($^{228}$Th) is, the better the PSD background rejection. The reason is that a broader $A/E$ distribution for DEP events results in a low $A/E$ cut position and hence more MSE events with $A/E\approx$1 are accepted.\\
\indent In a second study we investigated whether there are correlations between $b_{A/E}$($^{228}$Th) and other detector parameters, which would permit recursively a fast diagnosis about the expected PSD capability of a detector. Only one small correlation seems to exist: detectors with larger height/mass ratios prefer to populate the lower band of the PSD eff vs. $b_{A/E}$($^{228}$Th) plot. Hence, it seems that smaller net impurity concentration gradients are not only beneficial for obtaining a thicker and thus more massive diodes (and thus less channels in a low background experiment), but improve also the PSD performance, while the energy resolution deterioration will be minimal.
\paragraph{\bf Comparison with other sets of detectors:}
The PSD performance of the \textsc{Gerda} Phase~II BEGe detectors was compared with other sets of detectors operated in vacuum under similar conditions.\\
\indent First, they were compared with six \textsc{Gerda} prototype BEGe detectors that were produced from 
$^{76}$Ge depleted Ge, i.e. the left-over material of the $^{76}$Ge enrichment process \cite{bib:deplBEGe}. In the case of these prototype detectors, the average PSD efficiencies for the $^{228}$Tl SEP at 2104\,keV and for the FEPs at 2615\,keV and 1620\,keV were determined to be 5.4\%, 7.7\% and 10.0\% with SDs of 1.7\%, 2.7\% and 2.4\%, respectively. These were compared with the mean values obtained with all 30 \textsc{Gerda} Phase~II BEGe detectors: 9.1\%, 14.0\% and 13.9\% with SDs  of 3.3\%, 5.7\% and 3.7\%, accordingly. Only 10\% of the \textsc{Gerda} Phase~II BEGe detectors have comparable PSD efficiencies. However, this is due to the fact that also detectors affected by the `$A/E$ anomaly' were included. The latter effect did not appear for the prototype BEGe detectors and was most probably introduced afterwards by a change in the production process by the manufacturer Canberra Olen.\\ 
\indent Second, we also compared the \textsc{Gerda} Phase~II BEGe detectors with the P-PC detectors owned by the \textsc{Majorana} collaboration. For these detectors, the average efficiencies for the SEP at 2104\,keV and for the energy interval at [2000,2050]\,keV are 5.6\% and 43.5\% with SDs of 1.7\% and 7.3\%, respectively (extracted from Figure~3(b) in \cite{2015_mertens}). Under the assumption of almost equal experimental test conditions (and within similar energy ranges) these efficiencies can be compared with 9.1\% and 43.6\% and their corresponding SD values of 3.1\% and 5.1\% obtained with the \textsc{Gerda} Phase~II BEGe detectors. SEP events are almost exclusively of MSE type while continuum events have a large fraction of single Compton scattered $\gamma$'s which are SSE type. The on average better PSD performance of the \textsc{Majorana} P-PC detectors - as indicated by the SEP survival fractions - result therefore in a similar background suppression of $^{208}$Tl background events at $Q_{\beta\beta}$. 

\subsection{$^{241}${Am} source scan results}

\begin{figure}[h]
\begin{center}
  	\includegraphics[scale=.45]{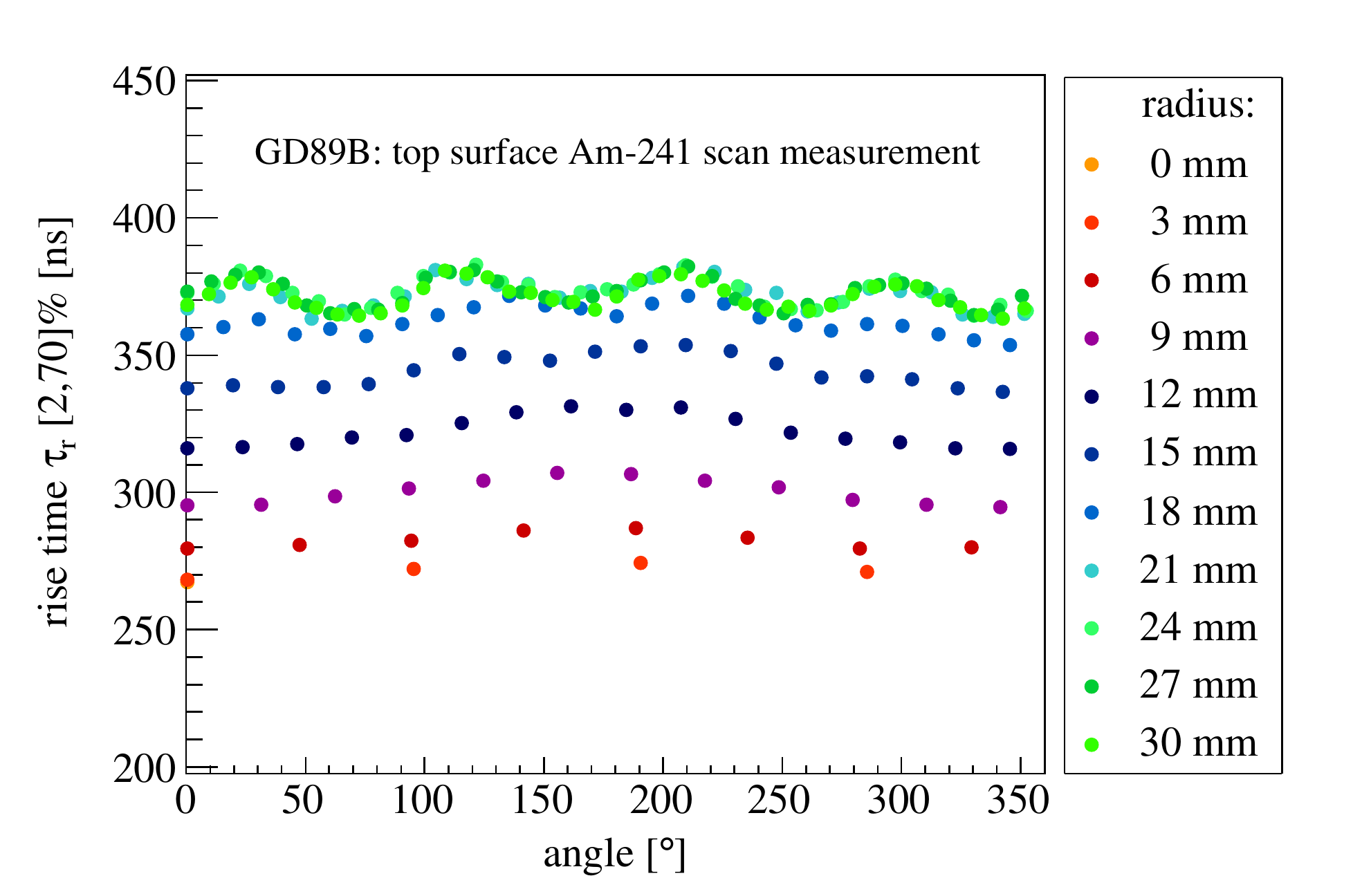}  
  	\vspace{-0.5 cm}
 	\caption{Detector GD89B: Measured rise time curves of circular scans on the top diode surface, with a $\tau_r$ definition in the [2,70]\% interval.}
 	\label{fig:GD89B_risetime_measurement} 
 	\end{center}
\end{figure}

\paragraph{\bf Rise time of $^{241}$Am source induced signals:}
For every scanned position on each detector surface, the mean rise time $\tau_r$ of pulses induced by the 60\,keV $\gamma$-rays was calculated. Among other possibilities, the [2,70]\% and [1,90]\% time intervals on the emerging pulses were favored to define the value of $\tau_r$. The obtained $\tau_r$ values determined at different scan positions were plotted as a function of either radius and angle for top surface scans, or of height and angle for side surface scans.\\ 
\indent The $\tau_r$([2,70]\%) evaluation of the top surface scan performed on detector GD89B is depicted exempla\-rily in Figure~\ref{fig:GD89B_risetime_measurement}. For circles with the largest radius, i.e. 30\,mm, a 90$^\circ$ oscillation pattern becomes clearly visible. The $\tau_r$ has an average of $\sim$370\,ns, while the difference between the minima and maxima is $\sim$20\,ns, i.e. 5\%. The observed 90$^\circ$ oscillation is due to different hole drift mobilities, which depend not only on the applied electric field and the doping levels, but also on the crystal axes, which are aligned along the three crystallographic $\braket{100}$, $\braket{110}$ and $\braket{111}$ directions in Ge crystals. These have a cubic diamond lattice single-crystalline structure throughout the entire volume. The observed `anisotropic mobility' is expected, since the electric field in a macroscopic electrically depleted Ge diode is already rather high and the hole temperature differs substantially from the lattice temperature. In the case of holes, the fast axis becomes $\braket{100}$, while the slowest is $\braket{111}$. A more detailed description can be found for instance in \cite{bib:2006_phd_bruyneel}.\\
\indent The entire \textsc{Gerda} Phase~II BEGe detector survey allowed to draw several conclusions about the observed `anisotropic mobility': 
\begin{itemize}
\item Collimated $^{241}${Am} source data from top detector scans are always suitable for the observation of the 90$^\circ$ oscillation, as long as circular scans with at least $>$25\,mm radius are performed. 
\item The $\tau_r$([2,70]\%) definition is more suitable to extract the 90$^\circ$ oscillation, while
the $\tau_r$([1,90]\%) values are closer to the charge collection process time.
Considering that the BEGe detectors have the same read-out electrode geometry and similar outer dimensions, the average $\tau_r$([1,90]\%) value for a top scan radius of [25,30]\,mm is (520$\pm$50)\,ns. The difference between the minimum and maximum rise time due to the oscillation pattern is typically $\sim$20\,ns.
\item At large radii of top surface scans, some detectors were found to have a second 180$^\circ$ oscillation that superimposes the 90$^\circ$ one (cf. Figure 16 in \cite{bib:first-hades-paper}). It turned out that these detectors are also affected by the `$A/E$ anomaly' in $^{228}${Th} source data, as described in Section \ref{228Th source results}. Moreover, problematic cases like GD02D have an irregular trend in the rise time curves.
\end{itemize}
\begin{figure}[h]
\begin{center}
\includegraphics[scale=0.44]{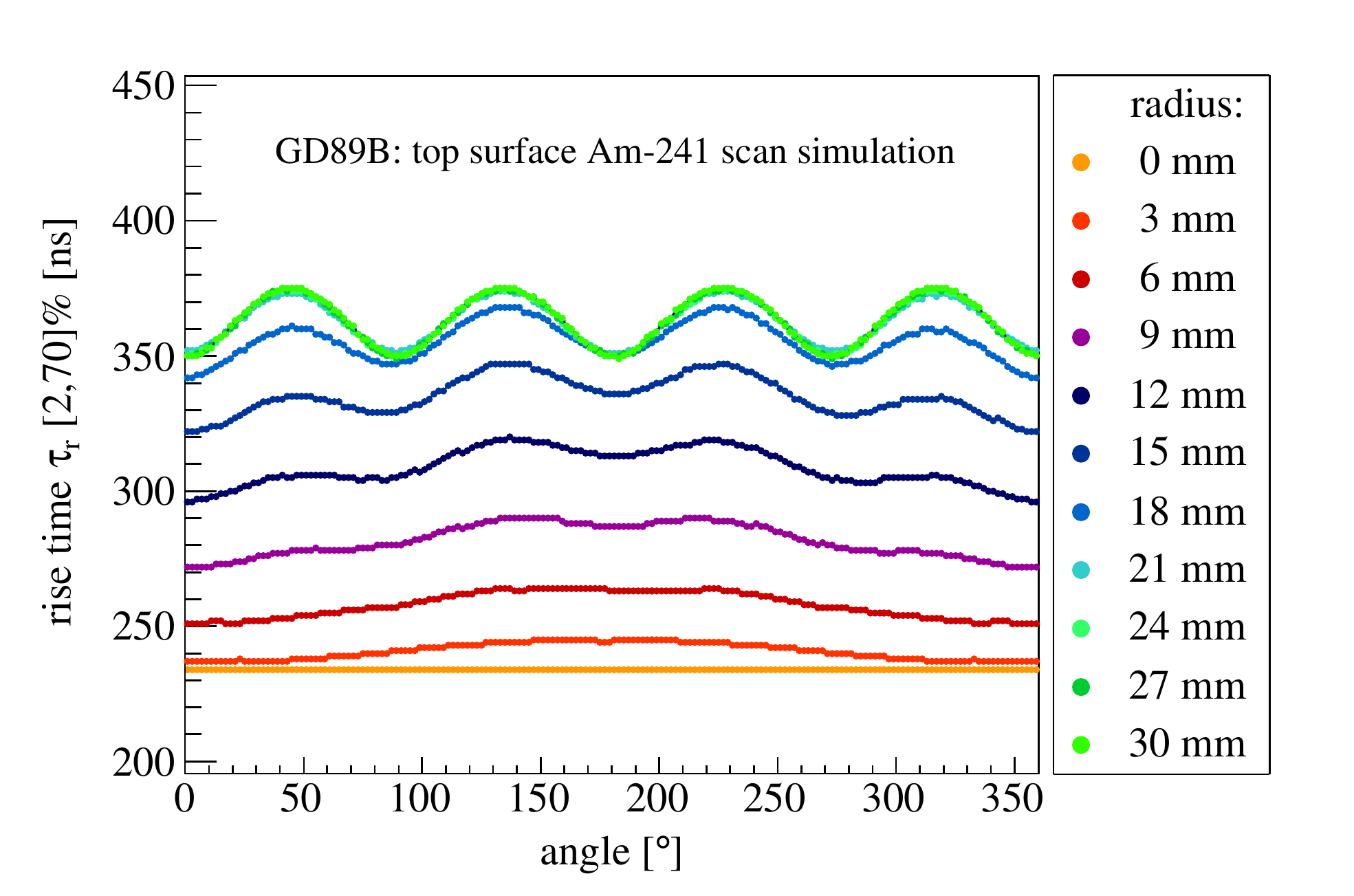}
\caption{Detector GD89B: {\texttt {ADL3}}-based simulation of rise time curves of circular scans on the top diode surface using the $\tau_r$([2,70]\%) definition.}
 	\label{fig:GD89B_risetime_simulation}
\end{center}
\end{figure}
\indent We performed {\texttt {ADL3}}-based simulations of the expected rise time from $^{241}${Am} source scans on several BEGe detectors. The results for the top surface scan of GD89B are depicted in Figure~\ref{fig:GD89B_risetime_simulation}. The simulation predicts a gradual increase in rise/drift time and the 90$^\circ$ oscillation with increasing radius. The calculated $\tau_r$([2,70]\%) values match the experimental data well.\\
\indent At small radii of top surface scans, some detectors were found to have a modulation of the normally rather flat rise/drift time. The simulation is able to reproduce such artifacts, if a misalignment of 1\,mm of the circular scans from the diode central axis in the simulation is assumed. This corresponds indeed to the achievable precision in the experimental alignment procedure (cf. \cite{bib:heroica_infra}).

\begin{figure}[h]
\begin{center}
\includegraphics[scale=0.42]{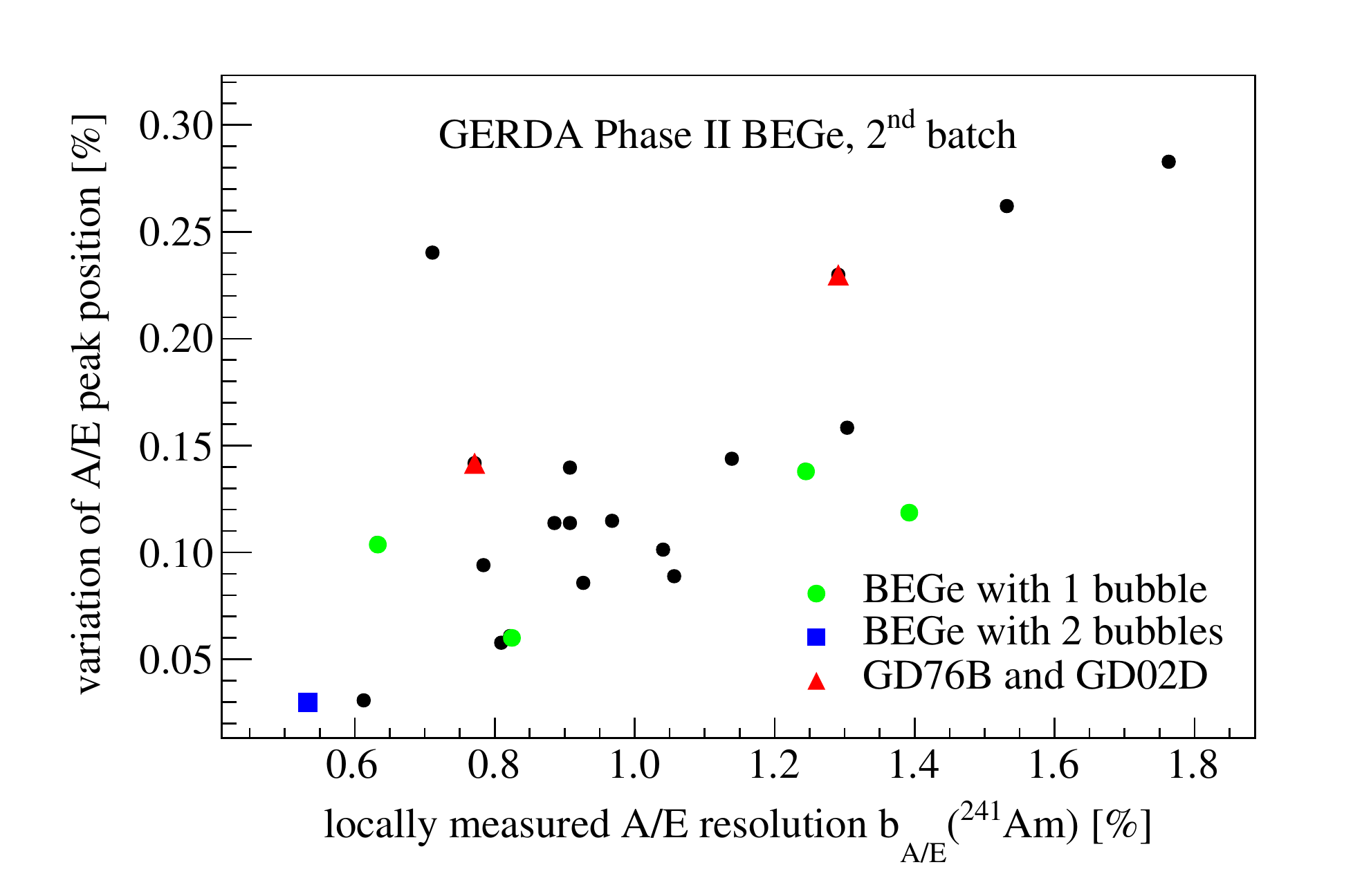}
\caption{Large vs. small scale pulse shape behavior in $^{241}$Am scan data: the global $A/E$ peak position variation along the top surface of BEGe detectors is plotted as a function of the averaged $A/E$ resolution of top surface points obtained with collimated $^{241}$Am source irradiations. Symbols denoted with black dots are appropriated to detectors with no appearance of the `bubble depletion' effect or which exclude the detectors GD76B and GD02D.}
\label{fig:am241_AEpeakposvar-vs-AEres}
\end{center}
\end{figure}

\begin{figure}[h]
\begin{center}
\includegraphics[scale=0.42]{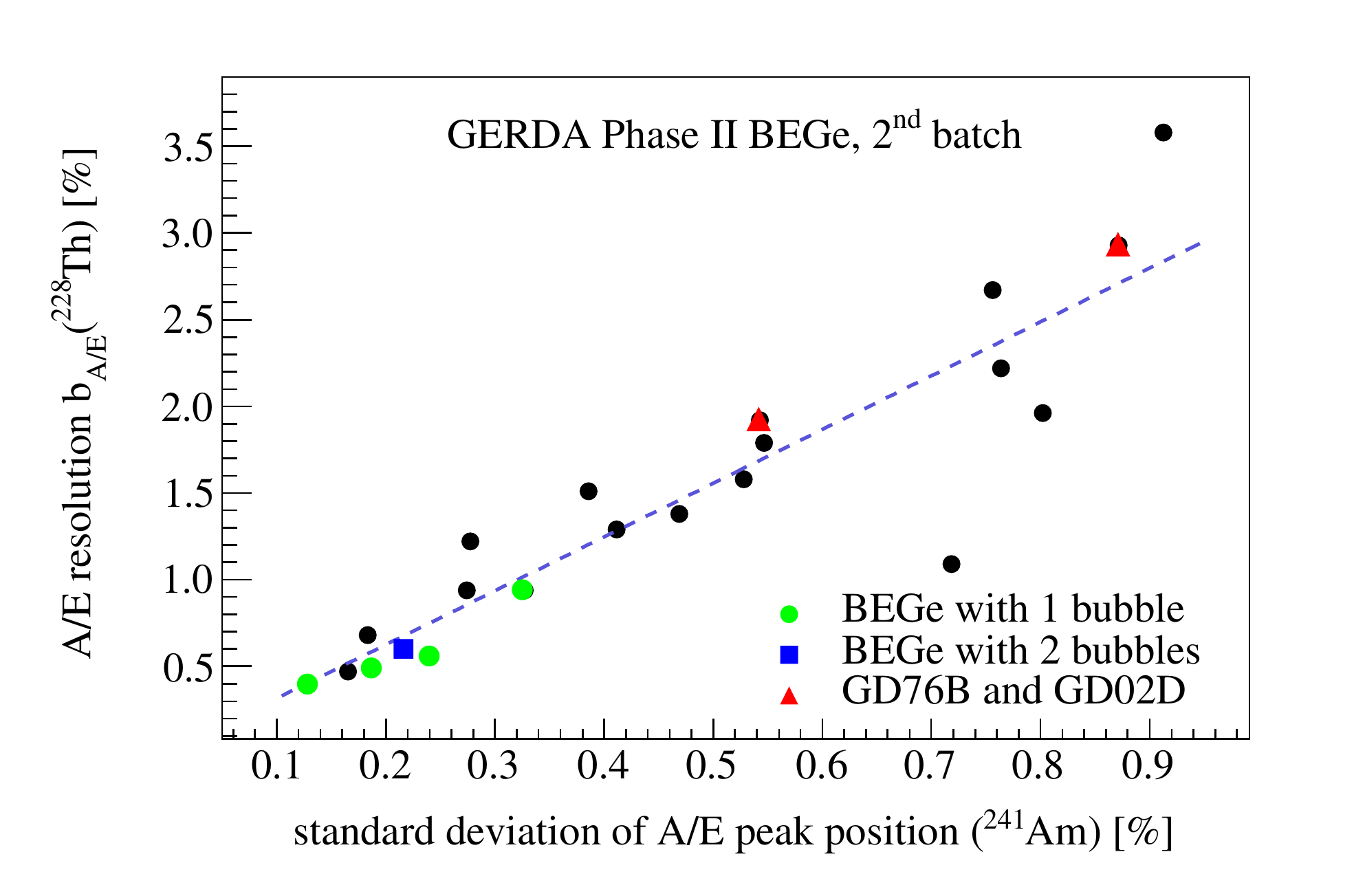}
\caption{High vs. low energy pulse shape behavior: Width of the $A/E$ distribution of 1592\,keV $\gamma$-rays plotted vs. the SD of $A/E$ peak positions from collimated 60\,keV $\gamma$-rays surface irradiations (along a circular scan of maximum radius). Symbols denoted with black dots are used like in Figure~\ref{fig:am241_AEpeakposvar-vs-AEres}.}
\label{fig:am241-AEres_vs_tl208-AEres}
\end{center}
\end{figure}

\paragraph{\bf $A/E$ distributions of 60\,keV $\gamma$-ray pulses in $^{241}$Am source scans:}
In a further detector-by-detector study, the $A/E$ distributions of 60\,keV $\gamma$-ray pulses were analyzed for each scanned position. In contrast to the $A/E$ distributions from $^{228}$Th source measurements, it was found that for $^{241}$Am the distributions do not show multiple structures like those observed for DEP events in non-collimated $^{228}$Th source irradiations (cf. Section \ref{228Th source results}). Thus, the $A/E$ distributions of the collimated 60\,keV $\gamma$-rays can be fitted with a Gaussian function plus a low-side energy tail. The latter term consi\-ders events affected by the not-fully efficient transition layer rather than by Compton-induced events. Quantities like the maximum or mean of the $A/E$ distributions ($A/E$ peak position) and the width, i.e. the FWHM of the $A/E$ distribution $b_{A/E}$($^{241}$Am), were extracted and compared among different positions on a top or side surface scan. The following observations were made:
\begin{itemize}
\item Anisotropic mobility: Next to the 90$^\circ$ oscillation observed for the rise time (cf. previous paragraph) there is an analogous oscillation pattern for the $A/E$ peak position. This, however, is not always as strongly pronounced as in the case of the rise time (cf. Figure 16 in \cite{bib:gerda_intro1}).
\item Large vs. small scale: The variation of the 60\,keV $A/E$ peak positions of all scanned points reflects the macroscopic pulse shape changes along a detector surface $A_d$ of a diode ($\mathcal{O}$($A_d$)=30\,cm$^2$). In a first comparison, the variation of the $A/E$ peak positions was compared with the width of the $A/E$ distributions $b_{A/E}$($^{241}$Am), generated by many points of an $^{241}$Am source scan. The irradiated areas $A_i$ are typically smaller than $\mathcal{O}$($A_i$)=0.5\,cm$^2$ and the energies are deposited nearby. As shown in Figure~\ref{fig:am241_AEpeakposvar-vs-AEres}, there is a correlation between the large and small scale behavior of the $A/E$ parameters. Detectors affected by the `bubble depletion' effect do not populate a specific region.
\item High vs. low energy: The large scale $A/E$ variation of low energy 60\,keV $\gamma$-ray events was compared with the behavior of high energy 1592\,keV $\gamma$-ray events obtained in non-collimated $^{228}$Th source measurements (cf. Section \ref{228Th source results}). Figure~\ref{fig:am241-AEres_vs_tl208-AEres} depicts the obtained width of $^{228}$Th DEP $A/E$ distribution $b_{A/E}$ ($^{228}$Th) versus the SD of the $A/E$ peak maxima from local 60\,keV $\gamma$-rays irradiations, that were obtained within a circular scan at maximum radius. A clear trend becomes visible, especially for the side surface scan: the broader the DEP $^{228}$Th $A/E$ distribution, the larger the SD of the 60\,keV $A/E$ peak positions. Therefore, the `$A/E$ anomaly' that causes larger $A/E$ distributions at high energies does ma\-nifest itself also at low energies. To conclude, detectors affected by the `bubble depletion' effect seem to perform better than most of the rest.
\end{itemize}

%%%%%%%%%%%%%%%%%%%%%%%%%%%%%%%%%%%%%%%%%%%%%%%%%%%%%%%%%%%%%%%%%%%%%%%%%%%%%%%%
\section{Summary and conclusions}
\label{sec:summary}

For Phase~II of the \textsc{Gerda} experiment, we have procured 30 new Broad Energy Ge (BEGe) detectors by the company Canberra. Prior to their integration at the experimental site, the detectors have been thoroughly tested in vacuum cryostats. This characterization campaign has led to the most detailed and extensive survey of high purity Ge (HPGe) detectors of the same design. These studies have allowed to search for correlations between different parameters and to test electric field calculations based on the \texttt{ADL3} and \texttt{siggen} codes. The most important experimental findings have been reported.\\
\indent First, the characterization tests confirmed the excellent energy resolution of the new detectors, with an average FWHM=(1.73$\pm$0.07)\,keV at the  reference \linebreak 1333\,keV $\gamma$-line. It was found that the energy resolution has a small dependence on the detector mass. The obtained energy resolutions do not only represent an improvement compared to the former semi-coaxial design, but are in general the best values obtained by a detector technology employed in 0$\nu\beta\beta$ search. The energy dependence of the energy resolution was also investigated. The related but not well known Fano factor was estimated to be (0.079$\pm$0.006). This is in agreement with recent results.\\
\indent Second, a careful examination of the full depletion voltage of the detectors via high voltage scans allowed us to revise the values recommended by the manufacturer. The new values turned out to be on average 600\,V lower than the recommended ones. This knowledge is used in \textsc{Gerda} Phase~II to prevent the development of prohibitive high leakage currents in a few proble\-matic channels. A correlation between full depletion voltage, net impurity concentration and diode dimensions could be established for the BEGe design. Moreover, the high voltage scans revealed that around 40\% of the new detectors are affected by the `bubble depletion' effect. In most of these cases a single `bubble' was observed, in two detectors even two independent `bubbles' for the first time. The `bubbles' appear only at well-defined high voltages. Thus, the successful reproduction of them by simulations is a unique validation test for the field calculation codes.\\  
\indent Third, a large effort was made to determine precisely the full charge collection depth (FCCD) and active volume (AV) of the detectors. The measurements were carried out with surface sensitive $^{241}$Am and $^{133}$Ba sources as well as with bulk sensitive $^{60}$Co $\gamma$-ray probes. Compared to $^{60}$Co, the results based on the first two sources turned out to have the smallest total uncertainty. Thus, these results were combined, then corrected for ageing effects caused by three years of storage at room temperature. Out of an initial crystal mass of (20.024$\pm$0.030)\,kg used for the construction of all 30 BEGe detectors, ($17.13^{+0.32}_{-0.29} \text{(uncorr)}^{+0.12}_{-0.06}\text{(corr)})\, \text{kg}$ are active and sensitive to 0$\nu\beta\beta$ events. Even though 34 systematic effects were considered, it remained unclear why the mean FCCD values from the higher energetic $^{60}$Co source are systematically larger. One of few remaining explanations might be related to the charge cloud size model  used in the simulation code. If the description is correct, then the FCCD/AV would be an energy-dependent quantity. However, if the models are incomplete, the discrepancy would be an artifact of the simulation.\\
\indent Fourth, the pulse shape behavior of the BEGe detectors was investigated. In a first study, non-collimated $^{228}$Th source tests allowed to define the discrimination efficiency of $\gamma$-induced radiation inside the bulk from signal-like events. While keeping 90\% of the signal-like proxies, $\gamma$-lines were suppressed on average by (86-91)\% and the Compton-continuum around the ROI by 56\%. This suppression is much better than for the former semi-coaxial Ge detector design, but is slightly deteriorated compared to the prototype BEGe detectors. A series of detector manipulations and tests as well as electric field calculations pointed towards a common origin of the observed anomaly, namely charges collected around the read-out electrodes during the ma\-nufacturing process. Finally, fine-grained scans with collimated $^{241}$Am sources allowed to test the pulse shape response of events generated close to the detector surface. In cases when a deteriorated $^{241}$Am pulse shape performance at small scale (via collimated irradiations) was found, a deteriorated behavior at large scale (non-collimated), as well as in $^{228}$Th calibration data was observed. Additionally, the scans allowed to visualize the crystal lattice orientation due to the expected hole drift anisotropy. Electric field calculations were able to reproduce the results.\\
\indent To summarize, the performed BEGe measurement campaign offered a unique possibility to collect a large variety of results, out of those several were incorporated in the standard \textsc{Gerda} Phase~II data collection and analysis procedure. In addition, the developed and improved electric field calculations advanced to valuable tools to interpret observed phenomena in HPGe detectors. The combination of dedicated measurements and validation via proper simulation codes has not only become useful for us, but will also be important for future HPGe-based experiments, that will face an even larger number of detectors.

%%%%%%%%%%%%%%%%%%%%%%%%%%%%%%%%%%%%%%%%%%%%%%%%%%%%%%%%%%%%%%%%%%%%%%%%%%%%%%%%
\section{Acknowledgements}
\label{sec:acknowlegments}

We are grateful to Canberra Industries Inc. in Oak Ridge, TN, USA and Canberra Semiconductor N.V. in Olen, Belgium, for the excellent cooperation and endu\-ring assistance. 
Furthermore, we express our gratitude to the team of EIG EURIDICE on the premises of the Nuclear Research Center SCK$\cdot$CEN in Mol, Belgium, for hosting the \textsc{Gerda} screening facility \textsc{Heroica} and for daily support. For technical and DAQ support we thank F.~Costa, K.~J\"anner, L.~Modenese, C.~Ur, M.~Reissfelder, M.~Turcato, and J.~Westermann. We acknowledge the radioprotection departments of MPIK, University of T\"ubingen, MPP, and SCK$\cdot$CEN, IRMM, as well as the company Eckert \& Ziegler Nuclitec GmbH for all items related to the radioactive sources used in this work.\\ 

The \textsc{Gerda} experiment is supported financially by the German Federal Ministry for Education and Research (BMBF), the German Research Foundation \linebreak (DFG) via the Excellence Cluster Universe, the Italian Istituto Nazionale di Fisica Nucleare (INFN), the Max Planck Society (MPG), the Polish National Science Centre (NCN), the Foundation for Polish Science (TEAM/2016-2/17), the Russian Foundation for Basic Research, and the Swiss National Science Foundation (SNF). The institutions acknowledge also internal financial support. This project has received fun\-ding/support from the European Union's \textsc{Horizon
\linebreak 2020} research and innovation programme under the Marie Sklodowska-Curie grant agreements No 690575 and No 674896.

%%%%%%%%%%%%%%%%%%%%%%%%%%%%%%%%%%%%%%%%%%%%%%%%%%%%%%%%%%%%%%%%%%%%%%%%%%%%%%
%%%%%%%%%%%%%%%%%%%%%%%%%%%%%%%%%%%%%%%%%%%%%%%%%%%%%%%%%%%%%%%%%%%%%%%%%%%%%%%%

%%%%%%%%%%%%%%%%%%%%%%%%%%%%%%%%%%%%%%%%%%%%%%%%%%%%%%%%%%%%%%%%%%%%%%%%%%%%%%%%
%%%%%%%%%%%%%%%%%%%%%%%%%%%%%%%%%%%%%%%%%%%%%%%%%%%%%%%%%%%%%%%%%%%%%%%%%%%%%%

\newpage
%\section{Appendix}

\begin{table*}[h]
\caption{Overall dimensions and masses of the 30 \textsc{Gerda} Phase~II BEGe diodes as defined by the \textsc{Gerda} collaboration. These values were used for the calculation of the full charge collection depths (FCCD) and active volumes (AV) in Section \ref{sec:full-charge-collection-depth-and-active-volume}. The definitions of the parameters have been introduced in Section \ref{sec:Dimensions-and-masses-of-BEGe-detectors}. Dimensions of diodes denoted with 1(2) asterisks require (special) attention. A mismatch of detector-to-mass assignment that occurred in Table~5 of \cite{bib:first-hades-paper} for 5 detectors of the 1$^{st}$ batch was corrected here.}
\vspace{2mm}
\begin{center}
\begin{tabular}{|l|ccccrrr|}
\hline
detector & H1 	& D1 	& H2 	& D2 	& $M_m$ 	&$\Delta M$		&$\Delta M'$	\\
ID         & [mm] & [mm]	& [mm] 	& [mm] 	& [g] 	& [\%]		& [\%]  	\\
\hline
GD32A	&	       24.90(14)             &       	   66.26(19)	  	 &	       19.40(11)	     	  &	      60.00(18) 		   &	   458     &	   1.33    &	   0.22    \\
GD32B	&	       32.16(09)             &       	   71.89(12)	  	 &				     	  &	    				   &	   716     &	   0.74    &	   0.13    \\
GD32C 	&	       33.15(04)             &       	   71.99(06)	  	 &				     	  &	    				   &	   743     &	   0.36    &	   0.06    \\
GD32D *	&	       32.12(14)             &       	   72.29(19)	  	 &				     	  &	    				   &	   720     &	   1.16    &	   0.20    \\
GD35A	&	       35.34(05)             &       	   73.54(08)	  	 &	       22.59(03)	     	  &	      58.25(06) 		   &	   768     &	   0.47    &	   0.12    \\
GD35B	&	       32.10(03)             &       	   76.33(04)	  	 &				     	  &	    				   &	   810     &	   0.23    &	   0.04    \\
GD35C	&	       26.32(10)             &       	   74.84(14)	  	 &				     	  &	    				   &	   634     &	   0.87    &	   0.09    \\
GD61A *	&	       33.57(10)             &       	   73.48(15)	  	 &	       17.44(05)	     	  &	      63.51(13) 		   &	   731     &	   0.91    &	   0.21    \\
GD61B *	&	       30.21(09)             &       	   75.95(12)	  	 &				     	  &	    				   &	   751     &	   0.71    &	   0.10    \\
GD61C	&	       26.45(07)             &       	   74.56(10)	  	 &				     	  &	    				   &	   634     &	   0.61    &	   0.07    \\
GD76B **&	       26.29(06)             &       	   58.27(09)	  	 &				     	  &	    				   &	   384     &	   0.65    &	   0.11    \\
GD76C 	&	       33.18(07)             &       	   75.84(10)	  	 &				     	  &	    				   &	   824     &	   0.54    &	   0.09    \\
GD79B *	&	       29.04(13)             &       	   76.84(19)	  	 &				     	  &	    				   &	   736     &	   1.1     &	   0.14    \\
GD79C	&	       30.22(09)             &       	   78.95(12)	  	 &				     	  &	    				   &	   812     &	   0.7     &	   0.09    \\
GD89A	&	       28.34(08)             &       	   68.63(12)	  	 &	       16.34(05)	     	  &	      50.50(09) 		   &	   524     &	   -0.83   &	   -0.20   \\
GD89B	&	       24.85(07)             &       	   76.05(09)	  	 &	        		     	  &	    				   &	   620     &	   0.56    &	   0.05    \\
GD89C	&	       24.75(08)             &       	   74.70(11)	  	 &				     	  &	    				   &	   595     &	   0.67    &	   0.06    \\
GD89D **&	       22.89(28)             &       	   73.43(40)	  	 &				     	  &	    				   &	   526     &	   2.48    &	   0.20    \\
GD91A *	&	       31.18(03)             &       	   70.53(04)	  	 &	       19.68(02)	     	  &	      56.00(03) 		   &	   627     &	   -0.28   &	   -0.07   \\
GD91B	&	       30.26(07)             &       	   70.58(10)	  	 &				     	  &	    				   &	   650     &	   0.61    &	   0.10    \\
GD91C	&	       29.79(08)             &       	   69.91(12)	  	 &				     	  &	    				   &	   627     &	   0.73    &	   0.12    \\
GD91D *	&	       31.88(17)             &       	   71.29(24)	  	 &				     	  &	    				   &	   693     &	   1.44    &	   0.25    \\
GD00A **&	       26.41(15)             &       	   70.33(21)	  	 &	       14.35(08)	     	  &	      46.50(14) 		   &	   496     &	   -1.53   &	   -0.38   \\
GD00B	&	       29.46(04)             &       	   73.96(06)	  	 &				     	  &	    				   &	   697     &	   0.33    &	   0.05    \\
GD00C *	&	       33.64(27)             &       	   75.52(38)	  	 &				     	  &	    				   &	   815     &	   2.15    &	   0.38    \\
GD00D	&	       32.28(07)             &       	   76.39(10)	  	 &				     	  &	    				   &	   813     &	   0.58    &	   0.09    \\
GD02A	&	       27.55(04)             &       	   70.46(06)	  	 &	       15.19(02)	     	  &	      57.50(05) 		   &	   545     &	   0.39    &	   0.08    \\
GD02B	&	       28.66(04)             &       	   71.01(05)	  	 &				     	  &	    				   &	   625     &	   0.31    &	   0.04    \\
GD02C	&	       32.59(08)             &       	   74.88(11)	  	 &				     	  &	    				   &	   788     &	   0.66    &	   0.11    \\
GD02D	&	       27.91(02)             &       	   74.59(03)	  	 &	       21.08(02)	     	  &	      68.50(03) 		   &	   662     &	   -0.21   &	   -0.03   \\
\hline
\end{tabular}
\end{center}
\label{table:dimensions-masses}
\end{table*}

%%%%%%%%%%%%%%%%%%%%%%%%%%%%%%%%%%%%%%%%%%%%%%%%%%%%%%%%%%%%%%%%%%%%%%%%%%%%%%

\begin{table*}[h]
\begin{center}
	\caption{\rm{High voltage scans of \textsc{Gerda} Phase~II BEGe detectors. Voltages $V$ at which the energy resolution, the peak position and the peak integral curves reach characteristic values close to the optimum/maximum are reported. The uncertainties are around $\pm$200\,V. The new operational voltages proposed for the usage in \textsc{Gerda} Phase~II are denoted with $V_r^G$. The full depletion and recommended voltages $V_r^C$, which were deduced from a peak position curve by Canberra Olen, are reported in the last two columns. No data is abbreviated with `n.d.'.}}
	\vspace{0.1 cm}
	\begin{tabular}{|l|cc|cc|cc|c|c|c|}	
	   \hline	
detector & \multicolumn{7}{c|}{V [kV] by \textsc{Gerda} - \textsc{Heroica}}	&\multicolumn{2}{c|}{V [kV] by Canberra Olen}	\\ 
 ID      & \multicolumn{2}{c}{best}	&\multicolumn{2}{|c|}{maximum}	&\multicolumn{2}{c|}{highest} &  & highest& \\
         &\multicolumn{2}{c}{energy resolution} &\multicolumn{2}{|c|}{peak integral}	&\multicolumn{2}{c|}{peak position}& $V_r^G$ &{peak position}	& $V_r^C$\\
	     \hline	
         &        95\% 	  & 99\%    &   95\%	& 99\%    &	  99\%    &   99.9\%	      & 	     &       n.d.    &   \\
        \hline
GD32A	&  	  2.0	  & 2.6     &	1.6	& 2.1	  &	  1.7	  &   {2.4}	      &       {2.9}  &       {2.5}   &       {3.0}   \\
GD32B	&  	  2.6	  & 3.0     &	1.6	& 2.1	  &	  2.5	  &   {2.7}	      &       {3.2}  &       {3.5}   &       {4.0}   \\
GD32C	&  	  3.1	  & 3.7     &	2.5	& 2.9	  &	  2.9	  &   {3.2}	      &       {3.7}  &       {3.5}   &       {4.0}   \\
GD32D	&  	  2.4	  & 2.8     &	1.8	& 2.2	  &	  2.2	  &   {2.7}	      &       {3.2}  &       {3.5}   &       {4.0}   \\
GD35A	&  	  2.5	  & 2.7     &	2.2	& 2.6	  &	  2.4	  &   {2.6}	      &       {3.1}  &       {3.0}   &       {4.0}   \\
GD35B	&  	  2.7	  & 3.5     &	2.1	& 2.5	  &	  2.5	  &   {2.9}	      &       {3.4}  &       {3.5}   &       {4.0}   \\
GD35C	&  	  2.7	  & 3.3     &	1.9	& 2.3	  &	  2.1	  &   {3.0}	      &       {3.5}  &       {3.0}   &       {3.5}   \\
GD61A	&  	  3.2	  & 4.2     &	2.3	& 2.7	  &	  2.9	  &   {3.1}	      &       {3.6}  &       {4.0}   &       {4.5}   \\
GD61B	&  	  3.1	  & 3.8     &	2.3	& 2.5	  &	  2.6	  &   {3.2}	      &       {3.7}  &       {3.5}   &       {4.0}   \\
GD61C	&  	  2.5	  & 3.7     &	1.7	& 2.0	  &	  2.7	  &   {3.2}	      &       {3.7}  &       {3.0}   &       {4.0}   \\
GD76B	&  	  2.2	  & 2.5     &	1.6	& 1.8	  &	  2.0	  &   {2.7}	      &       {3.2}  &       {3.0}   &       {3.5}   \\
GD76C	&  	  2.1	  & 3.2     &	1.5	& 1.6	  &	  1.9	  &   {2.1}	      &       {2.6}  &       {3.0}   &       {3.5}   \\
GD79B	&  	  2.5	  & 2.7     &	1.6	& 1.9	  &	  2.1	  &   {2.5}	      &       {3.0}  &       {3.0}   &       {3.5}   \\
GD79C	&  	  2.3	  & 3.1     &	1.4	& 1.6	  &	  1.8	  &   {2.4}	      &       {2.9}  &       {3.0}   &       {3.5}   \\
GD89A	&  	  3.0	  & 3.7     &	2.1	& 2.3	  &	  2.5	  &   {3.1}	      &       {3.6}  &       {3.5}   &       {4.0}   \\
GD89B	&  	  2.6	  & 2.8     &	2.0	& 2.5	  &	  2.3	  &   {2.7}	      &       {3.2}  &       {3.0}   &       {3.5}   \\
GD89C	&  	  2.6	  & 3.5     &	2.0	& 2.5	  &	  2.3	  &   {2.8}	      &       {3.3}  &       {3.5}   &       {4.0}   \\
GD89D	&  	  2.2	  & 3.4     &	1.6	& 1.8	  &	  1.7	  &   {2.5}	      &       {3.0}  &       {3.5}   &       {4.0}   \\
GD91A	&  	  2.5	  & 2.7     &	2.0	& 2.7	  &	  2.2	  &   {2.6}	      &       {3.1}  &       {3.0}   &       {3.5}   \\
GD91B	&  	  2.6	  & 2.7     &	2.1	& 2.5	  &	  2.4	  &   {2.8}	      &       {3.3}  &       {3.0}   &       {3.5}   \\
GD91C	&  	  3.1	  & 3.6     &	2.2	& 2.7	  &	  2.5	  &   {3.2}	      &       {3.7}  &       {3.5}   &       {4.0}   \\
GD91D	&  	  3.6	  & 3.9     &	2.8	& 3.2	  &	  3.3	  &   {4.0}	      &       {4.5}  &       {4.0}   &       {4.5}   \\
GD00A	&  	  1.3	  & 2.1     &	0.8	& 0.9	  &	  1.0	  &   {1.2}	      &       {1.7}  &       {1.5}   &       {2.5}   \\
GD00B	&  	  1.7	  & 3.0     &	1.0	& 1.1	  &	  1.2	  &   {1.6}	      &       {2.1}  &       {2.5}   &       {3.5}   \\
GD00C	&  	  2.5	  & 2.6     &	2.0	& 2.7	  &	  2.5	  &   {2.7}	      &       {3.2}  &       {3.0}   &       {3.5}   \\
GD00D	&  	  2.4	  & 3.2     &	2.3	& 2.3	  &	  2.4	  &   {2.5}	      &       {3.0}  &       {3.0}   &       {3.5}   \\
GD02A	&  	  1.5	  & 1.6     &	1.0	& 1.1	  &	  1.3	  &   {1.7}	      &       {2.2}  &       {2.0}   &       {2.5}   \\
GD02B	&  	  1.8	  & 2.1     &	1.2	& 1.2	  &	  1.5	  &   {1.8}	      &       {2.3}  &       {2.5}   &       {3.0}   \\
GD02C	&  	  2.2	  & 2.5     &	1.6	& 1.9	  &	  1.6	  &   {2.3}	      &       {2.8}  &       {2.5}   &       {3.5}   \\
GD02D	&  	  n.d.    & n.d.    &	n.d.	& n.d.    &	  n.d.    &   n.d.	      &       n.d.   &       {3.0}   &       {4.0}   \\ 	
\hline
average &  	  2.5	  & 3.0     &	1.8	& 2.1	  &	  2.2	  &   2.6	      &       3.1    &        3.1    &       {3.7}   \\        
\hline
    \end{tabular}
	\label{table:depl-voltages-all-detectors}	
\end{center}
\end{table*}

%%%%%%%%%%%%%%%%%%%%%%%%%%%%%%%%%%%%%%%%%%%%%%%%%%%%%%%%%%%%%%%%%%%%%%%%%%%%%%

\begin{table*}[h]
\caption{Energy resolution of the 30 \textsc{Gerda} Phase~II BEGe detectors at 1333\,keV. All detectors were operated at the voltages $V_r^C$. The Canberra values in the 4$^{th}$ column were relayed without uncertainties. The \textsc{Gerda}-\textsc{Heroica} values in the 2$^{nd}$ and 3$^{rd}$ column were rounded to the relevant number of digits, i.e. two decimal places. Detectors affected by the so-called `bubble depletion' effect are marked with a dagger ($\dagger$). Conical detectors are marked with an asterisk ($^{*}$). GD02D is the only detector with a remarkable low-side energy tail. Averaged values are given with their standard deviations.}
\begin{center}
\begin{tabular}{|l|ccc|}
\hline
detector&   HV scan based $\Delta E$  	&single meas. based  $\Delta E$ 	& $\Delta E$ \\
ID		&	{\scriptsize(\textsc{Gerda}-\textsc{Heroica})}&{\scriptsize(\textsc{Gerda}-\textsc{Heroica})}			& {\scriptsize(Canberra Olen)}\\
		& [keV]	& [keV]		& [keV] \\
\hline
GD32A$^{*}$	&	1.65(2)	 &	 1.71(1)    &	  1.695   \\
GD32B$\dagger$	&	1.73(1)	 &	 1.73(1)    &	  1.747   \\
GD32C		&	1.65(1)	 &	 1.65(1)    &	  1.658   \\
GD32D		&	1.73(1)	 &	 1.65(1)    &	  1.757   \\
GD35A$\dagger$ $^{*}$&	1.67(4)	 &	 1.73(1)    &	  1.785   \\
GD35B$\dagger$	&	1.77(1)	 &	 1.77(1)    &	  1.748   \\
GD35C		&	1.71(1)	 &	 1.68(2)    &	  1.643   \\
GD61A$\dagger$ $^{*}$&	1.89(1)	 &	 1.85(2)    &	  1.820	  \\
GD61B		&	1.68(1)	 &	 1.73(2)    &	  1.734   \\
GD61C$\dagger$	&	1.72(1)	 &	 1.72(2)    &	  1.708   \\
GD76B		&	1.67(2)	 &	 1.64(3)    &	  1.694   \\
GD76C$\dagger$	&	1.72(1)	 &	 1.73(2)    &	  1.710	  \\
GD79B$\dagger$	&	1.83(2)	 &	 1.86(2)    &	  1.820   \\
GD79C		&	1.87(1)	 &	 1.86(2)    &	  1.812   \\
GD89A$^{*}$	&	1.69(1)	 &	 1.68(2)    &	  1.720   \\
GD89B$\dagger$	&	1.72(3)	 &	 1.79(2)    &	  1.684   \\
GD89C		&	1.69(1)	 &	 1.75(2)    &	  1.686   \\
GD89D$\dagger$	&	1.65(1)	 &	 1.66(2)    	&	  1.721   \\
GD91A$\dagger$ $^{*}$&	1.65(3)	 &	 1.66(2)    	&	  1.746   \\
GD91B$\dagger$	&	1.72(2)	 &	 1.72(2)   	&	  1.708   \\
GD91C		&	1.74(2)	 &	 1.71(2)    	&	  1.708   \\
GD91D		&	1.68(2)	 &	 1.68(2)    	&	  1.742   \\
GD00A$^{*}$	&	1.77(1)	 &	 1.80(2)   	&	  1.724   \\
GD00B		&	1.76(1)	 &	 1.84(2)    	&	  1.745   \\
GD00C$\dagger$	&	1.70(1)	 &	 1.65(2)    	&	  1.762   \\
GD00D$\dagger$	&	1.71(1)	 &	 1.73(2)    	&	  1.782   \\
GD02A$^{*}$	&	1.74(2)	 &	 1.79(2)    	&	  1.749   \\
GD02B		&	1.70(1)	 &	 1.70(3)    	&	  1.720   \\
GD02C		&	1.71(1)	 &	 1.80(3)    	&	  1.748   \\
GD02D		&	1.84(11) &	 1.73(4)	&	  1.846   \\
\hline
average 	&	1.72(7)	 &   1.73(7)	  &   1.74(5)  \\
\hline
\end{tabular}
\label{tab:ene-res-all-results}
\end{center}
\end{table*}

%%%%%%%%%%%%%%%%%%%%%%%%%%%%%%%%%%%%%%%%%%%%%%%%%%%%%%%%%%%%%%%%%%%%%%%%%%%%%%

\begin{table*}
\centering
\caption{FCCD results of the 29 well working \textsc{Gerda} Phase~II BEGe detectors, as determined directly after diode production in vacuum croystats and using different probes. The spectra of the $^{241}${Am} and $^{133}${Ba} source measurements have been evaluated with a fitting method, while the $^{60}${Co} $\gamma$-lines have been determined with a counting method. The detectors for which the systematic uncertainty was kept small are denoted with a (+) sign. Those with less reliable FCCD values have a (-) sign. The uncertainties are separated into correlated and uncorrelated components. For comparison, the $^{241}${Am}-based FCCD values provided by the manufacturer are also reported. These were relayed without uncertainties.}
\label{tab:fccd-single-source-measurements}
\begin{tabular}{|l|cccc|c|}
%toprule
\hline
detector	&\multicolumn{4}{c|}{FCCD$^{+\text{ucorr}+\text{corr}}_{-\text{ucorr}-\text{corr}}$  by \textsc{Gerda}-\textsc{Heroica}} & FCCD by Canberra\\
ID			&  $^{241}${Am} 	& $^{133}${Ba} & $^{60}${Co} & $^{60}${Co}     &$^{241}${Am} \\
			&  [mm] & [mm]  & [mm] & [mm] & [mm]  \\
\hline 
GD32A &$0.59^{+0.03+0.02}_{-0.03-0.02}$ & $0.53^{+0.06+0.01}_{-0.06-0.01}$ & $0.78^{+0.20+0.14}_{-0.20-0.14}$&$0.79^{+0.19+0.14}_{-0.20-0.14}$&0.60\\
GD32B &$0.84^{+0.03+0.02}_{-0.03-0.02}$ & $0.73^{+0.06+0.00}_{-0.06-0.01}$ & $1.01^{+0.24+0.17}_{-0.24-0.17}$&$1.02^{+0.23+0.17}_{-0.24-0.17}$&0.90\\
GD32C$^{-}$ &$0.81^{+0.03+0.02}_{-0.04-0.02}$ & $0.54^{+0.06+0.01}_{-0.06-0.01}$ & $0.82^{+0.24+0.18}_{-0.24-0.18}$&$0.78^{+0.23+0.17}_{-0.24-0.18}$&0.70\\
GD32D &$0.61^{+0.03+0.03}_{-0.04-0.03}$ & $0.41^{+0.06+0.01}_{-0.06-0.01}$ & $0.75^{+0.24+0.17}_{-0.24-0.17}$&$0.76^{+0.23+0.17}_{-0.24-0.17}$&0.70\\
GD35A$^{+}$ &$0.62^{+0.03+0.01}_{-0.03-0.01}$ & $0.58^{+0.05+0.01}_{-0.04-0.01}$ & $0.68^{+0.25+0.18}_{-0.25-0.18}$&$0.68^{+0.24+0.18}_{-0.25-0.18}$&0.70\\
GD35B$^{+}$ &$0.58^{+0.03+0.06}_{-0.04-0.05}$ & $0.53^{+0.06+0.01}_{-0.06-0.01}$ & $0.72^{+0.24+0.17}_{-0.24-0.17}$&$0.74^{+0.23+0.16}_{-0.24-0.17}$&0.70\\
GD35C$^{+}$ &$0.58^{+0.03+0.02}_{-0.03-0.02}$ & $0.50^{+0.06+0.01}_{-0.06-0.01}$ & $0.78^{+0.21+0.15}_{-0.21-0.16}$&$0.73^{+0.20+0.15}_{-0.21-0.15}$&0.60\\
GD61A$^{-}$ &$0.84^{+0.04+0.05}_{-0.04-0.04}$ & $0.65^{+0.05+0.01}_{-0.05-0.01}$ & $1.07^{+0.25+0.19}_{-0.25-0.19}$&$1.12^{+0.24+0.19}_{-0.25-0.19}$&0.76\\
GD61B &$0.75^{+0.03+0.04}_{-0.04-0.04}$ & $0.69^{+0.06+0.01}_{-0.06-0.01}$ & $1.05^{+0.23+0.17}_{-0.23-0.17}$&$1.06^{+0.22+0.17}_{-0.23-0.17}$&0.80\\
GD61C$^{+}$ &$0.70^{+0.03+0.04}_{-0.03-0.04}$ & $0.67^{+0.06+0.01}_{-0.07-0.01}$ & $0.80^{+0.21+0.15}_{-0.21-0.15}$&$0.85^{+0.20+0.15}_{-0.21-0.15}$&0.76\\
GD76B$^{-}$ &$0.93^{+0.03+0.03}_{-0.04-0.03}$ & $0.76^{+0.06+0.01}_{-0.07-0.01}$ & $1.15^{+0.19+0.14}_{-0.20-0.14}$&$1.16^{+0.19+0.14}_{-0.19-0.14}$&1.00\\
GD76C$^{+}$ &$0.89^{+0.03+0.03}_{-0.03-0.03}$ & $0.81^{+0.06+0.01}_{-0.06-0.01}$ & $1.00^{+0.25+0.17}_{-0.25-0.18}$&$1.06^{+0.24+0.16}_{-0.25-0.17}$&0.92\\
GD79B$^{-}$ &$0.76^{+0.03+0.03}_{-0.04-0.03}$ & $0.68^{+0.06+0.01}_{-0.06-0.01}$ & $0.88^{+0.23+0.16}_{-0.23-0.16}$&$0.88^{+0.22+0.16}_{-0.23-0.16}$&0.85\\
GD79C &$0.90^{+0.03+0.03}_{-0.04-0.03}$ & $0.78^{+0.06+0.01}_{-0.06-0.01}$ & $1.17^{+0.24+0.17}_{-0.24-0.17}$&$1.22^{+0.23+0.17}_{-0.24-0.17}$&0.90\\
GD89A &$0.72^{+0.03+0.04}_{-0.03-0.03}$ & $0.64^{+0.05+0.01}_{-0.05-0.01}$ & $0.84^{+0.21+0.14}_{-0.21-0.15}$&$0.85^{+0.21+0.15}_{-0.21-0.15}$&0.80\\
GD89B &$0.85^{+0.03+0.02}_{-0.04-0.02}$ & $0.75^{+0.06+0.01}_{-0.06-0.01}$ & $0.97^{+0.20+0.14}_{-0.20-0.15}$&$1.00^{+0.19+0.14}_{-0.20-0.15}$&0.80\\
GD89C$^{+}$ &$0.71^{+0.03+0.03}_{-0.03-0.03}$ & $0.66^{+0.06+0.01}_{-0.07-0.01}$ & $0.91^{+0.20+0.14}_{-0.20-0.14}$&$0.91^{+0.19+0.14}_{-0.20-0.14}$&0.85\\
GD89D$^{-}$ &$0.83^{+0.03+0.03}_{-0.03-0.02}$ & $0.62^{+0.07+0.01}_{-0.07-0.01}$ & $1.13^{+0.19+0.13}_{-0.19-0.14}$&$1.13^{+0.18+0.13}_{-0.19-0.13}$&0.76\\
GD91A &$0.75^{+0.03+0.04}_{-0.03-0.03}$ & $0.65^{+0.05+0.01}_{-0.05-0.01}$ & $0.86^{+0.23+0.16}_{-0.23-0.16}$&$0.89^{+0.22+0.16}_{-0.23-0.17}$&0.80\\
GD91B &$0.73^{+0.03+0.03}_{-0.04-0.03}$ & $0.60^{+0.06+0.01}_{-0.06-0.01}$ & $0.96^{+0.23+0.16}_{-0.23-0.17}$&$0.88^{+0.22+0.16}_{-0.23-0.17}$&0.80\\
GD91C &$0.76^{+0.03+0.04}_{-0.04-0.04}$ & $0.60^{+0.06+0.01}_{-0.06-0.01}$ & $1.02^{+0.22+0.16}_{-0.23-0.16}$&$1.03^{+0.22+0.16}_{-0.22-0.16}$&0.76\\
GD91D &$0.72^{+0.03+0.04}_{-0.03-0.04}$ & $0.64^{+0.06+0.01}_{-0.06-0.01}$ & $0.97^{+0.24+0.17}_{-0.24-0.17}$&$0.97^{+0.23+0.17}_{-0.24-0.17}$&0.80\\
GD00A &$0.62^{+0.03+0.04}_{-0.03-0.03}$ & $0.64^{+0.05+0.01}_{-0.05-0.01}$ & $0.87^{+0.20+0.15}_{-0.20-0.15}$&$0.88^{+0.19+0.15}_{-0.20-0.15}$&0.75\\
GD00B$^{+}$ &$0.81^{+0.03+0.04}_{-0.04-0.04}$ & $0.71^{+0.06+0.01}_{-0.06-0.01}$ & $1.03^{+0.23+0.17}_{-0.23-0.17}$&$1.11^{+0.22+0.16}_{-0.23-0.17}$&0.76\\
GD00C &$0.75^{+0.03+0.02}_{-0.04-0.02}$ & $0.62^{+0.06+0.01}_{-0.06-0.01}$ & $1.03^{+0.25+0.19}_{-0.25-0.19}$&$1.03^{+0.24+0.19}_{-0.25-0.19}$&0.76\\
GD00D$^{+}$ &$0.77^{+0.03+0.02}_{-0.03-0.02}$ & $0.68^{+0.06+0.01}_{-0.06-0.01}$ & $0.99^{+0.24+0.17}_{-0.24-0.18}$&$0.96^{+0.23+0.17}_{-0.24-0.17}$&0.80\\
GD02A$^{-}$ &$0.75^{+0.03+0.03}_{-0.03-0.03}$ & $0.52^{+0.05+0.01}_{-0.05-0.01}$ & $1.06^{+0.21+0.15}_{-0.21-0.15}$&$1.05^{+0.20+0.14}_{-0.21-0.15}$&0.75\\
GD02B$^{+}$ &$0.77^{+0.03+0.04}_{-0.03-0.04}$ & $0.63^{+0.06+0.01}_{-0.06-0.01}$ & $1.00^{+0.22+0.15}_{-0.22-0.15}$&$0.99^{+0.21+0.15}_{-0.22-0.15}$&0.80\\
GD02C$^{+}$ &$0.79^{+0.03+0.04}_{-0.04-0.04}$ & $0.71^{+0.06+0.01}_{-0.06-0.01}$ & $0.97^{+0.24+0.18}_{-0.24-0.18}$&$0.91^{+0.24+0.18}_{-0.24-0.18}$&0.76\\
\hline
\end{tabular} 
\end{table*}

%%%%%%%%%%%%%%%%%%%%%%%%%%%%%%%%%%%%%%%%%%%%%%%%%%%%%%%%%%%%%%%%%%%%%%%%%%%%%%

\begin{table*}
\begin{center}
\caption{Official dataset of FCCD and AV fractions used in \textsc{Gerda} Phase~II physics analyses: The FCCD values were obtained by combining the $^{241}${Am} FCCD and $^{133}${Ba} FCCD values (1$^{st}$ column) and adding an offset that considers a FCCD growth at room temperature (2$^{nd}$ column). The offsets are individual for each detector and consider different storage periods. The AV fractions (3$^{rd}$ column) were deduced via subtraction of the FCCD volumes from the crystal masses. The active masses (4$^{th}$ column) were quoted by multiplying the crystal masses with the AV fractions.}
\label{tab:241Am133BaFCCDcombined-FCCDgrowing}
\vspace{0.5 cm}
\begin{tabular}{|l|cccc|}
\hline
%			&\multicolumn{3}{c}{Official dataset for \textsc{Gerda} Phase~II}\\	
detector	   	&\multicolumn{2}{c}{ FCCD$^{+\text{ucorr}+\text{corr}}_{-\text{ucorr}-\text{corr}}$ }    & {$f_{av}$}\,$^{+\text{ucorr}+\text{corr}}_{-\text{ucorr}-\text{corr}}$  & {$M_{av}$}\,$^{+\text{ucorr}+\text{corr}}_{-\text{ucorr}-\text{corr}}$  \\
ID			&w/o growth		&w growth			&	w growth  &w growth	\\
			&\multicolumn{2}{c} { [mm] }&    & [g]                        \\
\hline
GD32A	&0.57	$^{+	0.03	+	0.02	}_{-0.06	-	0.02	}$&	0.91	$^{+	0.17	+	0.03	}_{-	0.17	-	0.06	}$	&	0.882	$^{+	0.021	+	0.008	}_{-	0.021	-	0.004	}$	&	404	$^{+	10	+	4	}_{-	10	-	2	}$		\\
GD32B	&0.80	$^{+	0.03	+	0.02	}_{-0.06	-	0.02	}$&	1.05	$^{+	0.13	+	0.03	}_{-	0.13	-	0.06	}$	&	0.883	$^{+	0.014	+	0.006	}_{-	0.014	-	0.003	}$	&	632	$^{+	10	+	4	}_{-	10	-	2	}$		\\
GD32C	&0.70	$^{+	0.03	+	0.02	}_{-0.06	-	0.02	}$&	0.96	$^{+	0.13	+	0.03	}_{-	0.13	-	0.06	}$	&	0.895	$^{+	0.014	+	0.006	}_{-	0.014	-	0.003	}$	&	665	$^{+	10	+	4	}_{-	10	-	2	}$		\\
GD32D	&0.52	$^{+	0.03	+	0.03	}_{-0.06	-	0.03	}$&	0.77	$^{+	0.13	+	0.03	}_{-	0.13	-	0.06	}$	&	0.913	$^{+	0.014	+	0.007	}_{-	0.014	-	0.003	}$	&	657	$^{+	10	+	5	}_{-	10	-	2	}$		\\
GD35A	&0.61	$^{+	0.03	+	0.01	}_{-0.04	-	0.01	}$&	0.95	$^{+	0.17	+	0.03	}_{-	0.17	-	0.04	}$	&	0.902	$^{+	0.017	+	0.004	}_{-	0.017	-	0.003	}$	&	693	$^{+	13	+	3	}_{-	13	-	2	}$		\\
GD35B	&0.55	$^{+	0.03	+	0.06	}_{-0.06	-	0.05	}$&	0.78	$^{+	0.13	+	0.03	}_{-	0.13	-	0.06	}$	&	0.914	$^{+	0.014	+	0.006	}_{-	0.014	-	0.003	}$	&	740	$^{+	11	+	5	}_{-	11	-	2	}$		\\
GD35C	&0.55	$^{+	0.03	+	0.02	}_{-0.06	-	0.02	}$&	0.79	$^{+	0.12	+	0.03	}_{-	0.12	-	0.06	}$	&	0.902	$^{+	0.014	+	0.007	}_{-	0.014	-	0.004	}$	&	572	$^{+	9	+	4	}_{-	9	-	3	}$		\\
GD61A	&0.72	$^{+	0.04	+	0.05	}_{-0.05	-	0.05	}$&	1.01	$^{+	0.15	+	0.04	}_{-	0.15	-	0.05	}$	&	0.892	$^{+	0.016	+	0.005	}_{-	0.015	-	0.004	}$	&	652	$^{+	12	+	4	}_{-	11	-	3	}$		\\
GD61B	&0.72	$^{+	0.03	+	0.04	}_{-0.06	-	0.04	}$&	1.00	$^{+	0.15	+	0.03	}_{-	0.15	-	0.06	}$	&	0.887	$^{+	0.016	+	0.007	}_{-	0.016	-	0.003	}$	&	666	$^{+	12	+	5	}_{-	12	-	2	}$		\\
GD61C	&0.68	$^{+	0.03	+	0.04	}_{-0.07	-	0.04	}$&	0.92	$^{+	0.13	+	0.03	}_{-	0.13	-	0.07	}$	&	0.887	$^{+	0.015	+	0.008	}_{-	0.015	-	0.004	}$	&	562	$^{+	10	+	5	}_{-	10	-	3	}$		\\
GD76B	&0.86	$^{+	0.03	+	0.03	}_{-0.07	-	0.03	}$&	1.14	$^{+	0.14	+	0.03	}_{-	0.14	-	0.07	}$	&	0.848	$^{+	0.018	+	0.009	}_{-	0.018	-	0.004	}$	&	326	$^{+	7	+	3	}_{-	7	-	2	}$		\\
GD76C	&0.85	$^{+	0.03	+	0.03	}_{-0.06	-	0.03	}$&	1.14	$^{+	0.15	+	0.03	}_{-	0.15	-	0.06	}$	&	0.878	$^{+	0.015	+	0.006	}_{-	0.015	-	0.003	}$	&	723	$^{+	12	+	5	}_{-	12	-	2	}$		\\
GD79B	&0.73	$^{+	0.03	+	0.03	}_{-0.06	-	0.03	}$&	1.04	$^{+	0.16	+	0.03	}_{-	0.16	-	0.06	}$	&	0.881	$^{+	0.018	+	0.007	}_{-	0.018	-	0.003	}$	&	648	$^{+	13	+	5	}_{-	13	-	2	}$		\\
GD79C	&0.85	$^{+	0.03	+	0.03	}_{-0.06	-	0.03	}$&	1.10	$^{+	0.13	+	0.03	}_{-	0.13	-	0.06	}$	&	0.878	$^{+	0.014	+	0.006	}_{-	0.014	-	0.003	}$	&	713	$^{+	11	+	5	}_{-	11	-	2	}$		\\
GD89A	&0.67	$^{+	0.03	+	0.04	}_{-0.05	-	0.03	}$&	0.99	$^{+	0.16	+	0.03	}_{-	0.16	-	0.05	}$	&	0.882	$^{+	0.019	+	0.006	}_{-	0.018	-	0.003	}$	&	462	$^{+	10	+	3	}_{-	9	-	2	}$		\\
GD89B	&0.82	$^{+	0.03	+	0.02	}_{-0.06	-	0.02	}$&	1.13	$^{+	0.16	+	0.03	}_{-	0.16	-	0.06	}$	&	0.859	$^{+	0.019	+	0.007	}_{-	0.019	-	0.004	}$	&	533	$^{+	12	+	4	}_{-	12	-	2	}$		\\
GD89C	&0.69	$^{+	0.03	+	0.04	}_{-0.07	-	0.04	}$&	0.99	$^{+	0.16	+	0.03	}_{-	0.16	-	0.07	}$	&	0.874	$^{+	0.020	+	0.009	}_{-	0.019	-	0.004	}$	&	520	$^{+	12	+	5	}_{-	11	-	2	}$		\\
GD89D	&0.75	$^{+	0.03	+	0.03	}_{-0.07	-	0.03	}$&	1.02	$^{+	0.14	+	0.03	}_{-	0.14	-	0.07	}$	&	0.863	$^{+	0.018	+	0.009	}_{-	0.018	-	0.004	}$	&	454	$^{+	9	+	5	}_{-	9	-	2	}$		\\
GD91A	&0.69	$^{+	0.03	+	0.04	}_{-0.05	-	0.03	}$&	0.99	$^{+	0.16	+	0.03	}_{-	0.15	-	0.05	}$	&	0.889	$^{+	0.016	+	0.005	}_{-	0.017	-	0.003	}$	&	557	$^{+	10	+	3	}_{-	11	-	2	}$		\\
GD91B	&0.68	$^{+	0.03	+	0.03	}_{-0.06	-	0.03	}$&	0.96	$^{+	0.14	+	0.03	}_{-	0.14	-	0.06	}$	&	0.889	$^{+	0.016	+	0.007	}_{-	0.016	-	0.003	}$	&	578	$^{+	10	+	5	}_{-	10	-	2	}$		\\
GD91C	&0.68	$^{+	0.03	+	0.04	}_{-0.06	-	0.04	}$&	0.96	$^{+	0.15	+	0.03	}_{-	0.15	-	0.06	}$	&	0.887	$^{+	0.017	+	0.007	}_{-	0.017	-	0.003	}$	&	556	$^{+	11	+	4	}_{-	11	-	2	}$		\\
GD91D	&0.68	$^{+	0.03	+	0.04	}_{-0.06	-	0.04	}$&	0.99	$^{+	0.16	+	0.03	}_{-	0.16	-	0.06	}$	&	0.888	$^{+	0.017	+	0.007	}_{-	0.017	-	0.003	}$	&	615	$^{+	12	+	5	}_{-	12	-	2	}$		\\
GD00A	&0.63	$^{+	0.03	+	0.04	}_{-0.05	-	0.03	}$&	0.91	$^{+	0.15	+	0.03	}_{-	0.14	-	0.05	}$	&	0.886	$^{+	0.017	+	0.006	}_{-	0.018	-	0.004	}$	&	439	$^{+	8	+	3	}_{-	9	-	2	}$		\\
GD00B	&0.76	$^{+	0.03	+	0.04	}_{-0.06	-	0.04	}$&	1.04	$^{+	0.15	+	0.03	}_{-	0.15	-	0.06	}$	&	0.880	$^{+	0.017	+	0.007	}_{-	0.017	-	0.003	}$	&	613	$^{+	12	+	5	}_{-	12	-	2	}$		\\
GD00C	&0.70	$^{+	0.03	+	0.02	}_{-0.06	-	0.02	}$&	1.01	$^{+	0.16	+	0.03	}_{-	0.16	-	0.06	}$	&	0.892	$^{+	0.017	+	0.006	}_{-	0.016	-	0.003	}$	&	727	$^{+	14	+	5	}_{-	13	-	2	}$		\\
GD00D	&0.73	$^{+	0.03	+	0.02	}_{-0.06	-	0.02	}$&	1.02	$^{+	0.15	+	0.03	}_{-	0.15	-	0.06	}$	&	0.889	$^{+	0.016	+	0.006	}_{-	0.016	-	0.003	}$	&	723	$^{+	13	+	5	}_{-	13	-	2	}$		\\
GD02A	&0.62	$^{+	0.03	+	0.03	}_{-0.05	-	0.03	}$&	0.86	$^{+	0.12	+	0.03	}_{-	0.12	-	0.05	}$	&	0.896	$^{+	0.014	+	0.006	}_{-	0.014	-	0.003	}$	&	488	$^{+	8	+	3	}_{-	8	-	2	}$		\\
GD02B	&0.70	$^{+	0.03	+	0.04	}_{-0.06	-	0.04	}$&	0.97	$^{+	0.14	+	0.03	}_{-	0.14	-	0.06	}$	&	0.885	$^{+	0.016	+	0.007	}_{-	0.016	-	0.003	}$	&	553	$^{+	10	+	4	}_{-	10	-	2	}$		\\
GD02C	&0.75	$^{+	0.03	+	0.04	}_{-0.06	-	0.04	}$&	1.03	$^{+	0.15	+	0.03	}_{-	0.15	-	0.06	}$	&	0.888	$^{+	0.016	+	0.006	}_{-	0.016	-	0.003	}$	&	700	$^{+	13	+	5	}_{-	13	-	2	}$		\\
\hline
\end{tabular}
\end{center}
\end{table*}

%%%%%%%%%%%%%%%%%%%%%%%%%%%%%%%%%%%%%%%%%%%%%%%%%%%%%%%%%%%%%%%%%%%%%%%%%%%%%%
\begin{table*}
\begin{center}
\caption{%PSD efficiencies of all 30 \textsc{Gerda} Phase II BEGe detectors operated under vacuum conditions.
Gamma-ray background survival fractions and the DEP $A/E$ resolution $b_{A/E}$($^{228}$Th) for the 30 \textsc{Gerda} Phase~II BEGe detectors operated under vacuum conditions. In the case of the detector GD32D, the pulse shape performance was measured before (I) and after (II) reprocessing. The uncertainty of the survival fraction includes the statistical and the systematic uncertainty. Averaged values are given with their standard deviations.}
\label{tab:th228-pulse-shape-results}
\vspace{0.5 cm}
\begin{tabular}{|l|ccccc|c|}
\hline
detector 	&DEP     		&SEP			&FEP 			&FEP 		&ROI				&$b_{A/E}$ \\
ID 			&[\%]     		&[\%]			&[\%] 			&[\%] 		&[\%]				&[\%]\\
\hline	
GD32A	&	90.0(0.9)	&	12.0(1.0)&	16.8(0.7)	&	17.8(1.2)	&	43.4(0.4)	&1.32(1)	\\
GD32B	&	90.0(1.3)	&	5.6(0.8)	&	8.0(0.7)		&	10.0(0.9)	&	32.3(0.6)	&0.76(1)	\\
GD32C	&	90.0(2.0)	&	8.1(1.2)	&	13.8(0.7)	&	11.6(1.6)	&	42.4(0.7)	&1.82(1)	\\
GD32D-I	&	90.0(0.3)	&	11.8(1.6)&	17.4(0.9)	&	18.6(1.7)	&	43.9(0.6)	&1.42(3)	\\
GD32D-II&	90.0(0.9)	&	6.1(0.7)	&	8.7(0.5)		&	9.0(0.9)	&	39.6(0.4)	&0.32(1)	\\
GD35A	&	90.0(0.3)	&	9.0(1.0)	&	13.3(0.7)	&	12.9(1.2)	&	36.9(0.5)	&3.23(3)	\\
GD35B	&	90.0(1.2)	&	5.0(0.8)	&	6.4(0.6)		&	9.9(1.1)	&	32.3(0.6)	&0.30(1)	\\
GD35C	&	90.0(0.6)	&	11.2(0.9)&	16.3(0.6)	&	16.0(1.0)	&	41.7(0.4)	&2.95(3)	\\
GD61A	&	90.0(0.7)	&	7.0(0.5)	&	9.9(0.4)	&	11.5(0.9)	&	39.8(0.4)	&0.94(1)	\\
GD61B	&	90.0(0.8)	&	8.5(0.8)	&	13.7(0.5)	&	13.5(0.9)	&	45.4(0.3)	&1.22(1)	\\
GD61C	&	90.0(0.3)	&	7.4(0.4)	&	10.2(0.3)	&	12.4(0.7)	&	41.4(0.4)	&0.49(2)	\\
GD76B	&	90.0(1.0)	&	21.2(1.9)&	34.7(1.1)	&	21.9(2.7)	&	48.6(0.5)	&1.92(6)	\\
GD76C	&	90.0(0.8)	&	5.6(0.6)	&	7.0(0.6)		&	8.9(1.0)	&	37.1(0.7)	&0.40(3)	\\
GD79B	&	90.0(1.0)	&	11.9(0.7)&	16.4(0.4)	&	17.7(0.7)	&	48.4(0.3)	&1.77(2)	\\
GD79C	&	90.0(0.6)	&	7.4(0.5)	&	12.8(0.3)	&	13.0(0.6)	&	44.8(0.2)	&2.22(2)	\\
GD89A	&	90.0(0.9)	&	10.6(0.8)&	17.0(0.5)	&	16.5(1.0)	&	48.5(0.3)	&1.25(1)	\\
GD89B	&	90.0(1.1)	&	7.5(0.6)	&	12.4(0.4)	&	12.3(0.8)	&	43.8(0.3)	&1.51(4)	\\
GD89C	&	90.0(0.9)	&	9.1(0.6)	&	13.2(0.4)	&	14.3(1.0)	&	46.3(0.3)	&0.51(1)	\\
GD89D	&	90.0(0.5)	&	8.5(0.9)	&	14.8(0.9)	&	16.4(2.1)	&	47.4(0.6)	&1.38(2)	\\
GD91A	&	90.0(1.1)	&	8.0(0.8)	&	11.9(0.5)	&	11.9(0.7)	&	43.3(0.4)	&0.94(1)	\\
GD91B	&	90.0(0.8)	&	8.5(0.7)	&	12.2(0.4)	&	13.3(0.9)	&	43.8(0.3)	&1.09(2)	\\
GD91C	&	90.0(1.1)	&	11.6(0.6)&	17.4(0.4)	&	17.8(0.8)	&	49.5(0.2)	&2.67(1)	\\
GD91D	&	90.0(0.4)	&	7.0(0.5)	&	10.6(0.3)	&	11.7(0.7)	&	42.4(0.3)	&0.47(1)	\\
GD00A	&	90.0(0.6)	&	10.8(0.5)&	16.9(0.4)	&	16.6(0.9)	&	49.7(0.2)	&1.79(1)	\\
GD00B	&	90.0(0.9)	&	8.9(0.7)	&	12.5(0.6)	&	11.0(1.2)	&	44.1(0.5)	&1.29(2)	\\
GD00C	&	90.0(0.8)	&	6.4(0.6)	&	10.0(0.4)	&	10.3(0.6)	&	41.0(0.4)	&0.56(1)	\\
GD00D	&	90.0(0.8)	&	6.0(0.6)	&	9.1(0.4)		&	9.9(0.7)	&	38.9(0.4)	&0.60(1)	\\
GD02A	&	90.0(0.5)	&	11.8(0.7)&	18.6(0.4)	&	18.4(0.8)	&	49.9(0.2)	&3.58(1)	\\
GD02B	&	90.0(0.7)	&	7.4(0.8)	&	13.8(0.6)	&	11.4(0.9)	&	45.8(0.4)	&0.68(3)	\\
GD02C	&	90.0(0.9)	&	7.6(0.6)	&	11.2(0.4)	&	11.9(1.0)	&	42.3(0.4)	&1.96(3)	\\
GD02D	&	90.0(0.7)	&	15.0(1.7)&	27.6(1.1)	&	23.2(2.5)	&	56.1(0.8)	&2.93(12)\\
\hline															
average	&	90.0(0.2)	&	9.0(0.2)	&	13.9(0.1)	&	13.8(0.2)	&	43.6(0.1)	&1.42(2)\\
\hline
\end{tabular}
\end{center}
\end{table*}

%%%%%%%%%%%%%%%%%%%%%%%%%%%%%%%%%%%%%%%%%%%%%%%%%%%%%%%%%%%%%%%%%%%%%%%%%%%%%%%
\end{document}